\documentclass{LMCS}

\usepackage{graphicx}
\usepackage{color}
\usepackage{amssymb}
\usepackage{stmaryrd}
\usepackage{rotating}
\usepackage{cmll,enumerate}

\input{macros.sty}

\def\doi{5 (4:6) 2009}
\lmcsheading%
{\doi}
{1--61}
{}
{}
{Feb.~29, 2008}
{Dec.~22, 2009}
{}   

\begin{document}

\title[Interaction Combinators: Observational Equivalence and Full
  Abstraction]{Observational Equivalence and Full Abstraction in the
  Symmetric Interaction Combinators}

\author[D.~Mazza]{Damiano Mazza\rsuper*} 
\address{LIPN -- UMR 7030
  CNRS - Universit\'e Paris 13, F-93430 Villetaneuse, France}

\email{Damiano.Mazza@lipn.univ-paris13.fr}

\thanks{{\lsuper *}Partially supported by a post-doctoral fellowship
  of the \emph{Fondation Sciences Math\'ematiques de Paris}.}



\begin{abstract}
	\noindent The symmetric interaction combinators are an equally
        expressive variant of Lafont's interaction combinators. They
        are a graph-rewriting model of deterministic computation. We
        define two notions of observational equivalence for them,
        analogous to normal form and head normal form equivalence in
        the lambda-calculus. Then, we prove a full abstraction result
        for each of the two equivalences. This is obtained by
        interpreting nets as certain subsets of the Cantor space,
        called edifices, which play the same role as B\"ohm trees in
        the theory of the lambda-calculus.
\end{abstract}

\maketitle

\section*{Introduction}
\subsection*{A foundational study of interaction nets}
  Lafont's interaction nets \cite{Lafont:IN} are a powerful and
  versatile model of deterministic computation, derived from the proof
  nets of Girard's linear logic
  \cite{Girard:LL,Girard:ProofNets,Lafont:PNToIN}. Interaction nets
  are characterized by the atomicity and locality of their rewriting
  rules. As in Turing machines, computational steps are elementary
  enough to be considered as constant-time operations, but, unlike
  Turing machines, several steps can be executed in parallel, \ie,
  interaction nets actually model a kind of distributed computation.

  Several interesting applications of interaction nets exist. The most
  notable ones are implementations of optimal evaluators for the
  \lac\ \cite{Lamping,Mackie:EffEval}, but efficient evaluation of
  other functional programming languages using richer data structures
  is also possible with interaction nets~\cite{Mackie:AddMult}.

  However, so far the practical aspects of this computational model
  have arguably received much more attention than the strictly
  theoretical ones. With the exception of Lafont's work on the
  interaction combinators \cite{Lafont:INComb} and Fern\'andez and
  Mackie's work on operational equivalence
  \cite{FernandezMackie:OpEquiv}, no foundational study of interaction
  nets can be found in the existing literature. For example, until
  very recently \cite{Mazza:CombSem}, no denotational semantics had
  been proposed for interaction nets.

This work aims precisely at studying and expanding the theory of
interaction nets, in particular of the symmetric interaction
combinators. These latter are especially interesting because of their
\emph{universality}: any interaction net system can be translated in
the symmetric interaction combinators \cite{Lafont:INComb}. Therefore,
within the graph-rewriting paradigm given by interaction nets, the
symmetric interaction combinators stand out as a prototypical
language, just like the \lac\ is the prototypical language of the
functional paradigm.

More specifically, the contribution of this work is twofold, and can
be seen as a methodology for addressing the following two questions:
\begin{enumerate}[\hbox to8 pt{\hfill}]
\item\noindent{\hskip-12 pt\bf Observational equivalence:}\ Given two
  nets in the system of the symmetric interaction combinators, when
  can we say that they behave in the same way?
\item\noindent{\hskip-12 pt\bf Denotational semantics and full
  abstraction:}\ Any answer to the above question yields an
  equivalence on nets; can we denotationally characterize this
  equivalence? In other words, can we find an abstract interpretation
  of the syntax so that such equivalence becomes an equality? 
\end{enumerate}

\subsection*{Observations and contexts}
The first question is a central one in all programming languages. Indeed, any programmer is aware that, given two syntactically different programs, it may as well be that they ``do the same thing'', \ie, that one can be replaced by the other without anyone noticing the difference. Of course, the heart of the question lies in what differences we judge worth noticing: the final result of executing a program, the time it takes to obtain such result, etc. Different choices will of course lead to different notions of ``doing the same thing''.

In any case, the key notion is that of \emph{observation}: the program interacts with an environment, and we observe the outcome, based on our choice of what we consider relevant to be observed. If our two programs yield the same observations upon interacting with all possible environments, it is fair to say that, as far as we are concerned, they ``do the same thing''; more formally, we say that they are \emph{observationally equivalent}, according to our chosen notion of observation.

For functional programming languages, and in particular for the \lac, Morris~\cite{Morris} was the first to propose the now widely accepted idea that \emph{an environment is a context}. In this way, observations are internalized, \ie, they may be made directly on programs, because, given a program $P$ and a context $C$, $C[P]$ is still a program.

In interaction nets, there is a very natural notion of context for a net $\mu$: it is simply another net $C$ whose interface is big enough so that $\mu$ can be plugged into it, forming a new net $\ctxt\mu$. Morris' idea can therefore be straightforwardly applied in our framework.

\subsection*{Internal separation and observable axioms}
In order to choose what to observe in our nets, we draw inspiration from our previous work on internal separation~\cite{Mazza:CombSep}, in which we proved a result similar to the celebrated B\"ohm's theorem for the \lac~\cite{Bohm}. B\"ohm's theorem states the following: given two $\beta\eta$-normal \lat s $T,U$, $T\neq U$ implies that there exists a context $C$ such that $C[T]\red x$ and $C[U]\red y$, where $x$ and $y$ are two different variables.

A consequence of B\"ohm's result is that \emph{it is impossible to equate two distinct $\beta\eta$-normal forms, unless one equates all \lat s}. This makes us better understand the importance of B\"ohm's theorem, because it brings forth its negative content: as underscored for example by Giuseppe Longo~\cite{Longo:NegRes}, negative results are crucial in the development of a theory, since they witness the presence of a structure in the underlying objects. If ``everything is possible'', then the objects of our theory are shapeless, we can tamper with them at will, and the theory looses any scientific interest.

In the case of the symmetric interaction combinators, internal separation (\cf\ \refth{Separation}) cannot be realized using two arbitrary nets (by contrast, in the \lac, $x$ and $y$ may be replaced by two arbitrary distinct \lat s). Indeed, one of the two nets used contains a special kind of connection, which we call \emph{observable axiom}, while the other does not. Since identifying these two nets induces the identification of all nets, we are led to take the presence of observable axioms as the key phenomenon to observe.

\subsection*{Axiom-equivalences}
The discovery of observable axioms in interaction nets is the backbone of the development of our theory of observational equivalence. Indeed, we have several results hinting to the fact that observable axioms are analogous to \emph{head variables}; these occupy an arguably important place in the theory of the \lac, so it is perhaps not surprising that we give observable axioms a central role in interaction nets too.

Furthermore, observable axioms are related to a certain kind of paths of Girard's geometry of interaction~\cite{Girard:GoI1}, as reformulated by Lafont for the (symmetric) interaction combinators~\cite{Lafont:INComb}. In particular, it is possible to show (\cf\ \refsect{GoI}) that the observable axioms generated by a net in the course of its reduction correspond to its \emph{execution paths}~\cite{DanosRegnier:PNHilb}; these are the paths which are preserved by reduction, and are hence present in every reduct. In some sense, each execution path describes a portion of information produced by the computation of a net; in particular, if a net is normalizable, then its execution paths describe exactly its normal form. This idea of approximation, which is also already present in B\"ohm trees, is another way of looking at observable axioms as meaningful objects to study the behavior of a net (\cf\ \refsect{Meaningless}).

We thus introduce \emph{observable nets} and \emph{finitarily observable nets}: the first are nets which, in the course of their reduction, develop at least one observable axiom; the second are observable nets which develop only finitely many of them. It is useful to keep in mind an analogy with the \lac: observable nets are similar to \lat s having a head normal form, and finitarily observable nets are akin to normalizable \lat s. In the first case, we may additionally introduce the notion of \emph{solvable net}, and prove that solvable and observable nets coincide, just like solvable \lat s coincide with \lat s having a head normal form. In the second case, the correspondence is somewhat looser, because the symmetric combinators already have a notion of normalizable net, and it does not coincide with that of finitarily observable net. Nevertheless, there are several facts supporting this analogy.

The notions of observable and finitarily observable net can be used to define two observational equivalences on nets: \emph{axiom-equivalence} and \emph{finitary axiom-equivalence}. The first one is similar to head normal form equivalence (hnf-equivalence) in the \lac\ (two \lat s $T,U$ are hnf-equivalent iff, for every context $C$, $C[T]$ is head-normalizable iff $C[U]$ is). The second one is similar to normal form equivalence (nf-equivalence) in the \lac\ (two \lat s $T,U$ are nf-equivalent iff, for every context $C$, $C[T]$ is normalizable iff $C[U]$ is). By ``similar'' we mean that finitary axiom-equivalence is strictly included in axiom-equivalence, as nf-equivalence is strictly included in hnf-equivalence in the \lac, and that the examples proving strict inclusion are all related to a phenomenon similar to infinite $\eta$-expansion~\cite{Wadsworth}, as is the case for the \lac. Moreover, after transporting from the \lac\ to the symmetric interaction combinators the concepts of \emph{theory} and \emph{sensible theory} (\cf\ \refsect{Theories}), axiom-equivalence can be shown to be a maximal consistent theory, indeed the greatest consistent sensible theory, just like hnf-equivalence.

Our axiom-equivalences are not the only existing observational equivalences for the symmetric interaction combinators; in particular, Fern\'andez and Mackie~\cite{FernandezMackie:OpEquiv} proposed another notion of observational equivalence, based on \emph{visible nets}. This equivalence, which seems to correspond to \emph{weak} head normal form equivalence (whnf-equivalence) in the \lac, can be proved to be strictly stronger, \ie, more discriminative, than our finitary axiom-equivalence, and hence than axiom-equivalence (\cf\ \refsect{Comparison}). This is in accord with the \lac\ analogy: whnf-equivalence is strictly included in nf- and hnf-equivalence.

\subsection*{Equivalence as equality}
The second part of our work starts with the development of a denotational semantics of the symmetric interaction combinators. Denotational semantics originated in the late 1960's with the work of Scott and Strachey~\cite{ScottStr,Scott}. Its goal is to model the syntax of a programming language by means of a more abstract mathematical structure, on which a broader range of tools and proof techniques are available. In this way, one may be able to prove results about the language which would be very difficult, or even impossible, to prove by syntactic methods only.

In a nutshell, we could say that the ultimate goal of denotational semantics is \emph{to transform equivalences into equalities}. A typical example is precisely that of observational equivalence, as discussed above. If a denotational semantics gives the same interpretation to two programs exactly when they are observationally equivalent, then it is said to be \emph{fully abstract} with respect to the given observational equivalence. Finding a fully abstract denotational semantics can be a very hard problem: a notable example is that of $\mathrm{PCF}$, a \lac-like functional language for which completely new game-semantic models had to be developed to achieve full abstraction~\cite{AbramskyJagadeesanMalacaria,HylandOng}.

In the \lac, both nf- and hnf-equivalence have been abstractly characterized in several different ways: Hyland~\cite{Hyland} proved that two terms are nf-equivalent iff their B\"ohm trees are equal up to $\eta$-equivalence, and went on to prove that nf-equivalence coincides with equality in Plotkin's $P\omega$ model~\cite{Plotkin}; Wadsworth~\cite{Wadsworth} obtained similar results for hnf-equivalence, showing that two terms are hnf-equivalent iff their B\"ohm trees are equal up to \emph{infinite} $\eta$-expansion, and that this equivalence corresponds to equality in Scott's $D_\infty$ model~\cite{Scott}. Shortly after, Nakajima~\cite{Nakajima} introduced a similar characterization of hnf-equivalence in terms of what are now known as Nakajima trees.

\subsection*{Edifices, the Cantor topology, and full abstraction}
Besides the description of a new theory of observational equivalence for interaction nets, the other principal contribution of the present work is the introduction of \emph{edifices}, which play the same role as B\"ohm or Nakajima trees, in that they provide a fully abstract model of the two axiom-equivalences mentioned above.

The starting point for defining edifices is the same as that of B\"ohm trees, reflecting the analogy between observable axioms and head variables: just like the B\"ohm tree of a \lat\ $T$ is basically the collection of the head variables appearing in the reducts of $T$ (with the additional information concerning their hierarchical structure and the abstractions preceding them), the edifice of a net $\mu$ is built from the collection of all observable axioms appearing in the reducts of $\mu$.

Nevertheless, the parallelism of interaction nets, unmatched in the \lac, induces some fundamental differences between the two contructions: in fact, apart from collecting the information concerning the position of observable axioms within the net in which they appear (analogous to the abstractions preceding a head variable), no evident hierarchical structure emerges for observable axioms (although some kind of structure might be attached to them, as briefly discussed in \refsect{FurtherWork}). This is why edifices are not at all trees. Still, just like the B\"ohm tree of a \lat, the edifice of a net is an invariant of reduction in the symmetric interaction combinators (\refprop{BetaEtaEpsModel}).

To achieve full abstraction, we endow edifices with a topological structure, which turns them into subsets of the Cantor space. In the case of finitary axiom-equivalence, this is needed for technical purposes: in fact, edifices characterize this equivalence \emph{as plain sets}, \ie, two nets are finitarily axiom-equivalent iff their edifice is the same, independently of any topology attached to it; however, the only way we are able to prove this is through a topological property, namely the compactness of the edifices interpreting a certain class of nets (\refprop{Compactness}). In the case of axiom-equivalence, topology plays a more fundamental role: in fact, we prove that two nets are axiom-equivalent iff the \emph{topological closure} of their edifices is the same; obviously, the notion of closure is meaningless without referring to a topology. This last result is particularly nice, because the phenomenon of infinite \mbox{$\eta$-expansion} (which, as mentioned above, is also present in the symmetric interaction combinators) receives a precise topological explanation.

Another nice aspect of edifices is that they are quite interesting in their own right, independently of the symmetric interaction combinators. In fact, several results in our theory hold for wider classes of edifices than those which interpret nets. In particular, there is a notion of \emph{trace} defined on edifices (\refsect{Edifices}), which is completely general, and which reminds of the notion of composition of strategies in games semantics \cite{AbramskyJagadeesanMalacaria,HylandOng}. When applied to the special case of edifices which interpret nets, the trace can be seen as an extension of the execution formula of the geometry of interaction, which works in all cases (\cf\ \refprop{CutFreeTrace}); by contrast, Girard's original execution formula, and its rephrasing developed by Lafont for the symmetric interaction combinators, is only defined under certain normalizability assumptions.

\subsection*{Acknowledgments}
Many thanks to the anonymous referees for their useful comments and suggestions, and a special thanks to the editor Simona Ronchi della Rocca for her patience in waiting for the revised version of this paper.

\section{The Symmetric Interaction Combinators}
\label{sect:TheSymmIC}

\subsection{Nets}
\label{sect:NetsAndRed}
The symmetric interaction combinators, or, more simply, the symmetric combinators, are an interaction net system \cite{Lafont:IN,Lafont:INComb}. An interaction net is built out of \emph{cells} and \emph{wires}. Each cell has a number of \emph{ports}, exactly one of which is \emph{principal}, the other being \emph{auxiliary}. In the case of the symmetric combinators, there are three kinds of cells: cells of type $\D$ and $\Z$, which have two auxiliary ports, numbered by the integers $1$ and $2$, and cells of type $\E$, which have no auxiliary ports. Cells of the first two kinds are called \emph{binary}, while those of the latter kind are called \emph{nullary}.

Each wire has two extremities; each extremity may be attached to the port of a cell, so we can use wires to connect cells together. We also allow \emph{loops}, which are wires whose extremities are attached one to the other. Wires which are not loops are called \emph{proper}.

A \emph{net} is any configuration of cells and wires, such that each port of each cell is attached to the extremity of a wire. Note that a net may contain wires with one or both extremities not attached to any port of any cell; these unattached extremities will be called the \emph{free ports} of the net.

\begin{figure}[t]
	\begin{center}\scalebox{0.8}{\input{ANet.pstex_t}}\end{center}
	\caption{A net.}
	\label{fig:ANet}
\end{figure}
Nets are usually presented in graphical form, as in \reffig{ANet}. Binary cells are represented by triangles, nullary cells by circles; in both cases, the symbol denoting the kind of cell is written inside the figure representing it. For a binary cell, the principal port is depicted as one of the ``tips'' of the triangle representing it. The numbering of the auxiliary ports of binary cells is assigned clockwise: in particular, the auxiliary port number $1$ of a binary cell is the left one if the cell is drawn with its principal port pointing towards the bottom of the picture, and it is the right one if the cell is drawn with its principal port pointing ``up''. Wires and loops are represented as\ldots\ wires and loops, and the free ports appear as extremities of ``pending'' wires. For example, the net in \reffig{ANet} has 11 cells, of which 4 nullary, 1 loop, 16 proper wires, and 7 free ports.

The above description is precise enough to develop the rest of the paper, and almost all of the theory of interaction nets. However, a more formal definition can be given, by considering an interaction net as the union of two structures: a labelled, directed hypergraph, and an undirected graph. The idea is that labelled and directed hyperedges correspond to cells, and undirected edges to wires. In what follows, we fix a denumerably infinite set of \emph{ports}, which we assume contains the positive integers.
\begin{defi}[Wire, cell, net]
	\label{def:Net}
	A \emph{wire} is a set of ports of cardinality $1$ or $2$; in the first case, we speak of a \emph{loop}, in the second case of a \emph{proper wire}. We fix three \emph{symbols} $\D,\E,\Z$; we say that $\D$ and $\Z$ are binary, while $\E$ is nullary. A \emph{cell} is a tuple $(\alpha,p_0,p_1,\ldots,p_n)$ where $\alpha$ is a symbol, $p_0,p_1,\ldots,p_n$ are ports, and $n=2$ if $\alpha$ binary, or $n=0$ if $\alpha$ is nullary. In both cases, $p_0$ is the \emph{principal port} of the cell, while $p_1,\ldots,p_n$ are the \emph{auxiliary ports}.
	
	A \emph{net} $\mu$ is a couple $(\Cells{\mu},\Wires{\mu})$, where $\Cells{\mu}$ is a finite set of cells and $\Wires{\mu}$ is a finite set of wires, satisfying the following:
	\begin{enumerate}[$\bullet$]
		\item each port appears at most twice in $\Cells{\mu}\cup\Wires{\mu}$;
		\item if a port appears in $\Cells{\mu}\cup\Wires{\mu}$, then it appears in exactly one wire.
	\end{enumerate}
	The set of ports appearing in $\mu$ is denoted by $\Ports{\mu}$. A port appearing only once in $\Cells{\mu}\cup\Wires{\mu}$ is called \emph{free}; the set of all free ports of $\mu$ is referred to as its \emph{interface}. We shall always assume that if a net has $n$ free ports, then its interface is $\{1,\ldots,n\}$.
\end{defi}
\begin{defi}[$\alpha$-equivalence]
	A \emph{renaming} for a net with $n$ free ports is an injective function from ports to ports which is the identity on $\{1,\ldots,n\}$. Two nets are \emph{$\alpha$-equivalent} iff they are equal modulo a renaming.
\end{defi}
Nets are always considered modulo $\alpha$-equivalence. In fact, observe that graphical representations equate exactly $\alpha$-equivalent nets. As in most of the existing literature on interaction nets, we shall preferentially disregard \refdef{Net}, in favor of more intuitive graphical notations. This is especially convenient for treating the dynamic aspects of nets, such as reduction (\cf\ \refsect{Red}); however, for static aspects, it is sometimes quite convenient to use \refdef{Net}, because it gives succinct, formal descriptions of the components of a net (\eg\ ports, wires as sets of ports, etc.).

Let us introduce some remarkable nets, which will be useful in the sequel:
\begin{enumerate}[\hbox to8 pt{\hfill}]
\item\noindent{\hskip-12 pt\bf Wirings:}\ A net containing no cell and no loop is called a \emph{wiring}. Wirings are permutations of free ports; they are ranged over by $\omega$. We shall often use $\omega$ also to denote a single wire.
	\item\noindent{\hskip-12 pt\bf $\textrm E$-nets:}\ The $\E$-net with $n$ free ports, denoted by $\Epsilon{n}$, is the net consisting of $n$ $\E$ cells;
	\item\noindent{\hskip-12 pt\bf Trees:}\ A \emph{tree} is a net defined by induction as follows. A single $\E$ cell is a tree with no leaf, denoted by $\E$; a proper wire is a tree with one leaf (it is arbitrary which of the two extremities is the root and which is the leaf), denoted by $\OneTree$; if $\tau_1,\tau_2$ are two trees with resp.\ $n_1,n_2$ leaves, and if $\alpha$ is a binary symbol, the net
	\begin{center}\scalebox{0.8}{\input{Tree.pstex_t}}\end{center}
	is a tree with $n_1+n_2$ leaves, denoted by $\alpha(\tau_1,\tau_2)$.

	As the reader may have noticed, in the above picture we represented trees adopting the same graphical notations as cells. We shall avoid possible ambiguities by never using $\D,\E,\Z$ to denote trees, and by using $\alpha,\beta$ exclusively to range over cell symbols, so that a triangle annotated with $\alpha$ or $\beta$ will unambiguously represent a cell (or a tree consisting of a single cell, if the reader prefers).
	\begin{figure}[t]
		\begin{center}\scalebox{0.8}{\input{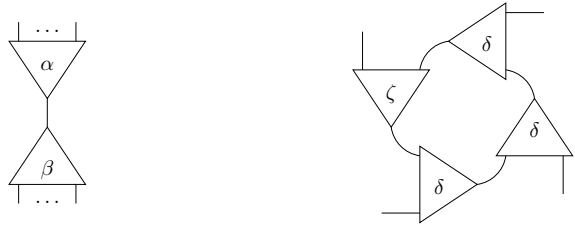}}\end{center}
		\caption{A generic active pair (left) and an example of vicious circle (right).}
		\label{fig:ActPairViciousCirc}
	\end{figure}
	\item\noindent{\hskip-12 pt\bf Active pairs:}\ An \emph{active pair} (\reffig{ActPairViciousCirc}, left) is a net consisting of two cells whose principal ports are connected by a wire.
	\item\noindent{\hskip-12 pt\bf Vicious circles:}| A \emph{vicious circle} is either a loop, or a net consisting of $n$ binary cells $c_1,\ldots,c_n$ such that, for all \mbox{$i\in\{1,\ldots,n-1\}$}, the principal port of $c_i$ is connected to an auxiliary port of $c_{i+1}$, and the principal port of $c_n$ is connected to an auxiliary port of $c_1$. An example is given in \reffig{ActPairViciousCirc} (right).
\end{enumerate}
It is also useful to identify two special sorts of wires in nets:
\begin{defi}[Axiom, cut, cut-free net]
	Let $\mu$ be a net, and $\omega=\{p,q\}$ a wire of $\mu$.
\begin{enumerate}[\hbox to8 pt{\hfill}]
\item\noindent{\hskip-12 pt\bf Proper axiom:}\ We say that $\omega$ is a \emph{proper axiom} if it is a proper wire and none of $p,q$ is the principal port of a cell of $\mu$.
		\item\noindent{\hskip-12 pt\bf Proper cut:}\ We say that $\omega$ is a \emph{proper cut} if it is a proper wire and both $p$ and $q$ are principal ports of cells of $\mu$.
		\item\noindent{\hskip-12 pt\bf Axiom-cut:}\ We say that $\omega$ is an axiom-cut if it is a loop, or if $\mu$ contains a tree $\tau$ such that $p$ is the root of $\tau$, and $q$ one of its leaves.
	\end{enumerate}
	An \emph{axiom} (resp.\ cut) of $\mu$ is either a proper axiom (resp.\ proper cut) or an axiom-cut; in the latter case, we refer to it as an \emph{improper axiom} (resp.\ \emph{improper cut}). We say that $\mu$ is \emph{cut-free} if it contains no cuts.
\end{defi}
As an example, consider the net in \reffig{ANet}, in which the reader should find 7 proper axioms, 2 proper cuts, and 2 axiom-cuts. Note that proper cuts are in one-to-one correspondence with active pairs. On the other hand, axiom-cuts are in many-to-one correspondence with vicious circles, \ie, an axiom-cut implies the presence of exactly one vicious circle, but a vicious circle implies the presence of at least one axiom-cut. Although the correspondence is one-to-one for the net in \reffig{ANet}, we have for instance that, despite having a single vicious circle, the net on the left in \reffig{ActPairViciousCirc} contains 4 axiom-cuts.

The following gives us a general understanding of the structure of cut-free nets:
\begin{lem}[Canonical form of a cut-free net]
	\label{lemma:CanForm}
	Let $\nu$ be a cut-free net with $n$ free ports. Then, for each $1\leq i\leq n$ there exist a unique tree $\tau_i$, and there exists a unique wiring $\omega$ such that
	\begin{center}\scalebox{0.8}{\input{CutFreeNet.pstex_t}}\end{center}
\end{lem}
\proof By induction on the number of cells of $\nu$.\qed
Note the wiring drawn as a rectangle; in the sequel, this graphical notation will be used also to represent generic nets, but $\omega$ will always denote a wiring. Observe that all wires in $\omega$ are proper axioms; in fact, the above is the shape of a generic cut-free multiplicative proof net, with the axiom links in $\omega$ and the logical links in $\tau_1,\ldots,\tau_n$, whence our terminology.

A fundamental notion for developing the rest of the paper is that of context:
\begin{defi}[Context, test, feedback]
	Let $\mu$ be a net with $n$ free ports. A \emph{context} for $\mu$ is a net $C$ with at least $n$ free ports. We denote by $\ctxt{\mu}$ the \emph{application} of $C$ to $\mu$, which is the net obtained by plugging the free port $i$ of $\mu$ to the free port $i$ of $C$, with $i\in\{1,\ldots,n\}$. A \emph{test} for $\mu$ is a particular context consisting of $n$ trees $\tau_1,\ldots,\tau_n$ such that the root of each $\tau_i$ is the free port~$i$. A \emph{feedback context} for $\mu$ is a context $\sigma$ consisting of a wiring connecting some of the free ports of $\mu$ between them.
\end{defi}
In the sequel, when we use the notation $\ctxt{\mu}$ we implicitly assume that $C$ is a context for $\mu$, \ie, that it has enough ports so that $\mu$ can be plugged into it. Moreover, we shall say that $\mu'$ is a \emph{subnet} of $\mu$ if there exists $C$ such that $\mu=\ctxt{\mu'}$. Using the above definitions, we can concisely formulate a decomposition result which, combined with \reflemma{CanForm}, uncovers the structure of a generic net:
\begin{lem}[Decomposition]
	\label{lemma:Decomposition}
	Let $\mu$ be a net. Then, there exists a cut-free net $\nu$ and a feedback context $\sigma$ such that $\mu=\Ctxt{\sigma}{\nu}$.
\end{lem}
\proof Simply let $\sigma$ contain all the proper cuts of $\mu$, plus one axiom-cut for each vicious circle of $\mu$, and let $\nu$ be the subnet of $\mu$ obtained by removing $\sigma$.\qed
Observe that the net $\nu$ of \reflemma{Decomposition} is unique as soon as $\mu$ does not contain vicious circles (or, equivalently, axiom-cuts).

\subsection{$\beta$-reduction and $\eta$-equivalence}
\label{sect:Red}
\begin{figure}[t]
	\begin{center}\scalebox{0.8}{\input{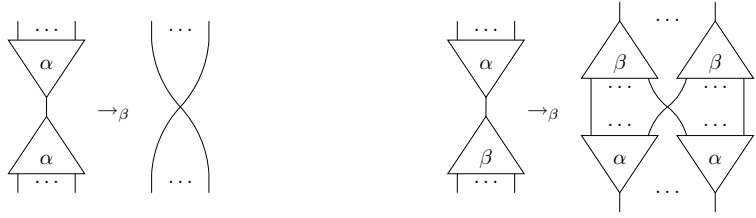}}\end{center}
	\caption{The interaction rules: annihilation (left) and commutation (right). In the annihilation, the right member is empty in case $\alpha=\E$.}
	\label{fig:Rules}
\end{figure}
The dynamics of interaction nets is based on the contextual rewriting of active pairs, which are thus analogous to redexes in the \lac. In the case of the symmetric combinators, the active pairs are rewritten according to the \emph{interaction rules} of \reffig{Rules}: the annihilations, concerning the interaction of two cells of the same type, and the commutations, concerning the interaction of two cells of different type.
\begin{defi}[$\beta$-reduction and $\beta$-equivalence]
	\emph{$\beta$-reduction} is the reflexive-transitive closure of the relation defined as follows: given two nets $\mu,\mu'$, we set $\mu\onered\mu'$ iff there exists $C$ such that $\mu=\ctxt{\mu_0}$, $\mu'=\ctxt{\mu_0'}$, and $\mu_0,\mu_0'$ match the left and right members of one of the rules of \reffig{Rules}, respectively. We define $\mu\BetaEq\nu$ iff there exists $o$ such that $\mu\red o$ and $\nu\red o$.
\end{defi} 
\begin{prop}[Strong confluence]
	\label{prop:StrongConfluence}
	If $\mu\onered\mu'$ and $\mu\onered\mu''$ with $\mu'\neq\mu''$, there exists $\nu$ such that $\mu'\onered\nu$ and $\mu''\onered\nu$. Hence, the relation $\red$ is confluent, and $\BetaEq$ is an equivalence relation.
\end{prop}
\proof Immediate: there are no critical pairs, because active pairs are always disjoint.\qed

We now give a few basic results concerning $\beta$-reduction. The first two are generalizations of the annihilation and commutation rules:
\begin{lem}
	\label{lemma:TreeAnnihilation}
	Let $\tau$ be a tree. Then, we have
	\begin{center}\scalebox{0.8}{\input{TauTauSymm.pstex_t}}\end{center}
\end{lem}
\proof By induction on the structure of $\tau$.\qed
\begin{lem}
	\label{lemma:TreeCommutation}
	Let $\alpha,\beta$ range over binary symbols, with $\alpha\neq\beta$. Let $A$ be a tree not containing $\beta$ cells, and let $B$ be a tree not containing $\alpha$ cells. Then, we have
	\begin{center}\scalebox{0.8}{\input{BigCommutation.pstex_t}}\end{center}
\end{lem}
\proof By double induction on the structures of $A$ and $B$.\qed
The following is an easy corollary of Lemmas~\ref{lemma:CanForm}, \ref{lemma:TreeAnnihilation}~and~\ref{lemma:TreeCommutation}:
\begin{lem}[Duplication]
	\label{lemma:Duplication}
	Let $\alpha$ be a binary symbol, let $\nu$ be a cut-free net containing no $\alpha$ cell, and let $\tau$ be a tree containing only $\alpha$ cells. Then, we have
	\begin{center}\scalebox{0.8}{\input{Duplication.pstex_t}}\end{center}
\end{lem}
Observe that the only cut-free net with an empty interface is the empty net. Then, the next result shows, as a special case, that cut-free nets can be freely erased:
\begin{lem}[Erasing]
	\label{lemma:Erasing}
	Let $\nu$ be a cut-free net, and let $\mu$ be any net obtained from $\nu$ by plugging any number of $\E$ cells to its free ports, as follows:
	\begin{center}\scalebox{0.8}{\input{Erasing.pstex_t}}\end{center}
	Then, there exists a cut-free net $\nu'$ such that $\mu\red\nu'$.
\end{lem}
\proof An immediate consequence of Lemmas~\ref{lemma:CanForm} and \ref{lemma:TreeCommutation}.\qed

\begin{figure}[t]
	\begin{center}\scalebox{0.8}{\input{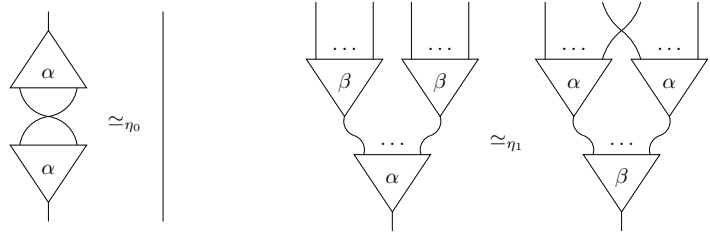}}\end{center}
	\caption{The equations defining $\eta$-equivalence. In both equations, $\alpha,\beta$ range over cell symbols. In the left equation, $\alpha$ is binary; in the right equation, $\alpha\neq\beta$.}
	\label{fig:Eta}
\end{figure}

We now introduce $\eta$-equivalence, which is similar to the homonymous relation in the \lac, with an essential difference: in the symmetric combinators $\eta$-equivalence cannot be presented as the symmetrization of a rewriting relation, like $\beta$-equivalence. In fact, one of the equations defining it (namely the $\eta_1$ equation applied to binary cells, \cf\ \reffig{Eta}) cannot be meaningfully oriented and transformed into a rewriting step. The $\eta_1$ equation was already known to Lafont~\cite{Lafont:INComb}; the $\eta_0$ equation was introduced by Fern\'andez and Mackie~\cite{FernandezMackie:OpEquiv}.
\begin{defi}[$\eta$-\ and $\beta\eta$-equivalence]
	We define the relations $\EtaEqZero$ and $\EtaEqOne$ as the reflexive, transitive, and contextual closure of respectively the left and right equation of \reffig{Eta}, which we call $\eta_0$ and $\eta_1$ equation, respectively. Then, we define $\eta$-\ and $\beta\eta$-equivalence respectively as \mbox{$\EtaEq=(\EtaEqZero\cup\EtaEqOne)^+$} and \mbox{$\BetaEtaEq=(\BetaEq\cup\EtaEq)^+$}.
\end{defi}

The following results are the counterparts of Lemmas~\ref{lemma:TreeAnnihilation} and \ref{lemma:TreeCommutation} for $\eta$-equivalence.
\begin{lem}
	\label{lemma:TreeAnnEta}
	Let $\tau$ be a tree without $\E$ cells. Then, we have
	\begin{center}\scalebox{0.8}{\input{TreeEta.pstex_t}}\end{center}
\end{lem}
\proof By induction on the structure of $\tau$.\qed
\begin{lem}
	\label{lemma:TreeCommEta}
	Let $\alpha,\beta$ range over binary symbols, with $\alpha\neq\beta$. Let $A$ be a tree not containing $\beta$ cells, and let $B$ be a tree not containing $\alpha$ cells. Then, we have
	\begin{center}\scalebox{0.8}{\input{BigEtaOne.pstex_t}}\end{center}
\end{lem}
\proof By double induction on the structure of $A$ and $B$.\qed
An easy corollary of \reflemma{TreeAnnEta} is that, modulo $\eta$-equivalence, the trees rooted at the free ports of a net in the decomposition given by Lemmas~\ref{lemma:CanForm} and \ref{lemma:Decomposition} can ``look like'' almost anything we want:
\begin{lem}
	\label{lemma:Equiv}
	For any net $\nu$ and for any trees without $\E$ cells $\tau_1,\ldots,\tau_n$, there exists a net $\nu'$ such that
	\begin{center}\scalebox{0.8}{\input{EquivLemma.pstex_t}}\end{center}
\end{lem}
\proof Simply ``$\eta$-expand'' the wires connected to the free ports of $\nu$ as in \reflemma{TreeAnnEta}.\qed

When oriented from right to left, the $\eta_0$ equation is formally identical to the standard $\eta$-expansion rule in multiplicative proof nets (see for example the work of Pagani~\cite{Pagani:Mult}), except that it may be applied to any wire in a net, while $\eta$-expansion in proof nets concerns only axioms. Actually, it is possible to show that, when combined with $\beta$-equivalence, the $\eta_0$ equation still generates $\beta\eta$-equivalence even if its application is limited to axioms.
\begin{lem}
	\label{lemma:EtaMinus}
	Let $\mu\etazerored^-\nu$ iff $\mu=\ctxt{\mu_0}$ and $\nu=\ctxt{\omega}$, where $\mu_0$ is a net matching the left member of the $\eta_0$ equation of \reffig{Eta}, and $\omega$ is an axiom of $\nu$. Let $\EtaEqZero^-$ be the reflexive-transitive closure of the symmetric closure of $\etazerored^-$, and let $\EtaEq^-=(\EtaEqZero^-\cup\EtaEqOne)^+$. Then, $(\BetaEq\cup\EtaEq^-)^+=\ \BetaEtaEq$.
\end{lem}
\proof Let us set $\BetaEtaEq^-\,=\,(\BetaEq\cup\EtaEq^-)^+$. The inclusion $\BetaEtaEq^-\,\subseteq\,\BetaEtaEq$ is obvious, because by definition $\EtaEq^-\,\subseteq\,\EtaEq$. For what concerns the reverse inclusion, it is enough to show that if $\nu$ is obtained from $\mu$ by a single application of the $\eta_0$ equation on a wire which is not an axiom, then $\mu\BetaEtaEq^-\nu$. We can assume witout loss of generality that such a wire is in $\nu$. Then, we must have $\mu=\ctxt{\mu_0}$ and $\nu=\ctxt{\tau}$, where $\tau$ is a tree and
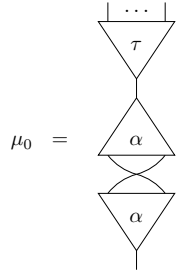
\begin{center}\scalebox{0.8}{\input{AxEta.pstex_t}}\end{center}
We need to prove that $\ctxt{\mu_0}\BetaEtaEq^-\ctxt{\tau}$. If $\tau=\E$, we leave it to the reader to check that $\ctxt{\mu_0}\onered\ctxt{\mu_0'}\EtaEqOne\ctxt\E$. Otherwise, we can assume $\tau$ to be ``maximal'', \ie, its leaves are either free or connected to an auxiliary port. In fact, this is not possible only if one of the leaves of $\tau$ is connected to the principal port of the $\alpha$ cell of $\mu_0$ shown at the bottom of the picture; but in this case the wire in $\nu$ obtained after applying the $\eta_0$ equation would be an axiom, against our hypothesis. Now, $\tau$ can always be decomposed as follows:
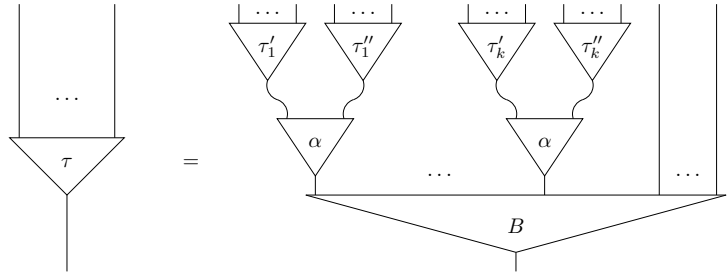
\begin{center}\scalebox{0.8}{\input{TreeDec.pstex_t}}\end{center}
where $B$ is a tree not containing $\alpha$ cells (we may have $B=\OneTree$), and $\tau_1',\ldots,\tau_k'$, $\tau_1'',\ldots,\tau_k''$ are trees ($k$ may be equal to zero). By \reflemma{TreeCommutation}, we have
\begin{center}\scalebox{0.8}{\input{AxEtaRed.pstex_t}}\end{center}
By \reflemma{TreeCommEta}, we have
\begin{center}\scalebox{0.8}{\input{AxEta1.pstex_t}}\end{center}
Now, if we apply the $\eta_0$ equation to the nets on the upper right of the above picture, the resulting wires will be axioms, because of the ``maximality'' of $\tau$. Hence, $\ctxt{\mu_0'}\EtaEq^-\ctxt{\tau}$, as desired.\qed

As recalled in the introduction, a fundamental result due to B\"ohm~\cite{Bohm} implies that no non-trivial congruence on \lat s may equate two distinct $\beta\eta$-normal forms. In the symmetric combinators there is a similar result~\cite{Mazza:CombSep}, except that one cannot speak of $\beta\eta$-normal forms, because, as discussed above, the symmetric combinators lack a notion of $\eta$-reduction.

There is actually a deeper difference between the symmetric combinators and the \lac, given by the existence of vicious circles. Observe that such configurations are stable under \mbox{$\beta$-reduction}, because cells can interact only through their principal port: they are sort of deadlocks. Although diverging computations certainly exist in the \lac, deadlocks are something completely new. Because of this, the notion of normalizable net (which, thanks to strong confluence, is the same as that of strongly normalizable net) does not play a central role in the theory of the symmetric combinators. Instead, cut-free nets are closer to a concept of ``true'' normal form: a net having no cut-free form represents either a diverging or an error-bound computation, \ie, one that generates deadlocks. Other interesting notions of convergence will be introduced in \refsect{ObsEq}, but none of them will coincide with simple normalization.

In the following, we say that a net is \emph{total} if it $\beta$-reduces to a cut-free net.
\begin{thm}[Separation \cite{Mazza:CombSep}]
	\label{th:Separation}
	Let $\mu,\nu$ be two total nets with the same interface, such that $\mu\not\BetaEtaEq\nu$. Then, there exists a test $\theta$ such that
	\begin{center}\scalebox{0.8}{\input{Separation.pstex_t}}\end{center}
	or vice versa.\qed
\end{thm}
The net on the right in \refth{Separation} is the one we denoted by $\Epsilon 2$; we denote the other, \ie, a single wire, by $\omega$. The following makes us understand the strength of the Separation Theorem:
\begin{prop}
	\label{prop:Collapse}
	Let $\GenEq$ be a congruence on nets, such that $\BetaEq\,\subseteq\,\GenEq$. Then, $\omega\GenEq\Epsilon 2$ implies that, for all $\mu,\nu$ with the same interface, $\mu\GenEq\nu$.
\end{prop}
\proof Let $\mu$ be a net with $n$ free ports. By the Decomposition \reflemma{Decomposition}, we have $\mu=\Ctxt{\sigma}{\nu}$ for some feedback context $\sigma$ and some cut-free net $\nu$. Since $\GenEq$ is a congruence, we can write
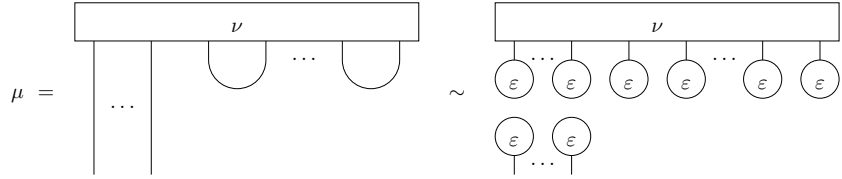
\begin{center}\scalebox{0.8}{\input{Collapse.pstex_t}}\end{center}
But by \reflemma{Erasing}, the net on the right $\beta$-reduces to $\Epsilon{n}$, where $\Epsilon{n}$ is a net with $n$ free ports containing $n$ $\E$ cells. Since $\GenEq$ contains $\beta$-reduction, we have proved that $\mu\GenEq\Epsilon{n}$ for any $\mu$ with $n$ free ports, and we may conclude by symmetry and transitivity of $\GenEq$.\qed
Therefore, the Separation Theorem implies that any non-trivial congruence containing $\beta$-equivalence cannot equate two \emph{total} $\beta\eta$-different nets. In particular, on total nets, such a congruence must be contained in $\BetaEtaEq$. The Separation Theorem will be fundamental in guiding us towards a definition of observational equivalence (\refsect{ObsEq}).

\subsection{Expressiveness}
\label{sect:Expressiveness}
In strictly computational terms, the interest of the symmetric combinators is given by the following result:
\begin{thm}[Lafont \cite{Lafont:INComb}]
	\label{th:Universality}
	Any interaction net system can be translated in the symmetric combinators.\qed
\end{thm}
The definitions of interaction net system and of the notion of translation are out of the scope of this paper. We shall only say that, modulo an encoding, Turing machines, cellular automata, and the \SK\ combinators are all examples of interaction net systems~\cite{Lafont:INComb,Mazza:CombSem}. An example of encoding of linear logic and the \lac\ in the symmetric combinators\footnote{Actually these encodings use the interaction combinators, but they can be adapted with very minor changes to the symmetric combinators.} is given by Mackie and Pinto~\cite{MackiePinto:LLComb}. We refer the reader to Lafont's paper \cite{Lafont:INComb} for a proper formulation and proof of \refth{Universality}.

However, to give an idea of the expressive power of the symmetric combinators, we shall show how general recursion can be implemented in the system, \ie, we shall see how all recursive relations of the form
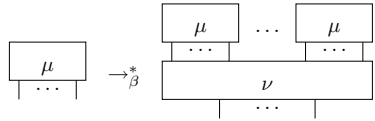
\begin{center}\scalebox{0.8}{\input{GeneralRecursion.pstex_t}}\end{center}
may be solved. In the \lac, the above relation would correspond to
$$M\red N[M/x]$$
where $x$ appears free in $N$. It is well known that a general solution can be given by resorting to a fixpoint combinator, \ie, a term $\Theta$ such that, for all $T$, $\Theta T\red T(\Theta T)$. Then, a solution to the above recursive relation would be $M=\Theta(\lambda x.N)$.

A necessary condition for having a fixpoint combinator is the ability of duplicating any term. In the symmetric combinators, we are only able to duplicate cut-free nets as in \reflemma{Duplication}, so we do not have a fixpoint combinator at our disposal. To compensate for this, we use a fundamental construction due to Lafont~\cite{Lafont:INComb}, and a generalization of it, first considered by Fernandez and Mackie~\cite{FernandezMackie:Packing}.

\begin{figure}[t]
	\begin{center}\scalebox{0.8}{\input{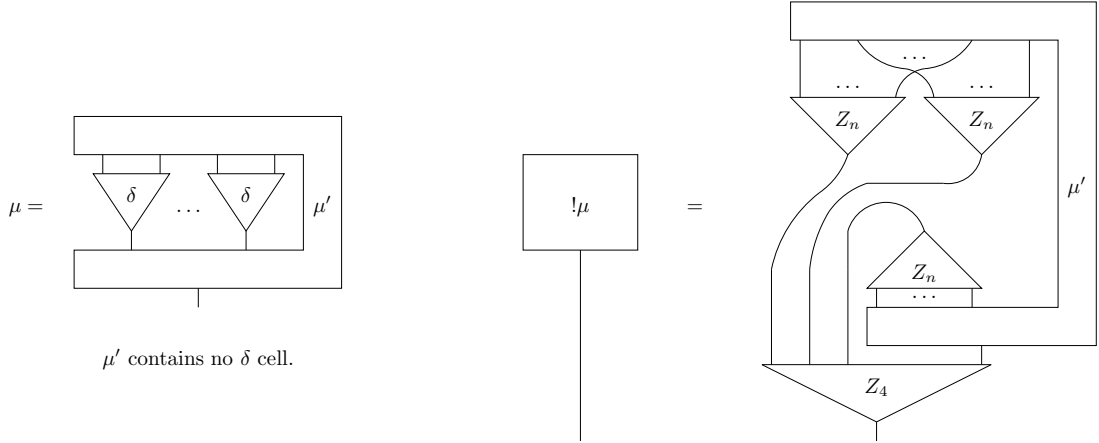}}\end{center}
	\caption{The Lafont code of a net.}
	\label{fig:Bang}
\end{figure}
\begin{figure}[t]
	\begin{center}\scalebox{0.8}{\input{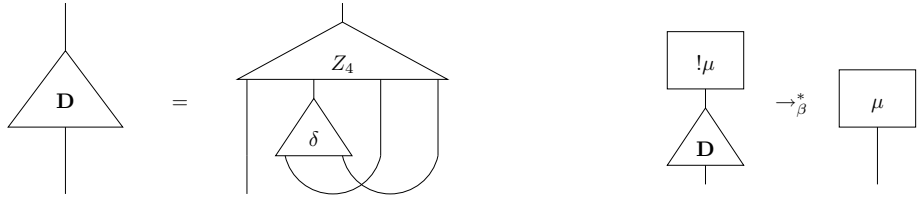}}\end{center}
	\caption{The universal decoder for the Lafont code.}
	\label{fig:Decoder}
\end{figure}
Given a net $\mu$ with one free port and containing $n$ cells of type $\D$, we build a net $\oc\mu$, called the \emph{Lafont code} of $\mu$, as in \reffig{Bang}. The $\Mux k$ are trees of $\Z$ cells, having $k$ leaves. We take $\Mux 0$ to be equal to one $\E$ cell; the actual shape of $\Mux k$ for $k>0$ is not important, as long as one tree is fixed for each $k$. Observe then that, by construction, a Lafont code never contains $\D$ cells. The net $\mu$ can be recovered from its Lafont code by means of a ``universal decoder'', \ie, independent of $\mu$, as in \reffig{Decoder}.

\begin{figure}[t]
	\begin{center}\scalebox{0.8}{\input{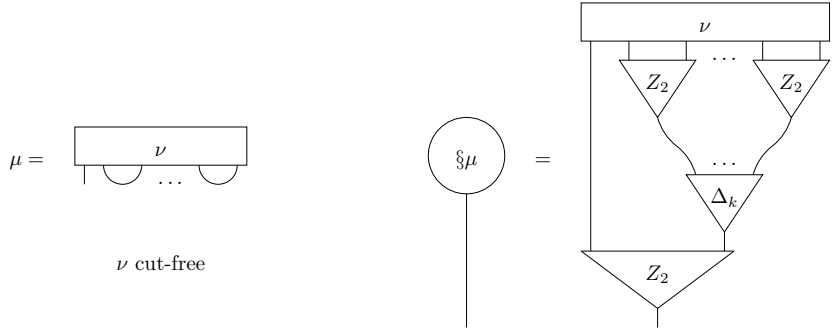}}\end{center}
	\caption{The cut-free code of a net.}
	\label{fig:Paragraph}
\end{figure}
\begin{figure}[t]
	\begin{center}\scalebox{0.8}{\input{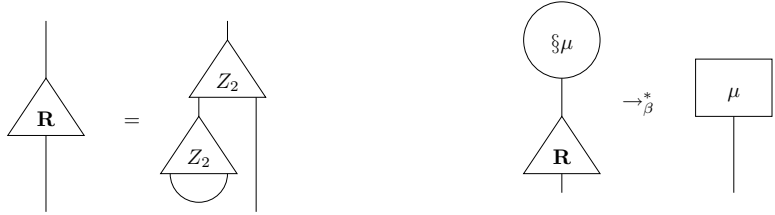}}\end{center}
	\caption{The universal decoder for the cut-free code.}
	\label{fig:Recover}
\end{figure}
A similar construction removes active pairs and vicious circles, and is given in \reffig{Paragraph}. This construction uses the Decomposition \reflemma{Decomposition}, and the trees $\Delta_k$ are defined as the trees $Z_k$, but using $\D$ cells instead of $\Z$ cells. The net $\S\mu$ is called the \emph{cut-free code} of $\mu$. Recovering $\mu$ from its cut-free code can be done as in \reffig{Recover}. Rigorously speaking, $\S\mu$ is not well defined because, given a net $\mu$, the cut-free net $\nu$ is not unique in general. However, the recovery process works regardless of the particular $\nu$ we chose for the cut-free code, so the abuse of notation is not problematic.

We shall take the net $\oc\S\mu$ as the \emph{code} of $\mu$. Decoding is done by composing the nets $\Decoder$ and $\Recover$ of Fig.~\ref{fig:Decoder}~and~\ref{fig:Recover}, respectively; we denote by $\FullDecoder$ the net resulting from their composition. Observe that the code of a net is cut-free and does not contain $\D$ cells. Hence, \reflemma{Duplication} applies to it, and it can be freely duplicated by means of trees of $\D$ cells.

\begin{figure}[t]
	\begin{center}\scalebox{0.8}{\input{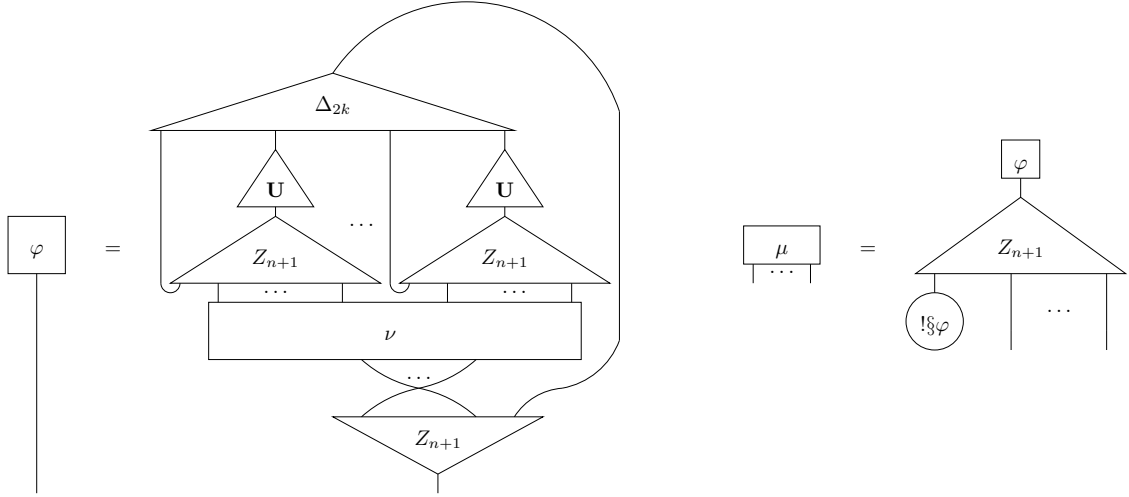}}\end{center}
	\caption{Solving recursive relations.}
	\label{fig:SolvingGenRec}
\end{figure}
We leave it as an instructive exercise for the reader to check that, using Lemmas~\ref{lemma:TreeAnnihilation}, \ref{lemma:TreeCommutation}, and~\ref{lemma:Duplication}, the net $\mu$ of \reffig{SolvingGenRec} is a solution to the recursive relation introduced above (we have supposed that $\mu$ has $n$ free ports and that there are $k$ copies of $\mu$ on the right hand side of the relation).

\section{Observational Equivalence}
\label{sect:ObsEq}

\subsection{Observable axioms}
\label{sect:ObsAx}
We already discussed in the introduction that in the \lac, and in functional programming in general, there is a standard way of defining an observational equivalence, which was first proposed by Morris~\cite{Morris}. The key idea is to put a term (\ie, a program) in a context (\ie, an environment) and to observe its behavior with respect to some interesting property (for example, termination).

In a more abstract way, once we have a language with an internal notion of context, we may take any set $S$ of objects of the language and define $S$-equivalence as follows: two objects $a,b$ of the language are $S$-equivalent iff, for every context $C$, $C[a]\in S$ iff $C[b]\in S$. In other words, we partition the set of all objects into two classes, and we observe the contextual behavior of objects with respect to these two classes. Usually, the property defining $S$ (and thus its complementary set) is referred to as an ``observable''.

As anticipated above, a typical observable is termination. In fact, Morris himself considered what we shall call \emph{nf-equivalence}, which is obtained by taking $S$ to be the set of normalizable \lat s. Another fundamental observable is head-termination, \ie, we may define \emph{hnf-equivalence} by taking $S$ to be the set of head-normalizable \lat s, which is the same as the set of solvable \lat s. For other examples of observables and for an account of Morris-like observational equivalences in the context of the \lac, we refer the reader to Dezani-Ciancaglini and Giovannetti's survey \cite{DezaniGiovannetti}.

We have shown that in the symmetric combinators, and in interaction nets in general, we have an internal notion of context, and we can therefore hope of applying the above ideas to generate interesting notions of observational equivalence: all we need is finding the right observables. We have already remarked in \refsect{Red} that termination is not an interesting observable in interaction nets, because of vicious circles. However, the Separation Theorem may give us a hint: in fact, it distinguishes two nets by sending one to a net presenting a direct connection between its free ports, and the other to a net in which no such direct connection will ever form. This points out that the appropriate notion of ``connection'' may be the right thing to observe.

Intuitively, an axiom in a net is observable when it can be ``extracted'' from the net through interaction, as in the Separation \refth{Separation}. We formalize this intuition as follows.
\begin{defi}[Observable axiom]
	Let $\mu$ be a net, and let $\omega=\{p,q\}$ be a proper axiom of $\mu$. We say that $\omega$ is \emph{observable} iff $\mu$ contains two trees $\tau_i,\tau_j$, of respective roots $i,j$, such that $p$ is a leaf of $\tau_i$, $q$ is a leaf of $\tau_j$, and $i,j$ are both free ports of $\mu$. We say that such observable axiom is \emph{based at} $i,j$.
\end{defi}
\begin{figure}[t]
	\begin{center}\scalebox{0.8}{\input{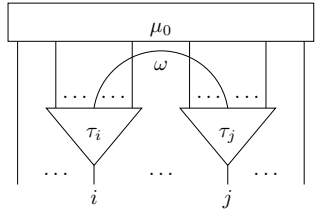}}\end{center}
	\caption{An observable axiom.}
	\label{fig:ObsAx}
\end{figure}
It is perhaps useful to visualize observable axioms. A net $\mu$ contains an observable axiom $\omega$ iff it is of the shape given in \reffig{ObsAx}. If $i=j$, then $\tau_i=\tau_j$, and $\omega$ connects two leaves of the same tree. Note also that one or both of $\tau_i,\tau_j$ may be equal to $\OneTree$; in particular, a wire whose both extremities are free (as in the Separation \refth{Separation}) is an observable axiom.

Observable axioms may be succinctly described by assigning them an \emph{address}. In the following, we let $a,b$ range over the set $\Words=\{\p,\q\}^\ast$ of finite binary words, and we denote by $\Id$ the empty word. Pairs of finite words are denoted by $\BiWord{a}{b}$, and ranged over by $s,t$. The concatenation of two finite words $a,b$ is denoted by simple juxtaposition, \ie, as $ab$. The concatenation of two pairs of finite words $\BiWord{a}{b},\BiWord{a'}{b'}$ is defined as $\BiWord{aa'}{bb'}$, and is also denoted by juxtaposition.
\begin{defi}[Address]
	\label{def:Address}
	An \emph{address} is an unordered pair of elements of $\Words\times\Words\times\Nat$, denoted by $\Arch{\Pillar{\BiWord{a}{b}}{i}}{\Pillar{\BiWord{c}{d}}{j}}$, and ranged over by $\fx,\fy$. Let $\tau$ be a tree, and $l$ a leaf of $\tau$. We associate with $l$ an element of $\Words\times\Words$, denoted by $\Br\tau l$, by induction on $\tau$:
	\begin{enumerate}[$\bullet$]
		\item $\tau=\OneTree$: $\Br\tau l=\BiWord\Id\Id$;
		\item $\tau=\D(\tau_1,\tau_2$): $\Br\tau l=(\BiWord\p\Id)\Br{\tau_1}{l}$ if $l$ is a leaf of $\tau_1$, $\Br\tau l=(\BiWord\q\Id)\Br{\tau_2}{l}$ if $l$ is a leaf of $\tau_2$;
		\item $\tau=\Z(\tau_1,\tau_2$): $\Br\tau l=(\BiWord\Id\p)\Br{\tau_1}{l}$ if $l$ is a leaf of $\tau_1$, $\Br\tau l=(\BiWord\Id\q)\Br{\tau_2}{l}$ if $l$ is a leaf of $\tau_2$.
	\end{enumerate}
	Let now $\omega=\{p,q\}$ be an observable axiom of a net $\mu$; we define its address to be
	$$\Addr{\mu}{\omega}=\Arch{\Pillar{\Br{\tau_i}{p}}{i}}{\Pillar{\Br{\tau_j}{q}}{j}},$$
	where $\tau_i,\tau_j$ are the trees among whose leaves there is $p,q$, respectively, and $i,j$ are the roots of $\tau_i,\tau_j$, respectively, which are free ports of $\mu$ (hence $i,j\in\Nat$ by our convention on free ports of \refdef{Net}).
\end{defi}

In the following, we denote by $\ObsPaths{\mu}$ the set of all addresses of the observable axioms of a net~$\mu$, and we define
$$\AllObsPaths{\mu}=\bigcup_{\mu\red\mu'}\ObsPaths{\mu'}.$$
\begin{prop}
	\label{prop:ObsStab}
	Let $\mu\red\mu'$. Then, $\ObsPaths\mu\subseteq\ObsPaths{\mu'}$, and $\AllObsPaths{\mu}=\AllObsPaths{\mu'}$.
\end{prop}
\proof Simply look at \reffig{ObsAx}, which represents the generic form of $\mu$: any active pair must be inside $\mu_0$, and, after reducing it, since interaction rules are completely local, the ``same'' axiom as $\omega$ is still present in $\mu'$, with the same address. The invariance of $\AllObsPaths{\cdot}$ is a consequence of the confluence of $\beta$-reduction.\qed

\begin{figure}[t]
	\begin{center}\scalebox{0.8}{\input{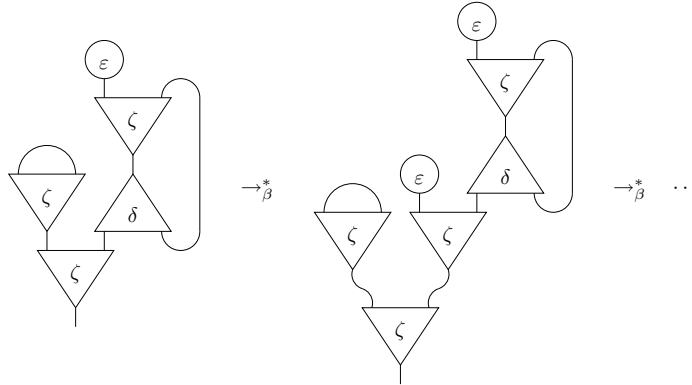}}\end{center}
	\caption{A non-normalizable net producing a finite number of observable axioms.}
	\label{fig:NonNormFinObsAx}
\end{figure}

Note that a net can only have a finite number of observable axioms; then, by \refprop{ObsStab}, $\AllObsPaths\mu$ is finite whenever $\mu$ is normalizable. However, normalizability is not necessary: non-normalizable nets producing a finite number of observable axioms exist, as shown by the example in \reffig{NonNormFinObsAx}: the only observable axiom ever to be found is the one already present in the net starting the reduction sequence, \ie, $\Arch{\Pillar{\BiWord{\Id}{\p\p}}{1}}{\Pillar{\BiWord{\Id}{\p\q}}{1}}$.

We now have two interesting notions of observable: the appearance of observable axioms during reduction, and the fact that these are only produced in finite number.
\begin{defi}[Observability predicates]
	\label{def:ObsPred}
	We say that $\mu$ is \emph{immediately observable}, and we write $\ImmObs\mu$, iff $\ObsPaths{\mu}\neq\emptyset$. We say that $\mu$ is \emph{observable}, and we write $\Obs\mu$, iff $\AllObsPaths{\mu}\neq\emptyset$, or, equivalently, $\mu\red\ImmObs{\mu'}$. We say that $\mu$ is \emph{finitarily observable}, and we write $\StrObs\mu$, iff $\AllObsPaths{\mu}$ is non-empty and finite. We write $\Blind\mu$ and $\WkBlind\mu$ for the negations of $\Obs\mu$ and $\StrObs\mu$, respectively. In particular, if $\Blind\mu$ we say that $\mu$ is \emph{blind}.
\end{defi}
\begin{defi}[Observational equivalences]
	\label{def:ObsEq}
	Two nets $\mu,\nu$ with the same interface are \emph{axiom-equivalent} (resp.\ \emph{finitarily axiom-equivalent}), and we write $\mu\AxEq\nu$ (resp.\ $\mu\FinAxEq\nu$), iff for all contexts $C$, $\Obs{\ctxt\mu}$ iff $\Obs{\ctxt\nu}$ (resp.\ $\StrObs{\ctxt\mu}$ iff $\StrObs{\ctxt\nu}$).
\end{defi}

It helps thinking of an immediately observable net as a head normal form in the \lac. This analogy can be made more precise: our definition of observable net can in fact be extended to any interaction net system~\cite{Mazza:PhDThesis}, in particular to sharing graphs~\cite{Lamping,GonthierAbadiLevy:GoI}; then, one can see that the head variable of a head normal form $T$ corresponds to an observable axiom in the sharing graph of $T$.

Observe that, once we think of observable axioms as head variables, we are naturally led to see $\AllObsPaths{\mu}$ as nothing but a sort of ``B\"ohm tree'' of $\mu$. For instance, $\AllObsPaths{\cdot}$ is an invariant of reduction (\refprop{ObsStab}), just like the B\"ohm tree of a \lat. In \refsect{Edifices}, we shall develop an abstract interpretation of nets starting from this intuition.

\subsection{Observable axioms and the geometry of interaction}
\label{sect:GoI}
The geometry of interaction (GoI) was introduced by Girard~\cite{Girard:GoI1} as a mathematical formulation, using functional analysis and operator algebras, of the cut-elimination process in linear logic. Later, it was reformulated using much less sophisticated tools by Danos and Regnier~\cite{DanosRegnier:PNHilb}, and it was also transported to the interaction combinators (symmetric and not) by Lafont~\cite{Lafont:INComb}. As a rough approximation, we can say that the GoI interprets nets as collections of paths which are stable under $\beta$-reduction; we shall see that these paths ultimately correspond to observable axioms.
\begin{figure}[t]
	\begin{displaymath}
		\begin{array}{ccccc}
			\opcs\opc\monored\Id &\qquad& \opcs\opf\monored\opf\opcs &\qquad& \opfs\opc\monored\opc\opfs \\
			\opds\opd\monored\Id &\qquad& \opcs\opg\monored\opg\opcs &\qquad& \opfs\opd\monored\opd\opfs \\
			\opfs\opf\monored\Id &\qquad& \opds\opf\monored\opf\opds &\qquad& \opgs\opc\monored\opc\opgs \\
			\opgs\opg\monored\Id &\qquad& \opds\opg\monored\opg\opds &\qquad& \opgs\opd\monored\opg\opds
		\end{array}
	\end{displaymath}
	\caption{Word rewriting on monomials.}
	\label{fig:Rewriting}
\end{figure}
\begin{defi}[Monomial, rewriting and value of monomials]
	An \emph{atom} is an element of $\{\opc,\opd,\opf,\opg,\opcs,\opds,\opfs,\opgs\}$; and atom is \emph{positive} if it belongs to $\{\opc,\opd,\opf,\opg\}$, otherwise it is \emph{negative}. A \emph{monomial} is a finite word on the set of atoms. The set of monomials is denoted by $\Monomials$, and ranged over by $A,B$; the empty monomial will be denoted by $\Id$. A monomial is \emph{positive} if it is empty or contains only positive atoms. We define an involution $(\cdot)^\ast$ on $\Monomials$ by setting $(\opc)^\ast=\opcs$, $(\opd)^\ast=\opds$, and similarly for $\opf,\opg$; $(\opcs)^\ast=\opc$, $(\opds)^\ast=\opd$, and similarly for $\opfs,\opgs$; $\Id^\ast=\Id$, and $(AB)^\ast=B^\ast A^\ast$. 

	We define a word rewriting relation $\monored$ on $\Monomials$ as in \reffig{Rewriting}. We say that $A'\in\Monomials$ is \emph{clash-free} iff $A'\monored^\ast AB^\ast$, where $A,B$ are positive monomials.

	Let $A\in\Monomials$, and let $A_0^\ast$ be an occurrence of negative atom in $A$, \ie, $A=A'A_0^\ast A''$ for some $A',A''\in\Monomials$. The \emph{value} of $A_0^\ast$ in $A$, denoted by $\AtomVal{A}{A_0^\ast}$, is the number of positive atoms in $A''$. We define the \emph{value} of $A$ to be $\Val A=\sum_{A_0^\ast}\AtomVal{A}{A_0^\ast}$, where $A_0^\ast$ ranges over the occurrences of negative atoms in $A$.
\end{defi}
\begin{lem}
	\label{lemma:MonoVal}
	Let $A\monored B$. Then, $\Val A>\Val B$.
\end{lem}
\proof By simple inspection of \reffig{Rewriting}.\qed 
\begin{prop}
	\label{prop:Rewriting}
	Rewriting of monomials is confluent and strongly normalizing.
\end{prop}
\proof For what concerns confluence, simply observe that there are no critical pairs. Strong normalization is a consequence of \reflemma{MonoVal}.\qed
\begin{defi}[Port graph]
	\label{def:PortGraph}
	The \emph{port graph} of a net $\mu$, denoted by $\PG\mu$, is an undirected multigraph whose edges are weighed in the positive monomials, defined as follows: its vertices are the elements of $\Ports{\mu}$, and there is an edge between two ports $p,q$ iff one of the following (non mutually exclusive) conditions hold:
\begin{enumerate}[\hbox to8 pt{\hfill}]
\item\noindent{\hskip-12 pt\bf external edge:}\ $\{p,q\}\in\Wires{\mu}$; the weight in this case is $\Id$;
		\item\noindent{\hskip-12 pt\bf internal edge:}\ $p$ and $q$ are principal and auxiliary ports of a cell of $\mu$; the weight in this case depends on the auxiliary port $q$: it is $\opc$ (resp.\ $\opd$) if $q$ is port number $1$ (resp.\ $2$) of a $\D$ cell, and it is $\opf$ (resp.\ $\opg$) if $q$ is port number $1$ (resp.\ $2$) of a $\Z$ cell.
	\end{enumerate}
\end{defi}
\begin{figure}[t]
	\begin{center}\scalebox{0.8}{\input{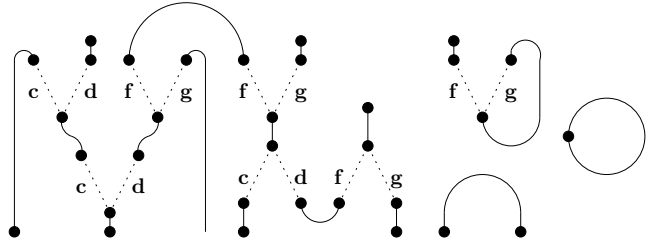}}\end{center}
	\caption{The port graph of the net of \reffig{ANet}. Internal edges are dotted, and their weight annotated beside them; the other weights are omitted, and are all equal to $\Id$.}
	\label{fig:PortGraph}
\end{figure}
As an example, in \reffig{PortGraph} we give the port graph of the net of \reffig{ANet}.
\begin{defi}[Paths, weights \cite{DanosRegnier:PNHilb,Lafont:INComb}, values]
	\label{def:PathWeight}
	A \emph{straight path} of a net $\mu$ is a path of $\PG\mu$ which does not contain two consecutive internal edges. We say that a straight path \emph{crosses an active pair} iff it contains an external edge corresponding to a proper cut. A \emph{maximal path} is a non-empty straight path connecting two free ports of~$\mu$. An \emph{observable path} is a maximal path crossing no active pair. We denote by $\MaxPathsP{i,j}{\mu}$ the set of maximal paths of a net $\mu$ from its free port $i$ to its free port $j$, and we denote by $\MaxPaths\mu$ the set of all maximal paths of $\mu$.

	Let $\phi\in\MaxPaths\mu$, and let $e$ be an internal edge of $\PG{\mu}$ used by $\phi$. Note that $e$ must correspond to a binary cell $c$ of $\mu$; hence, there is no ambiguity in saying that $\phi$ ``enters'' $c$, uses the internal edge, and then ``exits'' $c$. We say that $e$ is \emph{crossed downwards} (resp.\ \emph{upwards}) by $\phi$ if $\phi$ enters $c$ through an auxiliary port (resp.\ through its principal port) and exists $c$ through its principal port (resp.\ through one of its auxiliary ports). Let $A$ be the weight of $e$ in $\PG\mu$. The \emph{weight} of $e$ in $\phi$, denoted by $\EdgeWght{\phi}{e}$, is $A$ if $e$ is crossed downwards; otherwise, it is $A^\ast$. If $e$ is an external edge, we fix $\EdgeWght{\phi}{e}=\Id$.

	Now let $\phi=e_1,\ldots,e_n$ be a maximal path of a net; we define the \emph{weight} of $\phi$ to be the following monomial:
	$$\PathWght{\phi}=\EdgeWght{\phi}{e_n}\cdots\EdgeWght{\phi}{e_1}$$
	(note the reversal of the order of edges). Moreover, we define the \emph{value} of $\phi$, abusively denoted by $\Val\phi$, as $\Val{\PathWght{\phi}}$.
\end{defi}

In the sequel, we shall speak of paths in a net $\mu$ without explicitly referring to $\PG\mu$. This will not be a source of confusion, because all edges of $\PG\mu$ correspond to either wires of $\mu$ or ``internal connections'' represented by the cells of $\mu$; hence, given a straight path of $\mu$, we can easily trace it directly on its graphical representation, and vice versa.

Note that the presence of an observable path in a net $\mu$ implies the presence of exactly one observable axiom in $\mu$, and vice versa (modulo orientation of paths, \ie, an observable axiom actually induces two observable paths, which are the reversal of each other). In fact, if we look at \reffig{ObsAx}, we find an observable path by entering the net through its free 	port~$i$, ``going up'' $\tau_i$ through the branch leading to $\omega$, following $\omega$ itself, and ``descending'' $\tau_j$ through the only branch leading us to its root, which is the free port $j$. Conversely, it is easy to see that any observable path must be of this form (with $\tau_i,\tau_j$ which may be equal to $\OneTree$), because of the absence of active pairs.
\begin{defi}[Residue and lift of a maximal path]
	\label{def:Lift}
	Let $\mu\onered\mu'$, and let $\phi\in\MaxPaths\mu$. The \emph{residue} $\phi'$ of $\phi$ in $\mu'$ is, if it exists, the maximal path of $\mu'$ defined as follows. If $\phi$ does not cross the active pair reduced, then by the locality of interaction rules, ``the same'' path as $\phi$ is found in $\mu'$, and this is taken to be $\phi'$. Otherwise, we call $1,2$ and $3,4$ the auxiliary ports of the cells (which must be binary, because $\phi$ is maximal) composing the active pair reduced, and we distinguish two cases:
	\begin{enumerate}[$\bullet$]
		\item the two cells have the same symbol:
		\begin{enumerate}[$-$]
			\item if $\phi$ connects $1$ to $3$ or $2$ to $4$, then this connection becomes a wire in $\mu'$, so $\phi'$ is equal to what is left of $\phi$ with the active pair replaced by a wire;
			\item if $\phi$ connects $1$ to $4$ or $2$ to $3$, then $\phi$ has no residue;
		\end{enumerate}
		\item the two cells have different symbols; then, whatever ports are connected by $\phi$, the connection is still present in $\mu'$, so $\phi'$ is equal to what is left of $\phi$ with the active pair replaced by this new connection.
	\end{enumerate}
	Of course $\phi$ may cross the active pair more than once, but this is not problematic: its residue is still defined as above, replacing every crossing with the appropriate path described above. Conversely, if $\phi'\in\MaxPaths{\mu'}$, then it is the residue of exactly one maximal path of $\mu$, which is called the \emph{lift} of $\phi'$ in $\mu$.
\end{defi}
Remark that the notions of residue and lift can be extended to reductions of arbitrary length: if $\mu\red\mu'$, and if $\phi$ is a maximal path of $\mu$, we can look for its residue (if it exists) by tracing the successive residues of $\phi$ along the reduction; conversely, if $\phi'$ is a maximal path of $\mu'$, by successively lifting $\phi'$ along the reduction, we obtain the lift of $\phi'$ in $\mu$. Remark also that residues connect the same free ports as their lifts: if $\phi\in\MaxPathsP{i,j}{\mu}$, and if $\phi'$ is the residue of $\phi$ in a reduct $\mu'$ of $\mu$, then $\phi'\in\MaxPathsP{i,j}{\mu'}$.

The following shows that monomial rewriting is related to $\beta$-reduction:
\begin{prop}
	\label{prop:MonoRed}
	Let $\mu\onered\mu'$, let $\phi'\in\MaxPaths{\mu'}$, and let $\phi$ be the lift of $\phi'$ in $\mu$. Then, $\PathWght{\phi}\monored^\ast\PathWght{\phi'}$.
\end{prop}
\proof If $\phi$ does not cross the active pair reduced, we have $\phi'=\phi$, and the result trivially holds. So suppose that $\phi$ crosses the active pair reduced. We start by observing that, since $\phi$ is maximal, the active pair reduced must concern two binary cells $c,c'$. We set $i_1,i_2$ to be the auxiliary port number $1$ and number $2$ of $c$, respectively, and $i_1',i_2'$ to be the auxiliary port number $1$ and number $2$ of $c'$, respectively. We have two cases: an annihilation, or a commutation. Suppose we are in the first case. Since $\phi$ has a reduct in $\mu'$, each time $\phi$ crosses the active pair made by $c,c'$, it must do so by using the pairs of ports $i_1,i_1'$ or $i_2,i_2'$. Hence, if $\phi$ crosses the active pair $n$ times, there exist $n$ occurrences of positive atoms $A_1,\ldots,A_n$ such that $\PathWght{\phi}=\ldots A_1^\ast A_1\ldots A_n^\ast A_n\ldots$, \ie, the weight of $\phi$ contains a word $A_k^\ast A_k$ for each crossing of the active pair. Now, by \refdef{Lift}, all such crossings are replaced by wires in $\phi'$, so $\PathWght{\phi'}=\ldots\Id\ldots\Id\ldots$, and we thus clearly have $\PathWght{\phi}\monored^\ast\PathWght{\phi'}$. Suppose now that we are in the case of a commutation, with $\phi$ crossing $n$ times the active pair made of $c,c'$. This time, we must have $n$ negative atoms $A_1^\ast,\ldots,A_n^\ast$ and $n$ positive atoms $B_1,\ldots,B_n$ such that $\PathWght{\phi}=\ldots A_1^\ast B_1\ldots A_n^\ast B_n\ldots$, and, for each $1\leq k\leq n$, the atoms satisfy that, if $A_k^\ast\in\{\opcs,\opds\}$, then $B_k\in\{\opf,\opg\}$, and if $A_k^\ast\in\{\opfs,\opgs\}$, then $B_k\in\{\opc,\opd\}$. Hence, again by \refdef{Lift}, we have $\PathWght{\phi}=\ldots B_1A_1^\ast\ldots B_nA_n^\ast\ldots$, and, by looking at \reffig{Rewriting}, $\PathWght{\phi}\monored^\ast\PathWght{\phi'}$.\qed

\refprop{MonoRed} is the basis of the GoI. In fact, the above result basically transforms $\beta$-reduction into a word rewriting system; the idea then is to take a model of this rewriting system and build from it a model of $\beta$-reduction. By ``model'', we mean a function $\GoI\cdot$ interpreting monomials in some mathematical structure such that, for all $A,B\in\Monomials$, $A\monored B$ implies $\GoI A=\GoI B$. The following construction, given by Lafont~\cite{Lafont:INComb} and inspired by previous work of Girard~\cite{Girard:GoI1}, does precisely this.

In the following, an \emph{involutive monoid} is a couple $(M,(\cdot)^\ast)$ where $M$ is a multiplicative monoid, and $(\cdot)^\ast$ an involutive antiautomorphism of~$M$. A homomorphism $f$ between two involutive monoids $(M,(\cdot)^\ast),(M',(\cdot)^\dag)$ is a homomorphism of monoids preserving the involution, \ie, such that, for all $a\in M$, $f(a^\ast)=f(a)^\dag$. Similarly, an \emph{involutive unit semiring} is a couple $(S,(\cdot)^\ast)$ where $S$ is a unit semiring, whose additive and multiplicative units are denoted by $\Zero$ and $\Id$, respectively, and $(\cdot)^\ast$ is an involutive antiautomorphism of~$S$.

Let $\RingGoI$ be the involutive unit semiring generated by the set $\{\p,\q\}$ and by the relations
\begin{displaymath}
	\begin{array}{ccc}
		\p^\ast\p=\Id &\qquad& \p^\ast\q=\Zero, \\
		\q^\ast\q=\Id &\qquad& \q^\ast\p=\Zero.
	\end{array}
\end{displaymath}
Consider now the semiring $\RingGoI\otimes\RingGoI$; this is still an involutive unit semiring, the multiplicative unit being $\Id\otimes\Id$ and the involution being defined by $\left(\sum_{i=1}^n a_i\otimes b_i\right)^\ast=\sum_{i=1}^n a_i^\ast\otimes b_i^\ast$. Our model will interpret monomials in $\RingGoI\otimes\RingGoI$, as follows. By definition, the set $\Monomials$ is a free involutive monoid, and $\RingGoI\otimes\RingGoI$ may also be seen as an involutive monoid, by taking its multiplicative part; hence, we can define a homomorphism $\GoI{\cdot}$ by setting
\begin{displaymath}
	\begin{array}{ccc}
		\GoI\opc=\p\otimes\Id &\qquad& \GoI\opf=\Id\otimes\p, \\
		\GoI\opd=\q\otimes\Id &\qquad& \GoI\opg=\Id\otimes\q.
	\end{array}
\end{displaymath}
The fact that $\Monomials$ is free ensures that $\GoI{\cdot}$ is defined everywhere once it is defined on its generators.
\begin{prop}
	\label{prop:RewritingModel}
	For all $A,B\in\Monomials$, $A\monored B$ implies $\GoI A=\GoI B$.
\end{prop}
\proof The rules of \reffig{Rewriting} of the form $A^\ast A\monored\Id$ are modelled by the annihilation relations $a^\ast a=\Id$ defining $\RingGoI$; the rules of the form $A^\ast B\monored BA^\ast$ are modelled by the commutations $(a\otimes\Id)(\Id\otimes b)=(a\otimes b)=(\Id\otimes b)(a\otimes\Id)$ in $\RingGoI\otimes\RingGoI$.\qed

Now, given a cut-free net $\nu$ with $n$ free ports, the GoI assigns to it a formal $n\times n$ matrix $\GoI\nu$ with coefficients belonging to $\RingGoI\otimes\RingGoI$ and defined as follows:
\begin{displaymath}
	{\GoI\nu}_{j,i}=\sum_{\phi\in\MaxPathsP{i,j}{\nu}}\GoI{\PathWght{\phi}}
\end{displaymath}
Note that $\MaxPathsP{i,j}{\mu}$ coincides with the set of observable paths from $i$ to $j$, because $\nu$ is cut-free; there are obviously finitely many of these, so the above sum is finite and defines an element of $\RingGoI\otimes\RingGoI$.

If $\mu$ is a net, we know by the Decomposition \reflemma{Decomposition} that we can always write $\mu=\Ctxt{\sigma}{\nu_0}$ with $\nu_0$ cut-free and $\sigma$ a feedback context; the key result of the GoI is that, if $\mu$ is total of cut-free form $\nu$, we can compute $\GoI\nu$ starting from $\GoI{\nu_0}$. For this, we use a formal matrix associated with $\sigma$, which we denote by $\GoI\sigma$, defined as follows: if $\nu_0$ has $n+2k$ free ports, then $\GoI\sigma$ is a $(n+2k)\times(n+2k)$ matrix, such that ${\GoI\sigma}_{i,j}=\Id$ if $\sigma$ connects the free ports $i$ and $j$ of $\nu_0$, and $\llbracket\sigma\rrbracket_{i,j}=\mathbf{0}$ otherwise.
\begin{thm}
	\label{th:GoI}
	Let $\mu=\Ctxt\sigma{\nu_0}$ be a net with $n$ free ports, with $\nu_0$ having $n+2k$ free ports. Then, $\mu$ is total iff $\GoI{\sigma}\GoI{\nu_0}$ is nilpotent, and in that case, if $\nu$ is the cut-free form of $\mu$, we have
	\begin{displaymath}
		\GoI\nu=\pi^t\left(\sum_{h=0}^\infty\GoI{\nu_0}(\GoI{\sigma}\GoI{\nu_0})^h\right)\pi,
	\end{displaymath}
	where $\pi$ is the formal matrix of the inclusion morphism of $(\RingGoI\otimes\RingGoI)^n$ into $(\RingGoI\otimes\RingGoI)^{n+2k}$, and $\pi^t$ its transpose.
\end{thm}
\proof This was originally proved by Girard for linear logic\footnote{Actually, as formulated here, this result holds only for multiplicative linear logic; some technical constraints are needed in full linear logic, because the execution formula does not model cut-elimination in the general case.} \cite{Girard:GoI1}, with nilpotency only as a necessary condition for strong normalization (\ie, totality); Danos and Regnier later proved the converse~\cite{DanosRegnier:PNHilb}. For the symmetric combinators, the result with nilpotency as a necessary condition is due to Lafont~\cite{Lafont:INComb}; a proof that nilpotency is also sufficient for totality can be found in the author's Ph.D.\ thesis~\cite{Mazza:PhDThesis} (as the proof of Theorem~3.70).\qed

The formula for computing $\GoI\nu$ from $\GoI{\nu_0}$ and $\GoI{\sigma}$ given in \refth{GoI} is known as the \emph{execution formula}. For total nets, the execution formula is an invariant of $\beta$-reduction, and therefore gives a model of $\BetaEq$. However, this does not work for non-total nets, because in that case the sum in the execution formula has an infinite number of terms. To handle it, one solution would be to put some topology on $\RingGoI\otimes\RingGoI$ (or rather, on its algebra of formal matrices), and study the convergence of the execution formula in this topology. Another solution, which is the one we chose in this paper, is to deal directly with possibly infinite objects, just like one deals with possibly infinite B\"ohm trees in the \lac.

In fact, even when it ``diverges'', the execution formula is not completely meaningless: it computes the interpretations of the weights of those maximal paths which are never ``destroyed'' by reduction. We shall see that these correspond to the observable axioms generated during reduction. This gives a further justification for our definitions of \refsect{ObsAx}.
\begin{defi}[Execution path \cite{DanosRegnier:PNHilb}]
	\label{def:ExecPath}
	Let $\mu$ be a net. An \emph{execution path} of $\mu$ is a maximal path $\phi$ of $\mu$ such that, whenever $\mu\red\mu'$, $\phi$ has a residue in $\mu'$.
\end{defi}
Remark that, as an immediate consequence of the definition, any residue of an execution path is itself an execution path. As we said above, execution paths are those which are preserved by reduction, and which eventually generate observable paths (and hence observable axioms).
\begin{lem}
	\label{lemma:Val}
	Let $\mu$ be a net, and let $\phi$ be an execution path of $\mu$. Then:
	\begin{enumerate}[\em(1)]
		\item $\Val\phi=0$ implies that $\phi$ is observable;
		\item $\Val\phi>0$ implies that $\phi$ crosses an active pair, reducing which we obtain $\mu\onered\mu'$, and the residue $\phi'$ of $\phi$ in $\mu'$ is such that $\Val{\phi'}<\Val\phi$.
	\end{enumerate}
\end{lem}
\proof For part (1), observe that $\Val\phi=0$ implies $\PathWght{\phi}=AB^\ast$ for some positive monomials $A,B$, which is clearly the weight of an observable path. For part (2), note that $\Val\phi>0$ implies $\PathWght{\phi}=A'A_0^\ast B_0B'$ for some $A',B'\in\Monomials$ and some positive atoms $A_0,B_0$. Then, $\phi$ crosses an active pair, so we apply \refprop{MonoRed} (or rather its proof) and \reflemma{MonoVal}.\qed
\begin{lem}
	\label{lemma:ExecPaths}
	Let $\mu$ be a net, and let $\phi\in\MaxPaths\mu$. Then, the following are equivalent:
	\begin{enumerate}[\em(1)]
		\item $\phi$ is an execution path;
		\item $\mu\red\mu'$ such that the residue of $\phi$ in $\mu'$ is an observable path;
		\item $\PathWght{\phi}$ is clash-free;
		\item $\GoI{\PathWght\phi}\neq\Zero$.
	\end{enumerate}
\end{lem}
\proof (1) implies (2) is proved by induction on $\Val\phi$, using \reflemma{Val}.

For (2) implies (3), let $\phi'$ be the residue of $\phi$ in $\mu'$. Since it is an observable path, we have $\PathWght{\phi'}=AB^\ast$ for some positive monomials $A,B$, and we conclude by \refprop{MonoRed}.

For (3) implies (4), observe that normal form of $\PathWght{\phi}$ is by definition of the form $AB^\ast$, so by \refprop{RewritingModel} $\GoI{\PathWght{\phi}}=\GoI{AB^\ast}=\GoI A\GoI B^\ast$, which is never equal to $\Zero$ by definition of $\GoI{\cdot}$.

For (4) implies (1), we prove the contrapositive. Suppose $\phi$ is a maximal path of $\mu$ which is not an execution path. This means that $\mu\red\mu'\onered\mu''$, and the residue $\phi'$ of $\phi$ in $\mu'$ has no residue in $\mu''$. By \refdef{Lift}, the only possibility is that $\phi'$ crosses an active pair made of two binary cells $c,c'$ of the same kind, using auxiliary port $1$ of $c$ and auxiliary port $2$ of $c'$. Hence, we have, for some $A',B'\in\Monomials$, $\PathWght{\phi'}=A'A_0^\ast B_0 B'$, and either $A_0,B_0\in\{\opc,\opd\}$, or $A_0,B_0\in\{\opf,\opg\}$, but in both cases $A_0\neq B_0$. Hence, by \refprop{RewritingModel} and by definition of $\RingGoI$, we have $\GoI{\PathWght{\phi}}=\GoI{\PathWght{\phi'}}=\GoI{A'}\GoI{A_0}^\ast\GoI{B_0}\GoI{B'}=\GoI{A'}\Zero\GoI{B'}=\Zero$.\qed

Now, we invite the reader to check that, if a net $\nu$ has an observable axiom $\omega$ whose address is $\Arch{\Pillar{s}{i}}{\Pillar{t}{j}}$, and if $\phi$ is the observable path from $i$ to $j$ induced by $\omega$, we have $\GoI{\PathWght\phi}=ts^\ast$. Vice versa, if $\nu$ has an observable path $\phi$ from free port $i$ to free port $j$ such that $\PathWght\phi=BA^\ast$, then the corresponding observable axiom will have address $\Arch{\Pillar{\GoI A}{i}}{\Pillar{\GoI B}{j}}$. This is the reason behind our choice of notations for the addresses of observable axioms; moreover, it allows us to state the following:
\begin{prop}
	\label{prop:ObsAxAndGoI}
	For every net $\mu$, we have
	$$\AllObsPaths{\mu}=\left\{\Arch{\Pillar{s}{i}}{\Pillar{t}{j}}~\big|~\phi\in\MaxPathsP{i,j}{\mu},\GoI{\PathWght\phi}=ts^\ast\right\}.$$
\end{prop}
\proof For what concerns the inclusion from left to right, we have $\mu\red\mu'$ such that $\mu'$ contains an observable axiom of address $\Arch{\Pillar{s}{i}}{\Pillar{t}{j}}$. By the remark made above, we know that this observable axiom induces an observable path $\phi'$ of $\mu'$ from free port $i$ to free port $j$ such that $\GoI{\PathWght{\phi'}}=ts^\ast$; then, we take its lift $\phi$ in $\mu$, and conclude by Propositions~\ref{prop:MonoRed} and \ref{prop:RewritingModel}.

For the other inclusion, by hypothesis $\GoI{\PathWght{\phi}}\neq\Zero$, so by \reflemma{ExecPaths} we have $\mu\red\mu'$ such that the residue $\phi'$ of $\phi$ in $\mu$ is observable. By Propositions~\ref{prop:MonoRed} and \ref{prop:RewritingModel}, $\GoI{\PathWght{\phi'}}=\GoI{\PathWght{\phi}}$, and, as remarked above, we know that $\phi'$ induces an observable axiom of address $\Arch{\Pillar{s}{i}}{\Pillar{t}{j}}$, which is in $\AllObsPaths{\mu}$ by definition.\qed

We encourage the reader to compare \refprop{ObsAxAndGoI} with the
definition of $\GoI\nu$ given above: basically, we can see
$\AllObsPaths\cdot$ as the extension of $\GoI\cdot$ to arbitrary
nets. In fact, if we tried to define $\GoI\mu$ in the general case,
the sum ranging over $\MaxPathsP{i,j}{\mu}$ might be infinite; as
mentioned above, instead of introducing a topology to handle series,
we opted for a ``B\"ohm-tree'' approach: we reduce $\mu$, and collect
the observable axioms showing up along the way, accepting that there
may be an infinity of them. Using formal matrices does not make much
sense at this point, but we still need to retain the information
concerning the free ports connected by the observable paths: this is
the reason behind the presence of integers in addresses. The fact that
addresses are unordered pairs reflects the fact that an observable path carries the same information as its reversal; indeed, in the GoI interpretation of a cut-free net $\nu$ one can show that $\GoI{\nu}_{i,j}=\GoI{\nu}_{j,i}^\ast$ (all operators are ``Hermitian''~\cite{Girard:GoI1}).

The question of what can be done by taking the topological approach instead of the ``B\"ohm tree'' approach is left open, and is out of the scope of this work. We shall see that topology will eventually play a fundamental role in our approach too, but for quite different purposes.

\subsection{Solvability and $\E$-reduction}
\label{sect:Solvable}
\begin{figure}[t]
	\begin{center}\scalebox{0.8}{\input{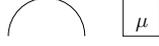}}\end{center}
	\caption{A quasi-wire; $\mu$ is an arbitrary net with an empty interface.}
	\label{fig:QuasiWire}
\end{figure}
The analogy ``an immediately observable net is like a \lat\ in head normal form'' can be given a justification within the theory of the symmetric combinators. In the following, we call a net of the form given in \reffig{QuasiWire} a \emph{quasi-wire}.
\begin{defi}[Solvable net]
	A net $\mu$ is \emph{solvable} iff there exist a test $\theta$ such that \mbox{$\Ctxt{\theta}{\mu}\red W$}, where $W$ is a quasi-wire.
\end{defi}

We shall see that, just as \lat s having a head normal form coincide with solvable \lat s, so in the symmetric combinators observable nets and solvable nets coincide. Furthermore, in \refsect{Meaningless} we shall give evidence supporting the fact that the above notion of solvable net is indeed analogous to that of a solvable \lat.

Let $C$ be a net, and let $I$ be a subset of its interface. We say that $C$ is \emph{relatively blind} on $I$ iff, whenever $\Arch{\Pillar{s}{i}}{\Pillar{t}{j}}\in\AllObsPaths{C}$, $i\in I$ implies $j\not\in I$, \ie, $C$ generates no observable axiom all based within $I$. A context $C$ for nets with $n$ free ports will be said to be relatively blind if its interface is $\{1,\ldots,n\}\uplus I$ and $C$ is relatively blind on $I$.
\begin{lem}
	\label{lemma:RelBlind}
	For every net $\mu$ and relatively blind context $C$, $\Blind\mu$ implies $\Blind{\ctxt\mu}$.
\end{lem}
\proof Let $\Blind\mu$, and suppose for the sake of absurdity that $\ctxt\mu\red\ImmObs{\mu'}$. The observable axiom in $\mu'$ induces an observable path $\phi'$ in $\mu'$, which has a lift $\phi$ in $\ctxt\mu$, connecting two free ports of the relatively blind interface of $C$. Suppose that $\phi\in\MaxPaths{C}$. By looking at the proof of \refprop{MonoRed}, we see that $\phi$ can be transformed into an observable path by reducing only the active pairs it crosses; hence, we would have $C\red C'$, with $\phi'$ an observable path of $C'$, which would contradict the fact that $C$ is relatively blind. Then, we must have the following situation
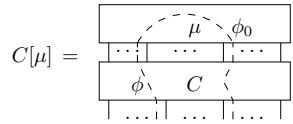
\begin{center}\scalebox{0.8}{\input{RelBlindCtxtPath.pstex_t}}\end{center}
where we drew $\phi$ as a dashed line. As a consequence, we can write $\phi=\phi'\phi_0\phi''$, where $\phi_0\in\MaxPaths{\mu}$. We thus have $\PathWght{\phi}=\PathWght{\phi''}\PathWght{\phi_0}\PathWght{\phi'}$, which means that $\PathWght{\phi_0}$ is a subword of $\PathWght{\phi}$. But, by \reflemma{ExecPaths}, $\PathWght{\phi}$ is clash-free, so $\PathWght{\phi_0}$ is clash-free too; if we apply \reflemma{ExecPaths} again, we obtain that $\phi_0$ has an observable residue, which, by the above remark, can be obtained by reducing only the active pairs it crosses, which are all within $\mu$. Therefore, we obtain $\Obs\mu$, a contradiction.\qed
\begin{lem}
	\label{lemma:Blindness}
	Let $\mu$ be a net. Then, $\Blind\mu$ implies $\Blind{\Ctxt\theta\mu}$ for any test $\theta$.
\end{lem}
\proof Simply observe that a test is a relatively blind context.\qed

Now, the forward implication of the equivalence ``observable iff solvable'' holds because observable axioms are defined precisely with the intuition that we can ``extract'' a wire from $\mu$ by means of a test; the converse is a consequence of \reflemma{Blindness}:
\begin{prop}
	\label{prop:Solvable}
	A net $\mu$ is observable iff it is solvable.
\end{prop}
\proof Let $\Obs\mu$; by \refdef{ObsPred}, we know that $\mu$ reduces to a net of the shape given in \reffig{ObsAx}. Then, consider the test
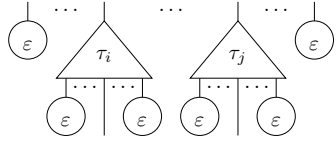
\begin{center}\scalebox{0.8}{\input{TestObsAx.pstex_t}}\end{center}
where we leave free exactly the two leaves of $\tau_i,\tau_j$ corresponding to those connected by the observable axiom $\omega$ of \reffig{ObsAx}. By \reflemma{TreeAnnihilation}, we have that $\Ctxt{\theta}{\mu}$ $\beta$-reduces to a quasi-wire, as desired. Suppose now $\Blind\mu$. Remark that quasi-wires are immediately observable, so if $\mu$ were solvable, we would have a test $\theta$ such that $\Obs{\Ctxt{\theta}{\mu}}$, contradicting \reflemma{Blindness}.\qed

Another interesting consequence of \reflemma{Blindness} is that finitary axiom-equivalence is stronger than axiom-equivalence:
\begin{prop}
	\label{prop:TotEqInObsEq}
	For all $\mu,\nu$, $\mu\FinAxEq\nu$ implies $\mu\AxEq\nu$.
\end{prop}
\proof Assume $\mu\FinAxEq\nu$, and let $C$ be a context such that $\Obs{\ctxt\mu}$. We need to show that $\Obs{\ctxt\nu}$; by symmetry of $\FinAxEq$, this will be enough to prove the result. By \refprop{Solvable}, $\ctxt{\mu}$ is solvable, so we have a test $\theta$ such that $\Ctxt{\theta}{\ctxt\mu}$ $\beta$-reduces to a quasi-wire. But quasi-wires generate exactly one observable axiom, so $\StrObs{\Ctxt{\theta}{\ctxt\mu}}$, which by $\mu\FinAxEq\nu$ implies $\StrObs{\Ctxt{\theta}{\ctxt\nu}}$, which by \refdef{ObsPred} implies $\Obs{\Ctxt{\theta}{\ctxt\nu}}$, which implies $\Obs{\ctxt\nu}$ thanks to \reflemma{Blindness}.\qed

The \lac\ has an interesting notion of \emph{$\Omega$-reduction}, which is defined by the reduction rule $M\rightarrow_\Omega\Omega$ iff $M$ is unsolvable and different from $\Omega$, where $\Omega$ is itself some fixed unsolvable term (usually one takes $\Omega=(\lambda x.xx)(\lambda x.xx)$). Added to other reductions, $\Omega$-reduction has interesting properties: $\beta\Omega$-reduction and $\beta\eta\Omega$-reduction characterize provability in the theory $\mathcal H$ (the smallest sensible theory, \cf\ \refsect{Theories}) and $\mathcal H\eta$, respectively~\cite{Barendregt}. This latter coincides with nf-equivalence, defined in \refsect{ObsAx}.

Since we have our own notion of unsolvable net, it may be interesting to study the behavior of the following rewriting rule, directly inspired by $\Omega$-reduction:
\begin{center}\scalebox{0.8}{\input{EpsilonRed.pstex_t}}\end{center}
where $\mu$ has $n$ free ports and $\mu\neq\Epsilon{n}$.
\begin{defi}[$\E$- and $\beta\E$-reduction]
	We write $\eonlyonered$ for the contextual closure of the above rule, and we define $\eonered=(\onered\cup\eonlyonered)$.
\end{defi}
Of course the rewriting rule defining $\E$-reduction is not recursive, because it is undecidable whether a net is blind (intuitively, this is a consequence of the Turing-completeness of the symmetric combinators, and of Rice's theorem---any non-trivial class of recursive functions, hence of nets, is undecidable). This is exactly the same situation as $\Omega$-reduction, where it is undecidable whether a \lat\ is unsolvable. We shall see that the interest of $\beta\E$-reduction is in its relationship with finitary axiom-equivalence (\refcor{BENormAndObsPaths}).

We say that a binary relation on nets $\genred$ has the \emph{quasi-diamond property} iff $\mu\genred\mu_1$ and $\mu\genred\mu_2$ implies that there exists $\nu$ such that $\mu_1\genred\nu$ or $\mu_1=\nu$, and $\mu_2\genred\nu$ or $\mu_2=\nu$.
\begin{lem}
	\label{lemma:PseudoDiamond}
	Let $\genred$ be a binary relation on nets satisfying the quasi-diamond property. Then, its reflexive transitive closure $\genred^\ast$ satisfies the diamond property, \ie, it is confluent.
\end{lem}
\proof A standard diagram-chasing argument.\qed
\begin{lem}
	\label{lemma:EOneRed}
	The relation $\eonered$ satisfies the quasi-diamond property.
\end{lem}
\proof Let $\mu\eonered\mu_1$ and $\mu\eonered\mu_2$. We may suppose $\mu_1\neq\mu_2$, otherwise there is nothing to prove. If the two reductions come from two active pairs, we conclude by applying \refprop{StrongConfluence}. Otherwise, suppose without loss of generality that $\mu\onered\mu_1$ and $\mu\eonlyonered\mu_2$, \ie, we have $\mu=\ctxt\nu$ and $\mu_2=\ctxt{\Epsilon{n}}$, where $\nu$ is a blind net with $n$ free ports. We have three cases:
\begin{enumerate}[$\bullet$]
	\item $\nu\onered\nu'$ and $\mu_1=\ctxt{\nu'}$, \ie, the active pair reduced to obtain $\mu_1$ is contained in $\nu$. In that case, we still have $\Blind{\nu'}$, so $\mu_1\eonlyonered\mu_2$.
	\item The active pair reduced to obtain $\mu_1$ is ``between'' $\nu$ and $C$, \ie, one of its cells is in $\nu$ and the other, call it $c$, is in $C$. We suppose $c$ to be binary; the nullary case is easier, and left to the reader. Then, the cell $c$ together with a suitable identity wiring (which may be empty in case $n=1$) forms a test $\theta$, and we can write $\mu=\Ctxt{C'}{\Ctxt{\theta}{\nu}}$ for a suitable context $C'$. Now $\Ctxt{\theta}{\nu}\onered\nu'$ and $\mu_1=\Ctxt{C'}{\nu'}$, while $\mu_2=\Ctxt{C'}{\Ctxt{\theta}{\Epsilon{n}}}$. By \reflemma{Blindness}, we have both $\Blind{\Ctxt{\theta}{\nu}}$ and $\Blind{\Ctxt{\theta}{\Epsilon{n}}}$, so both $\mu_1$ and $\mu_2$ $\E$-reduce in one step to $\Ctxt{C'}{\Epsilon{n+1}}$.
	\item The active pair reduced is completely disjoint from $\nu$. This case is trivial.
\end{enumerate}
We are left with the situation in which both $\mu_1$ and $\mu_2$ are obtained by means of $\E$-steps. Let $\nu_1,\nu_2$ be the blind subnets of $\mu$ reduced to obtain $\mu_1$ and $\mu_2$, respectively. If $\nu_1$ and $\nu_2$ are disjoint, then the diamond property holds trivially. Otherwise, we have $\mu=\ctxt\nu$, where $\nu$ is the net
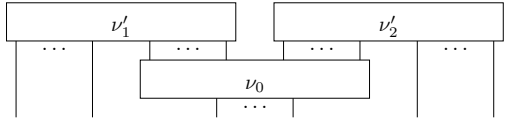
\begin{center}\scalebox{0.8}{\input{BlindNetsIntersect.pstex_t}}\end{center}
and $\nu_1$ is equal to $\nu_1'$ plus $\nu_0$, while $\nu_2$ is equal to $\nu_2'$ plus $\nu_0$. Now, if we put
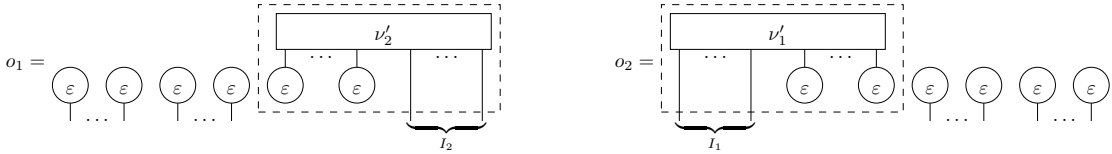
\begin{center}\scalebox{0.75}{\input{OoneAndOtwo.pstex_t}}\end{center}
we have $\mu_1=\ctxt{o_1}$ and $\mu_2=\ctxt{o_2}$. But $\nu_1'$ and $\nu_2'$ must be relatively blind on $I_1$ and $I_2$, respectively, because $\nu_1$ and $\nu_2$ are blind. Hence, by \reflemma{RelBlind}, the subnets marked by the dashed rectangles in the above picture are both blind, so $\mu_1$ and $\mu_2$ both reduce in at most one $\E$-step to $\ctxt{\Epsilon{n}}$, where $n$ is the number of free ports of $\nu$.\qed
\begin{prop}[Confluence of $\beta\E$-reduction]
	\label{prop:BetaEpsConfluence}
	The relation $\ered$ is confluent.
\end{prop}
\proof Apply Lemmas~\ref{lemma:PseudoDiamond}~and~\ref{lemma:EOneRed}.\qed

The confluence of $\beta\E$-reduction allows us to introduce the following congruences:
\begin{defi}[$\beta\E$- and $\beta\eta\E$-equivalence]
	\emph{$\beta\E$-equivalence} is defined by $\mu\BetaEpsilonEq\nu$ iff there exists $o$ such that $\mu\ered o$ and $\nu\ered o$; \emph{$\beta\E\eta$-equivalence} is defined by $\BetaEpsilonEtaEq\ =\ (\BetaEpsilonEq\cup\EtaEq)^+$.
\end{defi}

Note that $\beta\E$-normal forms are always cut-free. In particular, we have the following characterization, whose proof is left to the reader. In the following, an \emph{$\E$-tree} is a tree with no leaves; the $\E$-tree $\E$ is called \emph{trivial}.
\begin{prop}[$\beta\E$-normal forms]
	\label{prop:BetaEpsilonNormForm}
	A net $\mu$ is $\beta\E$-normal iff it is cut-free, and each $\E$-tree contained in $\mu$ is trivial.\qed
\end{prop}

The following result shows that $\beta\E$-reduction is related to finitary axiom-equivalence. We first need to extend \refprop{ObsStab} to $\beta\E$-reduction, which is unproblematic:
\begin{prop}
	\label{prop:ObsStabE}
	$\mu\ered\nu$ implies $\ObsPaths{\mu}\subseteq\ObsPaths{\nu}$ and $\AllObsPaths{\mu}=\AllObsPaths{\nu}$.\qed
\end{prop}
\proof It is enough to check one-step reductions. If $\mu\onered\nu$, we conclude by \refprop{ObsStab}. If $\mu\eonlyonered\nu$, we have that $\mu=\ctxt{\mu_0}$ and $\nu=\ctxt{\Epsilon{n}}$ for some context $C$ and blind net $\mu_0$ with $n$ free ports; precisely because $\mu_0$ is blind, the axioms disappearing from $\mu$ are not observable; moreover, $\nu$ does not contain new observable axioms, so $\ObsPaths{\mu}=\ObsPaths{\nu}$, and we conclude $\AllObsPaths{\mu}=\AllObsPaths{\nu}$ by confluence of $\beta\E$-reduction (\refprop{BetaEpsConfluence}).\qed
\begin{cor}
	\label{cor:BENormAndObsPaths}
	$\mu$ is $\beta\E$-normalizable iff $\AllObsPaths{\mu}$ is finite.
\end{cor}
\proof The forward implication is an immediate consequence of \refprop{ObsStabE}. For the converse, $\AllObsPaths{\mu}$ finite and \refprop{ObsStabE} imply that any reduction starting from $\mu$ stumbles upon a net $\mu'$ such that $\ObsPaths{\mu'}=\AllObsPaths{\mu}$. This means that all subnets of $\mu'$ containing active pairs are blind, \ie, they do not produce further observable axioms. There is of course at most a finite number of such subnets, so $\mu'$ reduces in finitely many $\E$-steps to a $\beta\E$-normal net.\qed

Up to now we have five congruences strictly extending $\beta$-equivalence: $\BetaEtaEq$, $\BetaEpsilonEq$, $\BetaEpsilonEtaEq$, $\FinAxEq$, and $\AxEq$ (the fact that these last two strictly extend $\BetaEq$ is an immediate consequence of \refdef{ObsEq}). We shall see that the last four congruences form a sequence of strictly stronger equivalences:
$$\BetaEq\ \varsubsetneq\ \BetaEpsilonEq\ \varsubsetneq\ \BetaEpsilonEtaEq\ \varsubsetneq\ \FinAxEq\ \varsubsetneq\ \AxEq.$$
The first two strict inclusions are obvious. We shall give a semantic proof of the third inclusion (\refcor{BetaEtaEpsInFinAxEq} of the full abstraction \refth{TotEqFullAbs}); \reffig{PingPong} shows that it is strict. We already established the fourth inclusion in \refprop{TotEqInObsEq}; \reffig{Iota} shows that the inclusion is strict, as an application of the full abstraction \refth{ObsEqFullAbs}.

\section{Denotational Semantics}
\label{sect:BigDenSem}

\subsection{Edifices}
\label{sect:Edifices}
In what follows, $\Cc=\{\p,\q\}^\Nat$ is the set of infinite binary words, ranged over by $x,y$. As in the case of finite words, the elements of $\Cc\times\Cc$ will be denoted by $\BiWord{x}{y}$, and ranged over by $u,v,w$. Given two words or pairs of words $s,u$, where $s$ is finite and $u$ may be infinite, we say that $s$ is \emph{prefix} of $u$ iff there exists $u'$ such that $u=su'$.
\begin{defi}[Pillar, arch, edifice, vault]
	\label{def:Edifice}
	Let $I\subseteq\Nat$, and set $\cP_I=\Cc\times\Cc\times I$. A \emph{pillar} is an element of $\cP=\cP_\Nat$.  Pillars are denoted by $\Pillar{u}{i}$, and are ranged over by $\xi,\upsilon$. The pillar $\Pillar{u}{i}$ is said to be \emph{based} at $i$.

	Set $\overrightarrow\cA_I=\cP_I\times\cP_I$, and, given $(\xi,\upsilon),(\xi',\upsilon')\in\overrightarrow\cA_I$, define $(\xi,\upsilon)\archsim(\xi',\upsilon')$ iff $\xi'=\upsilon$ and $\upsilon'=\xi$, or $\xi'=\xi$ and $\upsilon'=\upsilon$. We then set \mbox{$\cA_I=\overrightarrow\cA_I/\archsim$}. An \emph{arch} is an element of \mbox{$\cA=\cA_\Nat$}. Arches are denoted by $\Arch{\xi}{\upsilon}$ (which is the same as $\Arch{\upsilon}{\xi}$), and ranged over by $\fa,\fb$. An arch is said to be \emph{based} at the unordered pair where its two pillars are based.

	An \emph{edifice} is a set of arches; edifices are ranged over by $\fE,\fF$. A \emph{vault} is an edifice $\fA$ such that there exists an address $\fx=\Arch{\Pillar{s}{i}}{\Pillar{t}{j}}$ such that
	$$\fA=\left\{\Arch{\Pillar{sw}{i}}{\Pillar{tw}{j}}~|~w\in\Cc\times\Cc\right\}.$$
	The address $\fx$ is said to \emph{generate} $\fA$. We denoted by $\EdPath{\fx}$ the vault generated by $\fx$.
\end{defi}

We now introduce two special kinds of edifices, which will be useful in the sequel.
\begin{defi}[Uniform edifice]
	If $\fE$ is an edifice and $\fa\in\fE$, $\fa$ is said to be \emph{uniform in $\fE$} iff there exists a vault $\fA\subseteq\fE$ such that $\fa\in\fA$. An edifice is \emph{uniform} if all of its arches are uniform.
\end{defi}
\begin{prop}
	\label{lemma:UniformVaults}
	An edifice is uniform iff it is a union of vaults.
\end{prop}
\proof That a union of vaults is uniform is obvious. For the converse, take a uniform edifice $\fE$, and let $\fa\in\fE$. By definition, there exists a vault contained in $\fE$ which contains $\fa$; we call such vault $\fA_\fa$. Then, it is easy to check that $\fE=\bigcup_{\fa\in\fE}\fA_\fa$.\qed
\begin{defi}[Coherence, simple edifice]
	\label{def:SimpleEd}
	If $\fa=\Arch{\xi}{\upsilon}$ is an arch, we define its \emph{support} to be $\supp\fa=\{\xi,\upsilon\}$. Then, given $\fa,\fa'\in\cA$, we say that $\fa$ and $\fa'$ are \emph{coherent}, and we write $\fa\coh\fa'$, iff $\supp\fa\cap\supp{\fa'}$ is either empty or of cardinality $2$. An edifice $\fE$ is \emph{simple} iff it is a clique with respect to coherence, \ie, for all $\fa,\fa'\in\fE$, $\fa\coh\fa'$.
\end{defi}
Note that, although obviously symmetric, coherence is not reflexive: all arches of the form $\Arch{\xi}{\xi}$, which we may refer to as degenerated, are not coherent with themselves.

Edifices are naturally endowed with a \emph{trace} operation. We shall see that this operation closely corresponds to the execution formula of the GoI (\cf\ \refsect{GoI}); it is also reminiscent of the notion of composition of strategies in games semantics.
\begin{defi}[Feedback function, trace sequence]
	\label{def:TrSeq}
	A \emph{feedback function} $\sigma$ is a fixpoint-free partial involution on $\Nat$ of finite domain. In other words, $\sigma(i)$ is defined for finitely many $i\in\Nat$, and in that case $\sigma(i)\neq i$, and $\sigma^2(i)=i$. We denote by $\dom\sigma$ the domain of $\sigma$.

	Let $\sigma$ be a feedback function, let $\fE$ be an edifice, and let $\fs=\fs_1,\ldots,\fs_n$ be a non-empty finite sequence of arches of $\fE$, for which we set, for $1\leq k\leq n$, $\fs_k=\Arch{\Pillar{u_k}{i_k}}{\Pillar{v_k}{j_k}}$. We say that $\fs$ is a \emph{trace sequence of $\fE$ along $\sigma$} iff:
	\begin{description}
		\item[chain] for all $2\leq k\leq n$, $j_{k-1}\in\dom\sigma$ and $i_k=\sigma(j_{k-1})$;
		\item[match] for all $2\leq k\leq n$, $u_k=v_{k-1}$.
	\end{description}
	A trace sequence is \emph{visible} if it further satisfies $i_1,j_n\not\in\dom\sigma$.

	The \emph{length} of a trace sequence $\fs$ is denoted by $\Len\fs$. We denote by $\TrSeqFin{\sigma}{\fE}$ the set of trace sequences of $\fE$ along $\sigma$. If $\fs\in\TrSeqFin{\sigma}{\fE}$ such that $\Len\fs=n$, we define the \emph{arch generated by $\fs$} as $\fa(\fs)=\Arch{\Pillar{u_1}{i_1}}{\Pillar{v_n}{j_n}}$.
\end{defi}
Observe that, if $\fE$ is an edifice and $\sigma$ a feedback function, then $\fs=\fs_1,\ldots,\fs_n\in\TrSeqFin{\sigma}{\fE}$ implies $\fs'=\fs_n,\ldots,\fs_1\in\TrSeqFin{\sigma}{\fE}$, $\fa(\fs')=\fa(\fs)$, and $\fs'$ is visible iff $\fs$ is. Intuitively, this reflects the fact that a maximal path in a net can always be ``reversed''. In \refsect{NetsAsEd} we shall formalize the relation between visible trace sequences and maximal paths. We also remark that the role of non-visible trace sequences will be purely technical: their purpose is to allow proofs by induction on the length of sequences. In fact, visible trace sequences are not suitable for such proof technique, because an initial or final segment of a visible trace sequence is never visible.
\begin{defi}[Trace]
	Let $\fE$ be a set of arches, and $\sigma$ a feedback function. We define the \emph{trace} of $\fE$ along $\sigma$ as
	$$\Trace{\sigma}{\fE}=\{\fa(\fs)~|~\fs\in\TrSeqFin{\sigma}{\fE},\ \fs\textrm{ visible}\}.$$
\end{defi}

The trace is obviously monotonic:
\begin{lem}[Monotonicity of the trace]
	\label{lemma:TraceMono}
	Let $\fE,\fF$ be edifices, and let $\sigma$ be a feedback function. Then, $\fE\subseteq\fF$ implies $\Trace{\sigma}{\fE}\subseteq\Trace{\sigma}{\fF}$.
\end{lem}
\proof Obvious.\qed

In the following, if $\sigma,\sigma'$ are feedback functions of disjoint domain, we denote by $\sigma\uplus\sigma'$ the function defined by
$$(\sigma\uplus\sigma')(i)=\left\{
\begin{array}{l}
	\sigma(i) \textrm{ if } i\in\dom\sigma \\
	\sigma'(i) \textrm{ if } i\in\dom\sigma' \\
	\textrm{undefined otherwise}
\end{array}\right.,
$$
which is obviously a feedback function.
\begin{lem}[Associativity of the trace]
	\label{lemma:TraceAssoc}
	Let $\fE$ be an edifice, and let $\sigma,\sigma'$ be feedback functions of disjoint domain. Then,
	$$\Trace{\sigma}{\Trace{\sigma'}{\fE}}=\Trace{\sigma\uplus\sigma'}{\fE}.$$
\end{lem}
\proof We start by establishing the inclusion from left to right. Let $\fa\in\Trace{\sigma}{\Trace{\sigma'}{\fE}}$, and let $\fs\in\TrSeqFin{\sigma}{\Trace{\sigma'}{\fE}}$ be the trace sequence generating $\fa$, with $\Len\fs=m$. By definition, if we let $1\leq k\leq m$, each $\fs_k$ is generated by a trace sequence $\fs^k\in\TrSeqFin{\sigma'}{\fE}$, with $\Len{\fs^k}=n_k$. Then, it is not hard to see that $\fs'=\fs^1_1,\ldots,\fs^1_{n_1},\fs^2_1\ldots\fs^{m-1}_{n_{m-1}},\fs^m_1,\ldots,\fs^m_{n_m}$
is a visible trace sequence of $\fE$ along $\sigma\uplus\sigma'$, such that $\fa(\fs')=\fa(\fs)$.

For the reverse inclusion, let $\fa\in\Trace{\sigma\uplus\sigma'}{\fE}$ and let $\fs\in\TrSeqFin{\sigma\uplus\sigma'}{\fE}$ be the trace sequence generating $\fa$, with $\Len\fs=n$, and $\fs_k=\Arch{\Pillar{u_k}{i_k}}{\Pillar{v_k}{j_k}}$ for $1\leq k\leq n$. We say that $k$ is a \emph{breaking point} of $\fs$ if $j_k\not\in\dom\sigma'$. Let now $B=\{k_1<\cdots<k_m\}$ be the set of breaking points of $\fs$, ordered from the smallest to the greatest. We partition $\fs$ into $m$ sequences, as follows: $\fs^1=\fs_1,\ldots\fs_{k_1}$, and, for $2\leq h\leq m$, $\fs^h=\fs_{k_{h-1}},\ldots,\fs_{k_h}$. Once again, it is not hard to check that, for all $1\leq h\leq m$, $\fs^h\in\TrSeqFin{\sigma'}{\fE}$, and that $\fs'=\fa(\fs^1),\ldots,\fa(\fs^m)\in\TrSeqFin{\sigma}{\Trace{\sigma'}{\fE}}$, with $\fa(\fs')=\fa(\fs)$.\qed

The trace of a simple edifice has a very nice property, namely that each arch in it is generated by a unique trace sequence:
\begin{lem}
	\label{lemma:UniqueSeq}
	Let $\fE$ be a simple edifice, let $\sigma$ be a feedback function, and let $\fs,\fs'\in\TrSeqFin{\sigma}{\fE}$, such that $\fa(\fs)=\Arch{\zeta_1}{\zeta_2}$ and $\fa(\fs')=\Arch{\zeta'_1}{\zeta'_2}$. Then, $\zeta_1=\zeta'_1$ implies $\fs=\fs'$; in particular, $\fa(\fs)=\fa(\fs')$ implies $\fs=\fs'$.
\end{lem}
\proof Let $k$ be the smallest integer such that $\fs_k\neq\fs'_k$, and let $\fs_k=\Arch{\xi}{\upsilon}$, $\fs'_k=\Arch{\xi'}{\upsilon'}$. If $k=0$, $\xi=\zeta_1$ and $\xi'=\zeta'_1$, so $\xi=\xi'$ by hypothesis; if $k>0$, the chain and match conditions also imply $\xi=\xi'$, because we supposed $\fs_{k-1}=\fs'_{k-1}$. But $\fE$ is simple, so $\fs_k\coh\fs'_k$, which further implies $\upsilon=\upsilon'$. Then, $\fs_k=\fs'_k$, a contradiction.\qed
On the other hand, uniform edifices are preserved by the trace:
\begin{prop}
	\label{prop:UniformTrace}
	Let $\fE$ be a uniform edifice, and let $\sigma$ be a feedback function. Then, $\Trace{\sigma}{\fE}$ is uniform.
\end{prop}
\proof Let $\fF=\left\{\fa(\fs)~|~\fs\in\TrSeqFin{\sigma}{\fE}\right\}$. Let $\fs$ be a trace sequence; we shall prove, by induction on $\Len\fs$, that $\fa(\fs)$ is uniform in $\fF$, i.e., that there exists $\fx=\Arch{\Pillar{s}{i}}{\Pillar{t}{j}}$ such that $\fa\in\EdPath{\fx}\subseteq\fF$. This will be enough to prove the result; in fact, the trace sequences generating the arches of $\EdPath{\fx}$ are all visible if $\fs$ is visible, \ie, $\EdPath{\fx}\subseteq\Trace{\sigma}{\fE}$, because this depends only on $i$ and $j$, which are the same in all such sequences.

For the base case, $\Len\fs=1$, so $\fs$ consists of a single arch, and the result is a trivial consequence of the uniformity of $\fE$. The inductive case is $\Len\fs=n>1$. Let $\fs'=\fs_1,\ldots,\fs_{n-1}$. By induction hypothesis, $\fa(\fs')$ is uniform in $\fF$: we have an address $\fx'$ such that $\fa(\fs')\in\EdPath{\fx'}\subseteq\fF$. In particular, there exist $s,r_0\in\Words\times\Words$ and $w_0\in\Cc\times\Cc$ such that the first and last arches of $\fs'$ are respectively of the form
$$\Arch{\Pillar{sw_0}{i}}{\upsilon},$$
$$\Arch{\xi}{\Pillar{r_0w_0}{h}},$$
and, for all $w\in\Cc\times\Cc$, there exists a trace sequence $\fs^w$ whose first and last arches are respectively of the form
$$\Arch{\Pillar{sw}{i}}{\upsilon},$$
$$\Arch{\xi}{\Pillar{r_0w}{h}}.$$
Similarly, $\fE$ is uniform, so there exist $r_1,t\in\Words\times\Words$ and $w_1\in\Cc\times\Cc$ such that $$\fs_n=\Arch{\Pillar{r_1w_1}{l}}{\Pillar{tw_1}{j}},$$
with $\sigma(h)=l$, and, for all $w\in\Cc\times\Cc$, the arch
$$\fa^w=\Arch{\Pillar{r_1w}{l}}{\Pillar{tw}{j}}$$
is also in $\fE$. Now, by the match condition, we have $r_0w_0=r_1w_1$, which means that $r_0,r_1$ are one prefix of the other. Suppose $r_0=r_1r'$, and consider the trace sequences obtained by appending $\fa^{r'w}$ to $\fs^w$; these generate all the arches of the form $\Arch{\Pillar{r_0w}{i}}{\Pillar{tr'w}{j}}$, among which there is $\fa(\fs)$, which is therefore uniform in $\fF$. The other case is $r_1=r_0r'$; then, consider the sequences obtained by appending $\fa^w$ to $\fs^{r'w}$. As above, these prove that $\fa(\fs)$ is unifrom in $\fF$.\qed

\subsection{Nets as edifices}
\label{sect:NetsAsEd}
The basic idea to assign an edifice to a net is that arches model observable axioms/paths.\footnote{Graphically (\reffig{ObsAx}), observable axioms/paths look like arches, hence the terminology.} In fact, we have already seen that an observable axiom may be conveniently represented by an unordered pair of couples of the form $\Pillar{s}{i}$, where $s$ is the address of a leaf and $i$ a free port. A pillar contains the same information; the need for infinite words arises from $\eta$-expansion (the $\eta_0$ equation of \reffig{Eta}), which can be applied indefinitely, as in the pure \lac.
\begin{defi}[Edifice of a net]
	\label{def:EdNet}
	Let $\mu$ be a net. We associate an edifice with $\mu$, denoted by $\Ed\mu$, as follows:
	$$\Ed\mu=\bigcup_{\fx\in\AllObsPaths\mu}\EdPath\fx.$$
\end{defi}
The union of \refdef{EdNet} is actually disjoint, and the resulting edifice is simple:
\begin{lem}
	\label{lemma:DisjointVaults}
	Let $\mu$ be a net, and let $\fx,\fx'\in\AllObsPaths{\mu}$. Then
	\begin{enumerate}[\em(1)]
		\item $\EdPath{\fx}\cup\EdPath{\fx'}$ is simple, and hence $\Ed\mu$ is simple;
		\item $\fx\neq\fx'$ implies $\EdPath{\fx}\cap\EdPath{\fx'}=\emptyset$.
	\end{enumerate}
\end{lem}
\proof We start by observing that two arches in the same vault are coherent, because in an address $\Arch{\Pillar{s}{i}}{\Pillar{t}{j}}$ we always have $s\neq t$. This proves point (1) in case $\fx=\fx'$, so we may suppose $\fx\neq\fx'$. By confluence of $\beta$-reduction and by \refprop{ObsStab}, $\fx,\fx'\in\AllObsPaths{\mu}$ implies that there exists a net $\mu'$ such that $\mu\red\mu'$ and $\mu'$ contains two observable axioms of address $\fx,\fx'$, respectively, which are distinct because $\fx\neq\fx'$. Now, put $\fx=\Arch{\Pillar{s}{i}}{\Pillar{t}{j}}$ and $\fx'=\Arch{\Pillar{s'}{i'}}{\Pillar{t'}{j'}}$, and take $\fa\in\EdPath{\fx}$, $\fa'\in\EdPath{\fx'}$. By definition of vault, we have $\fa=\Arch{\Pillar{sw}{i}}{\Pillar{tw}{j}}$, $\fa'=\Arch{\Pillar{s'w'}{i'}}{\Pillar{t'w'}{j'}}$, for some $w,w'\in\Cc\times\Cc$. Suppose $\Pillar{sw}{i}=\Pillar{s'w'}{i'}$; we would obtain that $s$ is a prefix of $s'$, or vice versa. But this is absurd, because $s,s'$ are addresses of distinct leaves of $\mu'$. This  proves point (2); for point (1), simply note that the same holds for $t,t'$. To see that $\Ed\mu$ is itself is simple, note that a union of pairwise coherent simple edifices is obviously simple.\qed

Note that if $\mu$ has $n$ free ports and $\sigma$ is a feedback function whose domain is included in $\{1,\ldots,n\}$, then $\sigma$ defines a feedback context for $\mu$: it is the one connecting the free port $i$ to the free port $\sigma(i)$, or leaving it free if $\sigma(i)$ is undefined. Conversely, each feedback context for a net with $n$ free ports defines a feedback function of domain included in $\{1,\ldots,n\}$. Hence, we shall use $\sigma$ to range over both feedback functions and feedback contexts, and make no distinction between the two, speaking more generally of a ``feedback'' $\sigma$ for a net~$\mu$.

The next result shows how the trace construction is related to the execution formula of the GoI. In fact, in \refsect{GoI} we mentioned that this latter is invariant under reduction: if $\mu=\Ctxt{\sigma}{\nu}\onered\Ctxt{\sigma'}{\nu'}=\mu'$ with $\nu,\nu'$ cut-free and $\mu,\mu'$ total, then the formula of \refth{GoI} applied to $\GoI\sigma,\GoI\nu$ or $\GoI{\sigma'},\GoI{\nu'}$ yields the same result. Once again, our work generalizes this to non-total nets.
\begin{prop}[Invariance of the trace]
	\label{prop:StabTrace}
	Let $\mu\onered\mu'$, and let $\mu=\Ctxt{\sigma}{\nu}$ and $\mu'=\Ctxt{\sigma'}{\nu'}$ according to the Decomposition \reflemma{Decomposition}. Then, $\Trace{\sigma}{\Ed\nu}=\Trace{\sigma'}{\Ed{\nu'}}$.
\end{prop}
\proof The proof is a bit too long and not interesting enough to be included here. We prefer to give it in \refapp{TraceInv}.\qed

Let $\nu$ be a cut-free net, let $\sigma$ be a feedback for $\nu$, and let $\fa\in\Trace{\sigma}{\Ed\nu}$. Part (1) of \reflemma{DisjointVaults} and \reflemma{UniqueSeq} guarantee us that $\fa$ induces a unique $\fs\in\TrSeqFin{\sigma}{\Ed\nu}$ such that $\fa(\fs)=\fa$. If we let $\fs=\fs_1,\ldots,\fs_n$, we see that each $\fs_k$ determines an axiom of $\nu$, which is unique by part (2) of \reflemma{DisjointVaults}. Hence, $\fs$ induces a sequence of observable paths; thanks to the chain condition, these are all composable through $\sigma$, and, by the visibility condition, the first and last paths start and end at a free port of $\Ctxt{\sigma}{\nu}$. Therefore, their composition forms a maximal path of $\Ctxt{\sigma}{\nu}$.

To sum up, we found out that each arch $\fa$ of the trace of $\Ed\nu$ along $\sigma$ determines a unique maximal path of $\Ctxt{\sigma}{\nu}$; in what follows, we shall denote this path by $\phi(\fa)$.
\begin{lem}
	\label{lemma:VaultPath}
	Let $\nu$ be a cut-free net, let $\sigma$ be a feedback for $\nu$, let $\fx$ be an address, and let $\fa,\fa'\in\EdPath{\fx}\subseteq\Trace{\sigma}{\Ed\nu}$. Then, $\phi(\fa)=\phi(\fa')$.
\end{lem}
\proof Let $\fs_1,\ldots,\fs_n$ and $\fs'_1,\ldots,\fs'_{n'}$ be the trace sequences generating $\fa$ and $\fa'$, respectively. We decompose $\phi(\fa)$ and $\phi(\fa')$ into observable paths of $\nu$, following the $\fs_k,\fs'_k$, as described in the remarks above, and we obtain $\phi_1,\ldots,\phi_n$, $\phi'_1,\ldots,\phi'_{n'}$. Let $k$ be the smallest integer such that $\phi_k\neq\phi'_k$, and let $\fs_k=\Arch{\Pillar{u}{i}}{\Pillar{v}{j}}$, $\fs'_k=\Arch{\Pillar{u'}{i'}}{\Pillar{v'}{j'}}$. Suppose $k=0$; if we put $\fx=\Arch{\Pillar{s}{i}}{\Pillar{t}{j}}$, we have, for some $w,w'\in\Cc\times\Cc$, $u=sw$, $u'=sw'$, $i=i'$, and moreover for all $w_0\in\Cc\times\Cc$ we have an arch of the form $\Arch{\Pillar{sw_0}{i}}{\Pillar{tw_0}{j}}$ in $\Trace{\sigma}{\Ed\nu}$. By \refdef{TrSeq}, this latter implies that, for all $w_0\in\Cc\times\Cc$, there is some arch of the form $\Arch{\Pillar{sw_0}{i}}{\upsilon}$ in $\Ed\nu$. But, by looking at \refdef{EdNet}, we see that this is possible only if there is an axiom of $\nu$ of address $\Arch{\Pillar{s_0}{i}}{\Pillar{t'}{j'}}$, with $s_0$ a prefix of $s$. Then, the arches $\fs_0$ and $\fs'_0$ correspond to the same observable axiom, and $\phi_0=\phi'_0$. Therefore, we must have $k>0$. In this case, since $\phi_{k-1}=\phi'_{k-1}$, we have $\fs_{k-1}=\Arch{\xi}{\Pillar{v_{k-1}}{p}}$ and $\fs'_{k-1}=\Arch{\xi'}{\Pillar{v'_{k-1}}{p}}$, and the chain condition implies $i=i'=\sigma(p)$. But $v_{k-1},v'_{k-1}$ come from the same observable axiom, so they have a common prefix. By the match condition, $u,u'$ have a common prefix too; but this would be impossible if $\phi_k\neq\phi'_k$, because two distinct observable path generate pillars which have no common prefix (\cf\ the proof of \reflemma{DisjointVaults}). So we must conclude $\phi_k=\phi'_k$, a contradiction.\qed
\begin{lem}
	\label{lemma:MaxPathsAndTrace}
	Let $\nu$ be a cut-free net, let $\sigma$ be a feedback for $\nu$, and let $i,j$ be free ports of $\Ctxt{\sigma}{\nu}$. Then, the following are equivalent:
	\begin{enumerate}
		\item $\phi\in\MaxPathsP{i,j}{\Ctxt{\sigma}{\nu}}$, $\GoI{\PathWght\phi}=ts^\ast$;
		\item $\EdPath{\Arch{\Pillar{s}{i}}{\Pillar{t}{j}}}\subseteq\Trace{\sigma}{\Ed\nu}$.
	\end{enumerate}
\end{lem}
\proof We start with (1) implies (2), noting first that $\GoI{\PathWght\phi}=ts^\ast$ implies, by \reflemma{ExecPaths}, that $\phi$ is an execution path; therefore, we reason by induction on $\Val\phi$ (the value of $\phi$, \refdef{PathWeight}), using \reflemma{Val}.
\begin{enumerate}[$\bullet$]
	\item $\Val\phi=0$. We know that $\phi$ is an observable path. The situation can be schematically depicted as follows:
	\begin{center}\scalebox{0.8}{\input{ObsPathChain.pstex_t}}\end{center}
	In the above picture, $\phi$ goes ``from left to right'', and $\omega_1,\ldots,\omega_n$ are $n$ proper axioms of $\nu$. If, for $1\leq k\leq n$, we put $\Addr{\nu}{\omega_k}=\fx_k=\Arch{\Pillar{s_k}{i_k}}{\Pillar{t_k}{j_k}}$, from the above picture we deduce that $i_1=i$, $j_n=j$, $i,j\not\in\dom\sigma$, $i_{k+1}=\sigma(j_k)$ for all $1\leq k<n$, $s_k=\BiWord{\Id}{\Id}$ for all $2\leq k\leq n$, $s_1=s$, and $t_n\cdots t_1=t$. Now, given any $w\in\Cc\times\Cc$ and $1<k< n$, define $w_k=t_{k-1}\cdots t_1w$, and put
	\begin{eqnarray*}
		\fs_1 &=& \Arch{\Pillar{sw}{i}}{\Pillar{t_1w}{j_1}}, \\
		\fs_k &=& \Arch{\Pillar{w_k}{i_k}}{\Pillar{t_kw_k}{j_k}}, \textrm{ for $2\leq k\leq n$}.
	\end{eqnarray*}
	It is easy to check that $\fs=\fs_1,\ldots,\fs_n$ is a trace sequence of $\Ed\nu$ along $\sigma$ such that $\fa(\fs)=\Arch{\Pillar{sw}{i}}{\Pillar{tw}{j}}$, as desired.

	\item $\Val\phi>0$. We $\beta$-reduce $\Ctxt{\sigma}{\nu}$ along $\phi$ and obtain $\Ctxt{\sigma'}{\nu'}$ in which the residue $\phi'$ of $\phi$ is such that $\Val{\phi'}<\Val{\phi}$, so that the induction hypothesis applies to $\phi'$. By Propositions~\ref{prop:MonoRed} and \ref{prop:RewritingModel}, we have $\GoI{\PathWght{\phi'}}=ts^\ast$, so the induction hypothesis and \refprop{StabTrace} give us $\EdPath{\Arch{\Pillar{s}{i}}{\Pillar{t}{j}}}\subseteq\Trace{\sigma'}{\nu'}=\Trace{\sigma}{\nu}$.
\end{enumerate}

We now consider (2) implies (1). By \reflemma{VaultPath}, any arch of $\EdPath{\Arch{\Pillar{s}{i}}{\Pillar{t}{j}}}$ yields the same $\phi\in\MaxPathsP{i,j}{\Ctxt{\sigma}{\nu}}$. Hence, we need only check that $\GoI{\PathWght{\phi}}=ts^\ast$; we do this again by induction on $\Val\phi$.
\begin{enumerate}[$\bullet$]
	\item $\Val\phi=0$. We know that $\phi$ is an observable path of $\Ctxt\sigma\nu$, which we may decompose in several observable paths $\phi_1,\ldots,\phi_n$ of $\nu$, as above. Now, as we remarked in the proof of \reflemma{VaultPath}, $\EdPath{\Arch{\Pillar{s}{i}}{\Pillar{t}{j}}}\subseteq\Trace{\sigma}{\nu}$ is only possible if $\Pillar{s}{i}$ and $\Pillar{t}{j}$ are the addresses of the leaves of the trees resulting in the composition of $\phi_1,\ldots,\phi_n$, similarly to the drawing above used in the previous part of the proof. So $\GoI{\PathWght\phi}=ts^\ast$, as desired.
	\item $\Val\phi>0$. We know that $\phi$ crosses an active pair; however, this time we must prove that, by reducing it, we obtain a net in which $\phi$ has a residue. If this were not the case, by \reflemma{ExecPaths} and \refprop{MonoRed} the only possibility is that the active pair corresponds to a clash in $\PathWght{\phi}$, \ie, we have $\PathWght{\phi}=A'A_0^\ast B_0B'$ for some monomials $A',B'$ and some positive atoms $A_0,B_0$, such that, for example, $A_0=\opc$ and $B_0=\opd$. This would yield the presence of two observable axioms of $\nu$ of addresses $\fx=\Arch{\Pillar{s}{p}}{\Pillar{\BiWord{\q b}{b'}}{q}}$ and $\fx'=\Arch{\Pillar{\BiWord{\p a}{a'}}{q'}}{\Pillar{t}{p'}}$, such that $\sigma(q)=q'$, \ie, the free ports $q,q'$ of $\nu$ are connected by $\sigma$ in $\Ctxt{\sigma}{\nu}$. Then, it is easy to see that no arch of $\Ed\nu$ generated by $\fx$ and $\fx'$ could match: indeed, there are no $x,y,x',y'\in\Cc$ such that $\BiWord{\q bx}{b'y}=\BiWord{\p ax'}{a'y'}$. But this is absurd, because by hypothesis $\phi$ comes from an arch of $\Trace{\sigma}{\nu}$, which in turn comes from a unique trace sequence, and trace sequences satisfy the match condition.

	Therefore, we now know that we may reduce the active pair crossed by $\phi$, obtaining $\Ctxt{\sigma}{\nu}\onered\Ctxt{\sigma'}{\nu'}$ such that $\phi$ has a residue $\phi'$ in $\Ctxt{\sigma'}{\nu'}$. Now, by \refprop{StabTrace}, we have $\Trace{\sigma'}{\Ed{\nu'}}=\Trace{\sigma}{\Ed\nu}$, so $\EdPath{\Arch{\Pillar{s}{i}}{\Pillar{t}{j}}}\subseteq\Trace{\sigma'}{\Ed{\nu'}}$. 
	Thanks to \refprop{MonoRed} and \reflemma{MonoVal}, we can apply the induction hypothesis to $\phi'$, and obtain $\GoI{\PathWght{\phi'}}=ts^\ast$. But, by \refprop{RewritingModel}, $\GoI{\PathWght{\phi}}=\GoI{\PathWght{\phi'}}$, and we are done.
\end{enumerate}\qed
\begin{prop}
	\label{prop:CutFreeTrace}
	Let $\nu$ be a cut-free net, and $\sigma$ a feedback for $\nu$. Then
	$$\Ed{\Ctxt{\sigma}{\nu}}=\Trace{\sigma}{\Ed{\nu}}.$$
\end{prop}
\proof We start with the inclusion from left to right. Let $\fa\in\Ed{\Ctxt{\sigma}{\nu}}$. By \refdef{EdNet}, there is an address $\fx=\Arch{\Pillar{s}{i}}{\Pillar{t}{j}}\in\AllObsPaths{\Ctxt{\sigma}{\nu}}$ (which is also unique by \reflemma{DisjointVaults}) such that $\fa\in\EdPath{\fx}$; by \refprop{ObsAxAndGoI}, there exists $\phi\in\MaxPathsP{i,j}{\Ctxt{\sigma}{\nu}}$ such that $\GoI{\PathWght{\phi}}=ts^\ast$, and by \reflemma{MaxPathsAndTrace} $\EdPath{\fx}\subseteq\Trace{\sigma}{\Ed\nu}$.

For the inclusion from right to left, let $\fa\in\Trace{\sigma}{\Ed\nu}$. By \reflemma{UniformVaults}, $\Ed\nu$ is uniform, so by \refprop{UniformTrace} $\Trace{\sigma}{\Ed\nu}$ is also uniform. Then, there must be an address $\fx=\Arch{\Pillar{s}{i}}{\Pillar{t}{j}}$ such that $\fa\in\EdPath{\fx}\subseteq\Trace{\sigma}{\Ed\nu}$. We can thus apply \reflemma{MaxPathsAndTrace} and \refprop{ObsAxAndGoI}, which, by \refdef{EdNet}, give us $\EdPath{\fx}\subseteq\Ed{\Ctxt{\sigma}{\nu}}$.\qed

Compare \refprop{CutFreeTrace} with \refth{GoI}: basically, the trace construction can be seen as an extension of the execution formula, which works in all cases, even when $\Ctxt{\sigma}{\nu}$ is not total. Something similar happens when one formulates the GoI in categorical terms, using certain traced monoidal categories, as shown by Haghverdi and Scott~\cite{HaghverdiScott:CatGoI}. Thanks to the associativity of the trace, we can straightforwardly extend \refprop{CutFreeTrace} to arbitrary nets:
\begin{cor}
	\label{cor:Trace}
	For any net $\mu$ and feedback $\sigma$ for $\mu$, $\Ed{\Ctxt{\sigma}{\mu}}=\Trace{\sigma}{\Ed{\mu}}$.
\end{cor}
\proof By the Decomposition \reflemma{Decomposition}, we know that $\mu=\Ctxt{\sigma'}{\nu}$ for some cut-free net $\nu$ and feedback context $\sigma'$. Then, using \refprop{CutFreeTrace} and \reflemma{TraceAssoc}, we have
{\setlength\arraycolsep{2pt}\begin{eqnarray*}
	\Ed{\Ctxt{\sigma}{\mu}} & = & \Ed{\Ctxt{\sigma}{\Ctxt{\sigma'}{\nu}}}=\Ed{\Ctxt{(\sigma\uplus\sigma')}{\nu}}=\Trace{\sigma\uplus\sigma'}{\Ed\nu}=\\
	& = & \Trace{\sigma}{\Trace{\sigma'}{\Ed\nu}}=\Trace{\sigma}{\Ed{\Ctxt{\sigma'}{\nu}}})=\Trace{\sigma}{\Ed\mu},
\end{eqnarray*}}
where we used the notation $\sigma\uplus\sigma'$ also to denote the ``union'' of the nets $\sigma,\sigma'$.\qed

\subsection{A denotational semantics}
\label{sect:DenSem}
As explained in the introduction, a denotational semantics is an interpretation of the syntax transforming certain given syntactic equivalences into denotational equalities. In a syntax such as the \lac, or the symmetric combinators, the typical equivalence to be modelled is that induced by $\beta$-reduction. As proposed for example by Girard \cite{Girard:LC}, we may describe a denotational semantics of a syntax with a reduction relation $\genred$ and an internal notion of context as an interpretation satisfying at least the following:
\begin{description}
	\item[invariance] for any two syntactic objects $a,b$, $a\genred b$ implies that $a$ and $b$ are denotationally equal;
	\item[contextuality] the semantics induces a congruence on the syntax, \ie, if two syntactic objects $a,b$ are denotationally equal, then for every context $C$, $C[a]$ and $C[b]$ are also denotationally equal.
\end{description}
In this section, we shall prove that our interpretation of nets as edifices satisfies these two requirements. Actually, we shall see that, instead of just $\beta$-reduction, edifices model~$\BetaEpsilonEtaEq$.

\begin{lem}
	\label{lemma:BetaEtaEpsInEd}
	Let $\etazerored$ be the relation of \reflemma{EtaMinus}, \ie, a single application of the $\eta_0$ equation (\reffig{Eta}), oriented from left to right, in which the wire on the right member is an axiom. Moreover, let $\EtaEqOne^1$ be the contextual closure of the $\eta_1$ equation (\reffig{Eta}), \ie, the restriction of $\EtaEqOne$ to just one application of the equation.
	For all nets $\mu,\nu$, we have:
	\begin{enumerate}
		\item $\mu\eonered\nu$ implies $\Ed\mu=\Ed\nu$;
		\item $\mu\etazerored\nu$ implies $\Ed\mu=\Ed\nu$;
		\item $\mu\EtaEqOne^1\nu$ implies $\Ed\mu=\Ed\nu$.
	\end{enumerate}
\end{lem}
\proof Point (1) is an immediate consequence of \refprop{ObsStabE} and of \refdef{EdNet}.

For what concerns point (2), considering the Decomposition \reflemma{Decomposition} we obtain $\mu=\Ctxt{\sigma}{\mu_0}$ and $\nu=\Ctxt{\sigma}{\nu_0}$ such that $\mu_0,\nu_0$ are cut-free and $\mu_0=\ctxt{o}$, $\nu_0=\ctxt{\omega}$, where $o$ is a net matching the left member of the $\eta_0$ equation of \reffig{Eta}, and $\omega$ is an axiom of $\nu_0$. We suppose that $o$ consists of two $\D$ cells, the case of two $\Z$ cells being perfectly analogous. Now, remark that, if $\fx_0=\Addr{\nu_0}{\omega}=\Arch{\Pillar{\BiWord{a}{b}}{i}}{\Pillar{\BiWord{c}{d}}{j}}$, we have
$$\ObsPaths{\mu_0}=\{\fx\in\ObsPaths{\nu_0}~|~\fx\neq\fx_0\}\cup\{\Arch{\Pillar{\BiWord{a\p}{b}}{i}}{\Pillar{\BiWord{c\p}{d}}{j}},\Arch{\Pillar{\BiWord{a\q}{b}}{i}}{\Pillar{\BiWord{c\q}{d}}{j}}\}.$$
Moreover, note that, since $\mu_,\nu_0$ are cut-free, $\AllObsPaths{\mu_0}=\ObsPaths{\mu_0}$ and $\AllObsPaths{\nu_0}=\ObsPaths{\nu_0}$. So take $\fa\in\Ed{\mu_0}$. If $\fa$ is generated by an address in the left term of the above union, then we clearly have $\fa\in\Ed{\nu_0}$. Otherwise, we have for example $\fa=\Arch{\Pillar{\BiWord{a\p x}{by}}{i}}{\Pillar{\BiWord{c\p x}{dy}}{j}}$ for some $x,y\in\Cc$. But in this case too $\fa\in\Ed{\nu_0}$, because $\fx_0\in\ObsPaths{\nu_0}$. Of course a similar reasoning applies if we had chosen $\q$ instad of $\p$, so $\Ed{\mu_0}\subseteq\Ed{\nu_0}$. Conversely, let $\fa\in\Ed{\nu_0}$. Again, if $\fa$ is generated by $\fx\neq\fx_0$, then clearly $\fa\in\Ed{\mu_0}$. Otherwise, we have $\fa=\Arch{\Pillar{\BiWord{ax}{by}}{i}}{\Pillar{\BiWord{cx}{dy}}{j}}$ for some $x,y\in\Cc$. But $x$ must be of the form $\p x'$ or $\q x'$; in either case, we see that $\ObsPaths{\mu_0}$ contains an address generating $\fa$, so $\Ed{\nu_0}\subseteq\Ed{\mu_0}$. Point (2) can now be obtained by applying \refprop{CutFreeTrace}: $\Ed\mu=\Ed{\Ctxt{\sigma}{\mu_0}}=\Trace{\sigma}{\Ed{\mu_0}}=\Trace{\sigma}{\Ed{\nu_0}}=\Ed{\Ctxt{\sigma}{\nu_0}}=\Ed\nu$.

For point (3), we apply again the Decomposition \reflemma{Decomposition} and write $\mu=\Ctxt{\sigma}{\mu_0}$, $\nu=\Ctxt{\sigma}{\nu_0}$ with $\mu_0,\nu_0$ cut-free and such that $\mu_0\EtaEqOne^1\nu_0$. Observe now that the addresses of leaves are invariant under the $\eta_1$ equation of \reffig{Eta}; hence, $\ObsPaths{\mu_0}=\ObsPaths{\nu_0}$. Again, $\AllObsPaths{\mu_0}=\ObsPaths{\mu_0}$ and $\AllObsPaths{\nu_0}=\ObsPaths{\nu_0}$, because $\mu_0,\nu_0$ are cut-free. So we have $\Ed{\mu_0}=\Ed{\nu_0}$, and we can conclude once more by applying \refprop{CutFreeTrace}.\qed
\begin{prop}
	\label{prop:BetaEtaEpsModel}
	For all nets $\mu,\nu$, $\mu\BetaEpsilonEtaEq\nu$ implies $\Ed\mu=\Ed\nu$.
\end{prop}
\proof A straightforward consequence of \reflemma{BetaEtaEpsInEd}, using \reflemma{EtaMinus}.\qed
\begin{prop}
	\label{prop:EdCong}
	Let $\mu,\nu$ be two nets such that $\Ed\mu=\Ed\nu$. Then, for every context $C$, $\Ed{\ctxt\mu}=\Ed{\ctxt\nu}$.
\end{prop}
\proof Observe that applying a context $C$ to net $\mu$ with $n$ free ports can be done in two steps: first, we juxtapose $C$ and $\mu$, forming the net which we denote by $C\bullet\mu$. We stipulate that, in $C\bullet\mu$, the free ports of $\mu$ are labelled by $1,\ldots,n$, whereas the free ports of $C$ are ``shifted'' by $n$, \ie, they are labelled starting from $n+1$. Then, we consider the feedback $\sigma$ such that $\sigma(i)=i+n$ for $i\in\{1,\ldots,n\}$, $\sigma(i)=i-n$ for $i\in\{n+1,\ldots,2n\}$, and $\sigma$ is undefined everywhere else. We clearly obtain $\Ctxt{\sigma}{C\bullet\mu}=\ctxt\mu$. Note furthermore that $\Ed{C\bullet\mu}=\Ed{C}\cup\Ed{\mu}$, since the two nets are disjoint and do not share free ports by our assumption. The result is then an easy consequence of \refcor{Trace}:
{\setlength{\arraycolsep}{2pt}\begin{eqnarray*}
	\Ed{\ctxt\mu} &=& \Ed{\Ctxt{\sigma}{C\bullet\mu}}=\Trace{\sigma}{\Ed{C\bullet\mu}}=\Trace{\sigma}{\Ed{C}\cup\Ed{\mu}}=\\
	&=& \Trace{\sigma}{\Ed{C}\cup\Ed{\nu}}=\Trace{\sigma}{\Ed{C\bullet\nu}}=\Ed{\Ctxt{\sigma}{C\bullet\mu}}=\Ed{\ctxt\nu}.
\end{eqnarray*}}\qed

\section{Full Abstraction}
\label{sect:FullAbs}

\subsection{Edifices and the Cantor topology}
\label{sect:Cantor}
Our aim now is to show that edifices are able to fully characterize the observational equivalences introduced in \refsect{ObsAx}. For this, we shall take the sets introduced in \refdef{Edifice} and equip them with topological structures based on the Cantor topology. This will be needed for two reasons: first, to characterize the edifices which are interpretations of $\beta\E$-normalizable nets, a result which will be fundamental in characterizing finitary axiom-equivalence; second, to obtain a characterization of axiom-equivalence itself.

The idea of using topology for semantic purposes is of course far from being new: it is enough to think that the very basis of the denotational semantics of the \lac\ is Scott's intuition that computability should be interpreted by topological continuity~\cite{Scott}. Moreover, also non-Scott topologies have been attached to \lat s to obtain various kinds of results (Visser's topology is an example~\cite{Visser}). Closer to our work, we can mention the work of Kennaway et al.~\cite{KennawayEtAl:InfinitaryLambda}, who also used a Cantor-like topology, very similar to our own, to define the infinitary \lac.
\begin{defi}[Arch topology]
	The set $\Cc=\{\p,\q\}^\Nat$ may be equipped with the Cantor topology. This is well known to be metrizable, with the distance defined for example by $d_\Cc(x,y)=2^{-k}$, where $k$ is the length of the longest common prefix of $x,y$. We denote by $\OpenBall{x}{r}$ the open ball of center $x$ and radius $r$.

	As well known, $\Cc\times\Cc$ is also a Cantor space; if we equip $\Nat$ with the discrete topology, we can endow the set $\cP$ of pillars with the product topology. This is also metrizable: if $\xi=\Pillar{\BiWord x y}{i}$ and $\upsilon=\Pillar{\BiWord{x'}{y'}}{i'}$, we shall consider the distance
	$$d(\xi,\upsilon)=\max\{d_\Cc(x,x'),d_\Cc(y,y'),d_{\mathrm{disc}}(i,i')\},$$
	where $d_\mathrm{disc}$ is the discrete metric, defined as $d_\mathrm{disc}(i,i')=0$ if $i=i'$, and $d_\mathrm{disc}(i,i')=2$ if $i\neq i'$. Therefore, to be ``close'', two pillars must be based at the same integer.

	Similarly, we equip $\overrightarrow{\cA}$ with the product topology; the \emph{arch topology}, applied on the set $\cA$ of arches, is the quotient topology with respect to the relation $\archsim$ of \refdef{Edifice}.
\end{defi}
The following helps understanding the arch topology:
\begin{prop}
	\label{prop:Metrizable}
	The space $\cA$ is metrizable; if $\fa=\Arch{\xi}{\upsilon}$ and $\fa'=\Arch{\xi'}{\upsilon'}$, the function $D(\fa,\fa')=\min\{\max\{d(\xi,\xi'),d(\upsilon,\upsilon')\},\max\{d(\xi,\upsilon'),d(\upsilon,\xi')\}\}$ is a distance inducing its topology.\qed
\end{prop}
In other words, to compare two arches, we overlap them in both possible ways, and we take the way that ``fits best''. The distance $D$ is in fact the standard quotient metric; in this case, it collapses to this simple form.

The space $\cA$ is not a Cantor space, because it is not compact. In fact, we can give a characterization of its compact subsets. Recall from \refdef{Edifice} that, if $I\subseteq\Nat$, $\cA_I$ is the set of arches based within $I$. Then, we have
\begin{prop}
	\label{prop:Compactness}
	$\fE\subseteq\cA$ is compact iff it is a closed subset of $\cA_I$ for some finite $I$.
\end{prop}
\proof If $\fE$ is compact, then it must be closed; suppose however that $\fE\not\subseteq\cA_I$ for all finite $I$. Then, let $\fa_{i,j}$ be a sequence of arches spanning all of the $i,j$ where the arches of $\fE$ are based, and set $U_{i,j}=\fE\cap\OpenBall{\fa_{i,j}}{2}$. These are all open sets in the relative topology, and since, for all $i,j$, $D(\fa_{i,j},\fa)<2$ iff $\fa$ is based at $i,j$, they form an open cover of $\fE$. Now observe that, by the same remark on the distance, if we remove any $U_{m,n}$ we loose all arches of $\fE$ based at $m,n$. But we have supposed the sequence $\fa_{i,j}$ to be infinite, so $U_{i,j}$ is an infinite open cover of $\fE$ admitting no finite subcover, in contradiction with the compactness of $\fE$.

For the converse, $I$ being finite, it is not hard to show that $\cP_I$ is homeomorphic to $\Cc$. Therefore, $\cP_I$ is a Cantor space, hence compact. So $\cA_I$ is compact, because it is the quotient of a product of compact spaces. But a closed subset of a compact space is compact, hence the result.\qed

It can be shown that each $\cA_I$ is also perfect and totally disconnected, which means that actually these are all Cantor spaces whenever $I$ is finite. What really matters to us though is compactness, which implies completeness (with respect to the metric $D$ of \refprop{Metrizable}): when $I$ is finite, there is identity between closed, compact, and complete subsets of $\cA_I$.

Vaults are examples of compact edifices:
\begin{lem}
	\label{lemma:EdPathClosed}
	Vaults are compact.
\end{lem}
\proof Let $\fA=\EdPath{\fx}$ be a vault, with $\fx=\Arch{\Pillar{s}{i}}{\Pillar{t}{j}}$. Clearly $\EdPath{\fx}\subseteq\cA_{\{i,j\}}$. Now take an arch $\fa=\Arch{\Pillar{u}{i'}}{\Pillar{v}{j'}}$ not belonging to $\EdPath{\fx}$. If $i'\neq i$ or $j'\neq j$, then obviously $\OpenBall{\fa}{1}$ is all outside of $\EdPath{\fx}$. Otherwise, either $s$ is not a prefix of $u$, or $t$ is not a prefix of $v$; suppose we are in the first situation, and let $k$ be the length of the longest common prefix between $u$ and $s$. Then, it is easy to see that $\OpenBall{\fa}{2^{-(k+1)}}$ is all outside of $\EdPath{\fx}$. So $\EdPath{\fx}$ is a closed subset of $\cA_{\{i,j\}}$, and we conclude by \refprop{Compactness}.\qed
Observe that vaults are not open: given an arch $\fa\in\EdPath{\Arch{\Pillar{s}{i}}{\Pillar{t}{j}}}$, any non-empty open ball centered at $\fa$ contains arches of the form $\Arch{\Pillar{su}{i}}{\Pillar{tv}{j}}$, with $u\neq v$.

It turns out that the edifice of a net is compact exactly when the net is $\beta\E$-normalizable. This result, which we shall now prove, shows why we are interested in the compact sets of the arch topology. If $a$ is a finite binary word, we denote by $\Len a$ its length. If $k\geq\Len a$, we denote by $\Cmp k a$ the set of words $b$ of length $k$ such that $b=ab'$ for some word $b'$, \ie, all possible ``extensions'' of $a$ to length $k$. Let now $\fx=\Arch{\Pillar{\BiWord{a_0}{b_0}}{i}}{\Pillar{\BiWord{c_0}{d_0}}{j}}$ be an address, and let $k_\fx=\max\{\Len{a_0},\Len{b_0},\Len{c_0},\Len{d_0}\}$. We define the set of \emph{centers} of $\fx$ as
\begin{eqnarray*}
	\Centers{\phi} &=& \{\Arch{\Pillar{\BiWord{ax_0}{bx_0}}{i}}{\Pillar{\BiWord{cx_0}{dx_0}}{j}}~|\\
	&& a\in\Cmp{k_\fx}{a_0},\ b\in\Cmp{k_\fx}{b_0},\ c\in\Cmp{k_\fx}{c_0},\ d\in\Cmp{k_\fx}{d_0}\},
\end{eqnarray*}
where $x_0$ is some fixed infinite word, whose value is irrelevant. Then we set
$$\PathOpen{\fx}=\bigcup_{\fa\in\Centers{\fx}}\OpenBall{\fa}{2^{-{k_\fx+1}}}.$$
The set $\PathOpen{\fx}$ is clearly open; additionally, we have
\begin{lem}
	\label{lemma:EdInOpen}
	For every address $\fx$, $\EdPath{\fx}\subseteq\PathOpen{\fx}$.\qed
\end{lem}
\proof Let $\fx=\Arch{\Pillar{\BiWord{a}{b}}{i}}{\Pillar{\BiWord{c}{d}}{j}}$. We assume without loss of generality that $a$ is the longest of $a,b,c,d$. Then, given $\fa=\Arch{\Pillar{\BiWord{ax}{by}}{i}}{\Pillar{\BiWord{cx}{dy}}{j}}\in\EdPath{\fx}$, we can always write, for some $a_1,a_2,a_3\in\Words$, $y_1,x_2,y_3\in\Cc$, $by=ba_1y_1$, $cx=ca_2x_2$, $dy=da_3y_3$, such that $\Len{ba_1}=\Len{ca_2}=\Len{da_3}=\Len{a}=k_\fx$. By definition, $\fa_0=\Arch{\Pillar{\BiWord{ax_0}{ba_1x_0}}{i}}{\Pillar{\BiWord{ca_2x_0}{da_3x_0}}{j}}$ is a center of $\fx$, and we have $D(\fa_0,\fa)=2^{-k_\fx}<2^{-k_\fx+1}$, so $\fa\in\OpenBall{\fa_0}{2^{-{k_\fx+1}}}$.\qed
A version of part (2) of \reflemma{DisjointVaults} can be given for the sets $\PathOpen{\fx}$:
\begin{lem}
	\label{lemma:EdsDisjoint}
	Let $\mu$ be a net, and let $\fx,\fx'\in\ObsPaths{\mu}$, with $\fx\neq\fx'$. Then, $\PathOpen{\fx}\cap\PathOpen{\fx'}=\emptyset$.
\end{lem}
\proof Let $\fx=\Arch{\Pillar{s}{i}}{\Pillar{t}{j}}$, $\fx=\Arch{\Pillar{s'}{i'}}{\Pillar{t'}{j'}}$, and let $\fx,\fx'\in\ObsPaths\mu$. If $i\neq i'$ or $j\neq j'$, then the result is obvious. Otherwise, by the same arguments given in the proof of \reflemma{DisjointVaults}, $s$ and $t$ cannot be prefixes of $s'$ or $t'$, and vice versa. Now, the sets $\PathOpen{\fx},\PathOpen{\fx'}$ are built precisely so that, whenever $\Arch{\Pillar{u}{i}}{\Pillar{v}{j}}\in\EdPath{\fx}$ and $\Arch{\Pillar{u'}{i}}{\Pillar{v'}{j}}\in\EdPath{\fx'}$, $s,t$ are prefixes of resp.\ $u,v$, and $s',t'$ are prefixes of resp.\ $u',v'$; hence, the two sets cannot have any arch in common.\qed
\begin{prop}
	\label{prop:CompactEd}
	For all $\mu$, $\Ed{\mu}$ is compact iff $\mu$ is $\beta\E$-normalizable.
\end{prop}
\proof The backward implication is a straightforward consequence of \refcor{BENormAndObsPaths} and \reflemma{EdPathClosed} (a finite union of compact sets is compact). Suppose now that $\mu$ is not $\beta\E$-normalizable. Again thanks to \refcor{BENormAndObsPaths}, we know that $\AllObsPaths{\mu}$ is infinite. Consider now the family of sets $\PathOpen{\fx}\cap\Ed{\mu}$ as $\fx$ varies over $\AllObsPaths{\mu}$; by \reflemma{EdInOpen}, this forms an infinite open cover of $\Ed{\mu}$. By \reflemma{EdsDisjoint}, removing any of these sets causes the family not to cover $\Ed{\mu}$ anymore; hence, $\Ed{\mu}$ is not compact.\qed

\subsection{Closed edifices}
\label{sect:Closure}
Since the edifice of a net is always a subset of $\cA_I$ with $I$ finite, by \refprop{Compactness} we have a standard way to ``compactify'' it: we simply take its topological closure, denoted by $\Closure{(\cdot)}$.
\begin{prop}
	\label{prop:ClosureCompact}
	Let $I$ be a finite subset of $\Nat$. Then, for every $\fE\subseteq\cA_I$, $\Closure\fE$ is compact.
\end{prop}
\proof The arches based outside of $I$ are ``too far'' to be adherent to $\fE$, therefore its closure is still in $\cA_I$. By \refprop{Compactness}, this is enough to ensure compactness.\qed
\begin{defi}[Closed edifice of a net, closed trace]
	Let $\mu$ be a net. The \emph{closed edifice} of $\mu$ is defined as $\ClosedEd{\mu}=\Closure{\Ed\mu}$. Similarly, if $\fE$ is an edifice and $\sigma$ is a feedback, the \emph{closed trace} of $\fE$ along $\sigma$ is defined as $\ClosedTrace\sigma\fE=\Closure{\Trace{\sigma}{\fE}}$.
\end{defi}

Closed edifices also define a denotational semantics of the symmetric combinators. The fact that they model $\BetaEpsilonEtaEq$ is an immediate consequence of \refprop{BetaEtaEpsModel}, because $\Ed\mu=\Ed\nu$ quite obviously implies $\ClosedEd\mu=\ClosedEd\nu$. What is left to prove, is that they yield a congruence, which we do next.

The following result, which is proved by a slightly tricky induction, tells us that if the closure of a uniform edifice $\fF$ contains a uniform edifice $\fE$, then the trace sequences of $\fE$ along any feedback may be arbitrarily approximated by trace sequences of $\fF$ along the same feedback. The hypothesis that $\fE,\fF\subseteq\cA_I$ for $I$ finite is needed so that \refprop{ClosureCompact} can be tacitly applied.
\begin{lem}
	\label{lemma:ApproxComp}
	Let $\fE,\fF\subseteq\cA_I$ be uniform edifices, with $I$ finite, such that $\fE\subseteq\Closure\fF$, and let $\sigma$ be a feedback. Then, for all $\fa\in\Trace{\sigma}{\fE}$ and for all $\epsilon>0$, there exists $\fb\in\Trace{\sigma}{\fF}$ such that $D(\fa,\fb)<\epsilon$.
\end{lem}
\proof Let $\widetilde\fE=\{\fa(\fs)~|~\fs\in\TrSeqFin{\sigma}{\fE}\}$, and similarly $\widetilde\fF=\{\fa(\fs)~|~\fs\in\TrSeqFin{\sigma}{\fF}\}$. Given $\epsilon>0$ and a trace sequence $\fs=\fs_1,\ldots,\fs_n$ of $\fE$ along $\sigma$, we shall prove by induction on $n$ that there exists a vault $\fA\subseteq\widetilde\fF$ such that $\fA\subseteq\OpenBall{\fa(\fs)}{\epsilon}$. This will be enough to conclude, because whenever $\fs$ is visible, we have $\fA\subseteq\Trace{\sigma}{\fF}$, and any arch in $\fA$ satisfies the thesis.

In the base case, $n=1$, so $\fs$ consists of a single arch $\fa\in\fE\subseteq\Closure\fF$. By definition, $\fa$ can be arbitrarily approximated in $\fF$, \ie, given arbitrarily long $s,t$ such that $\fa=\Arch{\Pillar{su}{i}}{\Pillar{tv}{j}}$, there exists $\fb\in\fF$ such that $\fb=\Arch{\Pillar{su'}{i}}{\Pillar{tv'}{j}}$. It will then be enough to show that $\fF$ contains a vault contained in $\EdPath{\Arch{\Pillar{s}{i}}{\Pillar{t}{j}}}$, because $s,t$ are arbitrarily long, and $\fF\subseteq\widetilde\fF$. Now, by uniformity of $\fF$, we must have $\fb\in\EdPath{\Arch{\Pillar{s_0}{t}}{\Pillar{t_0}{j}}}\subseteq\fF$ for some $s_0,t_0$, and $\fb=\Arch{\Pillar{s_0w}{i}}{\Pillar{t_0w}{j}}$ for some $w\in\Cc\times\Cc$. Observe that $s_0,s$ and $t_0,t$ must then be prefixes of each other. Therefore, we have four cases, depending on the possible combinations of which is prefix of which:
\begin{enumerate}[$\bullet$]
	\item $s=s_0s'$ and $t=t_0t'$. Then both $s'$ and $t'$ are prefixes of $w$, which means that they are prefixes of each other. Suppose that $t'=s't''$; then, $\fb=\Arch{\Pillar{s_0s'w'}{i}}{\Pillar{t_0s't''w'}{j}}$ for some $w'$, and $\EdPath{\Arch{\Pillar{s}{i}}{\Pillar{t}{j}}}\subseteq\fF$. The case $s'=t's''$ is symmetric.
	\item $s=s_0s'$ and $t_0=tt'$. We have $w=s'w'$, and $\fb=\Arch{\Pillar{s_0s'w'}{i}}{\Pillar{tt's'w'}{j}}$, which means that $\EdPath{\Arch{\Pillar{s}{i}}{\Pillar{tt's'}{j}}}\subseteq\fF$.
	\item $s_0=ss'$ and $t=t_0t'$. This case is symmetric to the one above.
	\item $s_0=ss'$ and $t_0=tt'$. Then we may conclude, because $\EdPath{\Arch{\Pillar{s_0w}{i}}{\Pillar{t_0w}{j}}}\subseteq\EdPath{\Arch{\Pillar{s}{i}}{\Pillar{t}{j}}}$.
\end{enumerate}

Let now $n\geq 1$, and let $\fs=\fs_1,\ldots,\fs_n,\fs_{n+1}$. If we put $\fs'=\fs_1,\ldots,\fs_n$, by \refprop{UniformTrace}, we have that $\fa(\fs')$ is uniform in $\widetilde\fE$; furthermore, by hypothesis, $\fs_{n+1}$ is uniform in $\fE$. Hence, there exist $r_0,s_0,s_1,t_0$ and $h,i,j,l$ such that
\begin{eqnarray*}
	&&\fa(\fs')\in\EdPath{\Arch{\Pillar{r_0}{h}}{\Pillar{s_0}{i}}}\subseteq\widetilde\fE,\\
	&&\fs_{n+1}\in\EdPath{\Arch{\Pillar{s_1}{j}}{\Pillar{t_0}{l}}}\subseteq\fE,
\end{eqnarray*}
with $\sigma(i)=j$. Actually, we may always suppose $s_1=s_0$. In fact, by the match condition, $s_0,s_1$ are prefixes of each other; suppose for instance that $s_0=s_1s'$; then, we have $\fa(\fs')\in\EdPath{\Arch{\Pillar{r_0s'}{h}}{\Pillar{s_1}{i}}}$, which is still contained in $\widetilde\fE$. A symmetric argument applies in case $s_1=s_0s'$.

Let us now apply the induction hypothesis, obtaining $\EdPath{\Arch{\Pillar{r}{h}}{\Pillar{s}{i}}}\subseteq\widetilde\fF$, with $r=r_0r'$ and $s=s_0s'$. The length of $r',s'$ depends on how much we want to approximate $\fa(\fs')$; in fact, we know that the induction hypothesis allows us to use \emph{any} approximation, as precise as we want. Now, observe that, thanks to the match condition, $s_0s'$ and $t_0s'$ are prefixes of the words in $\fs_{n+1}$; therefore, because our ultimate goal is to approach $\fa(\fs)$ within distance $\epsilon$, we apply the induction hypothesis with an $\epsilon'$ small enough so that $r$ and $t_0s'$ are long enough prefixes of the words contained in $\fa(\fs)$ to satisfy the requirement, \ie, so that, for all $u,v\in\Cc\times\Cc$, we have $D(\fa(\fs),\Arch{\Pillar{ru}{h}}{\Pillar{t_0s'v}})<\epsilon$.

Now, observe that $\EdPath{\Arch{\Pillar{s_0}{j}}{\Pillar{t_0}{l}}}\subseteq\fE\subseteq\Closure\fF$ means that all of the arches of this vault can be arbitrarily approximated in $\fF$; this applies in particular to the arches of the vault $\EdPath{\Arch{\Pillar{s_0s'}{j}}{\Pillar{t_0s'}{l}}}$, which implies, by uniformity of $\fF$, that there exist $s'',t''$ such that $\fA_1=\EdPath{\Arch{\Pillar{s_0s's''}{j}}{\Pillar{t_0s't''}{l}}}\subseteq\fF$. But remark now that we have $\fA_0=\EdPath{\Arch{\Pillar{rs''}{h}}{\Pillar{ss''}{i}}}=\EdPath{\Arch{\Pillar{rs''}{h}}{\Pillar{s_0s's''}{i}}}\subseteq\widetilde\fF$, so, given any $w\in\Cc\times\Cc$, a trace sequence $\ft'$ generating the arch $\Arch{\Pillar{rs''w}{h}}{\Pillar{s_0s's''w}{i}}$ of $\fA_0$ may be extended with the arch $\Arch{\Pillar{s_0s's''w}{j}}{\Pillar{t_0s't''w}{l}}$ of $\fA_1$, yielding a trace sequence $\ft$ of $\fF$ along $\sigma$, such that
$$\fa(\ft)=\Arch{\Pillar{rs''w}{h}}{\Pillar{t_0s't''w}{l}},$$
which satisfies $D(\fa(\fs),\fa(\ft))<\epsilon$, because we chose $r,s'$ appropriately when we applied the induction hypothesis.\qed

Thanks to \reflemma{ApproxComp}, we can prove that if two uniform edifices have the same closure, then their closed traces coincide, with respect to any feedback. The congruence property of closed edifices is obtained as an easy corollary, with the help of \refcor{Trace}.
\begin{prop}
	\label{prop:ClosedTrace}
	Let $\fE,\fF\subseteq\cA_I$ be uniform edifices, with $I$ finite, and let $\sigma$ be a feedback. Then, $\Closure\fE=\Closure\fF$ implies $\ClosedTrace{\sigma}{\fE}=\ClosedTrace{\sigma}{\fF}$.
\end{prop}
\proof By symmetry, and since $\fE\subseteq\Closure\fE$, it is enough to show that $\fE\subseteq\Closure\fF$ implies $\ClosedTrace{\sigma}{\fE}\subseteq\ClosedTrace{\sigma}{\fF}$. So suppose that $\fE$ is a subset of the closure of $\fF$, and let $\fa\in\ClosedTrace{\sigma}{\fE}$. By definition, there exists a sequence $(\fa_n)_{n\in\Nat}\in\Trace{\sigma}{\fE}$ such that $\fa_n\rightarrow\fa$. Let now $\epsilon_m=2^{-m}$, for $m\in\Nat$. If we apply \reflemma{ApproxComp} to each $\fa_n$ and for each $\epsilon_m$, we obtain a sequence $(\fb^n_m)_{m,n\in\Nat}\in\Trace{\sigma}{\fF}$ such that, for all $m,n\in\Nat$, $D(\fa_n,\fb_m^n)<\epsilon_m$. Consider now the diagonalization of such sequence, \ie, the sequence $(\fa'_n)_{n\in\Nat}\in\Trace{\sigma}{\fF}$ defined by setting $\fa'_n=\fb_n^n$, for all $n\in\Nat$. We contend that $\fa'_n\rightarrow\fa$, which is enough to conclude. So let $\epsilon>0$. Since $\fa_n\rightarrow\fa$, there exists $N\in\Nat$ such that, for all $n\geq N$, $D(\fa,\fa_n)<\epsilon/2$. Similarly, let $M$ be smallest integer such that $\epsilon_M<\epsilon/2$, and let $K=\max(M,N)$. We then have, for all $n\geq K$,
$$D(\fa,\fa'_n)\leq D(\fa,\fa_n)+D(\fa_n,\fa'_n)=D(\fa,\fa_n)+D(\fa_n,\fb_n^n)<\epsilon,$$
which proves that $\fa'_n$ tends to $\fa$ as $n$ grows to infinity, as desired.\qed
\begin{cor}
	\label{cor:ClosedEdCong}
	Let $\mu,\nu$ be two nets such that $\ClosedEd\mu=\ClosedEd\nu$. Then, for every context $C$, $\ClosedEd{\ctxt\mu}=\ClosedEd{\ctxt\nu}$.
\end{cor}
\proof We use the notation of the proof of \refprop{EdCong} for juxtaposing nets, \ie, we denote by $C\bullet\mu$ the juxtaposition of $C$ and $\mu$, so that $\ctxt\mu$ may be written as $\Ctxt{\sigma}{C\bullet\mu}$ for a suitable feedback $\sigma$. Now, observe that the arches of $\Ed C$ and $\Ed\mu$ are based at different integers, so that the closure of these two edifices is completely disjoint. The same remark applies to $\Ed C$ and $\Ed\nu$; hence, we have $\ClosedEd{C\bullet\mu}=\ClosedEd C\cup\ClosedEd\mu=\ClosedEd C\cup\ClosedEd\nu=\ClosedEd{C\bullet\nu}$. Then, using \refcor{Trace} and \refprop{ClosedTrace}, we may write
$$\ClosedEd{\ctxt\mu}=\ClosedEd{\Ctxt{\sigma}{C\bullet\mu}}=\ClosedTrace{\sigma}{\Ed{C\bullet\mu}}=\ClosedTrace{\sigma}{\Ed{C\bullet\nu}}=\ClosedEd{\Ctxt{\sigma}{C\bullet\nu}}=\ClosedEd{\ctxt\nu}.
$$\qed

\subsection{Characterizing observational equivalence}
\label{sect:Charact}
Obtaining a semantic characterization of an observational equivalence involves two results: the proof of a first statement, usually referred to as the \emph{adequacy} of the semantics, establishing that denotational equality implies observational equivalence; and the proof of the converse, \ie, that observational equivalence implies denotational equality, which is usually known as \emph{full abstraction}.

The presence of both results is often simply referred to as a ``full abstraction result'', because the second property is in most cases harder to obtain, and is thus the fundamental one. In fact, adequacy is an immediate consequence of the following two properties:
\begin{description}
	\item[contextuality] the semantics induces a congruence on the syntax, \ie, if two syntactic objects $a,b$ are denotationally equal, then for every context $C$, $C[a]$ and $C[b]$ are also denotationally equal;
	\item[discrimination] the semantics is able to discriminate between the two classes of syntactic objects used to define the observational equivalence, \ie, if the set $S$ is the basis of the equivalence, as described in \refsect{ObsAx}, then for any $a\in S$ and $b\not\in S$, one must have that $a$ and $b$ are denotationally different.
\end{description}
To see that adequacy follows from the above two properties, consider the contrapositive statement: let $a,b$ be observationally different, \ie, suppose there exists $C$ such that $C[a]\in S$ and $C[b]\not\in S$; by discrimination, we obtain that $C[a]$ and $C[b]$ are denotationally different, so we conclude by contextuality. Note that this latter property is usually taken as a basic property of denotational semantics, \ie, all semantics are assumed to verify it, as described in \refsect{DenSem}. Hence, all that is left to verify is the discrimination property, which is often not so hard to obtain. This is why, in most cases, full abstraction receives all the attention. Nevertheless, this does not mean that adequacy itself is banal: for instance, our proofs of the contextuality property for edifices and closed edifices (\refprop{EdCong} and \refcor{ClosedEdCong}, respectively) are far from being trivial.

So, we start by ensuring that edifices enjoy the discrimination property with respect to finitary axiom-equivalence. For this, we use the topological characterization based on compactness (\refprop{CompactEd}):
\begin{lem}
	\label{lemma:ObsSepFin}
	For all nets $\mu,\nu$, $\StrObs{\mu}$ and $\WkBlind{\nu}$ implies $\Ed\mu\neq\Ed\nu$.
\end{lem}
\proof Simply observe that, by \refprop{CompactEd}, $\Ed\mu$ is non-empty and compact, while $\Ed\nu$ is either empty, or not compact.\qed

We now have our first full abstraction result:
\begin{thm}[Full abstraction for $\FinAxEq$]
	\label{th:TotEqFullAbs}
	For all nets $\mu,\nu$, $\mu\FinAxEq\nu$ iff $\Ed\mu=\Ed\nu$.
\end{thm}
\proof As discussed above, the implication from right to left, or the adequacy property, is a consequence of \refprop{EdCong} and \reflemma{ObsSepFin}; so let us examine directly the converse, or rather its contrapositive. Suppose that $\Ed{\mu}\neq\Ed{\nu}$, and let $\fa\in\Ed{\mu}\setminus\Ed{\nu}$ (we are supposing without loss of generality that $\Ed{\mu}$ is not contained in $\Ed{\nu}$). We then have
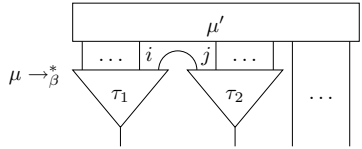
\begin{center}\scalebox{0.8}{\input{MuRedObsPath.pstex_t}}\end{center}
where the observable axiom shown generates $\fa$, whereas, by \reflemma{Equiv},
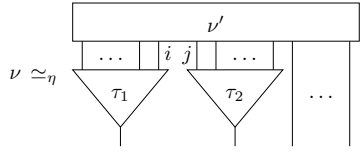
\begin{center}\scalebox{0.8}{\input{NuEtaEqNoPath.pstex_t}}\end{center}
and no reduct of $\nu'$ develops a connection between the ports $i,j$ generating $\fa$. Now consider the test
\begin{center}\scalebox{0.8}{\input{TestTotEq.pstex_t}}\end{center}
By \reflemma{TreeAnnihilation}, $\test{\mu}$ $\E$-reduces to a quasi-wire (\cf\ \reffig{QuasiWire}), so $\StrObs{\test{\mu}}$. On the contrary, $\test\nu$ reduces to a net with $2$ free ports which cannot be $\beta\eta$-equivalent to a wire, otherwise we would have $\fa\in\Ed{\nu}$. We have two possibilities: either $\test{\nu}$ is $\beta\E$-normalizable, or it is not. In the latter case, by \refcor{BENormAndObsPaths}, we have $\WkBlind{\test\nu}$, so we are done. In the former case, we take the $\beta\E$-normal forms of $\test{\mu}$ and $\test{\nu}$, which are cut-free by \refprop{BetaEpsilonNormForm}, and conclude by applying the Separation Theorem~\ref{th:Separation}.\qed

%

The fact that closed edifices enjoy the discrimination property with respect to axiom-equivalence is trivial, and no special topological property is needed to prove it:
\begin{lem}
	\label{lemma:ObsSep}
	For all nets $\mu,\nu$, $\Obs\mu$ and $\Blind\nu$ implies $\ClosedEd\mu\neq\ClosedEd\nu$.
\end{lem}
\proof $\Ed\mu$ is non-empty (it contains at least one vault), and the closure of a non-empty set is non-empty; on the contrary, $\Ed\nu$ is empty, and so is its closure.\qed

By contrast, in the case of axiom-equivalence, compactness (and hence completeness) becomes essential for yielding a fully-abstract denotational semantics. It is crucial in the proof of the following result:
\begin{lem}
	\label{lemma:ComplDiff}
	Let $\mu,\nu$ be such that $\ClosedEd\mu\neq\ClosedEd\nu$. Then, one of the following holds:
	\begin{enumerate}[$\bullet$]
		\item there exists an observable axiom $\fx\in\AllObsPaths\mu$ such that $\EdPath{\fx}\subseteq\Ed\mu\setminus\ClosedEd\nu$;
		\item there exists an observable axiom $\fy\in\AllObsPaths\nu$ such that $\EdPath{\fy}\subseteq\Ed\nu\setminus\ClosedEd\mu$.
	\end{enumerate}
\end{lem}
\proof Let $\mu$ have $n$ free ports, and suppose, without loss of generality, that there exists $\fa\in\ClosedEd\mu\setminus\ClosedEd\nu$, based at $i,j\in\{1,\ldots,n\}$. Remember that $\ClosedEd\mu$ and $\ClosedEd\nu$ are defined as the closures of resp.\ $\Ed\mu$ and $\Ed\nu$, and that by \refprop{ClosureCompact} they are both compact, hence complete. Then, if $\fa\in\ClosedEd\mu\setminus\Ed\mu$, $\fa$ must be a ``missing limit'' of a Cauchy sequence $(\fa_n)_{n\in\Nat}$ of $\Ed\mu$. Since a subsequence of a Cauchy sequence is still a Cauchy sequence, there must exists an integer $m$ such that, for all $n\geq m$, \mbox{$\fa_n\in\Ed\mu\setminus\ClosedEd\nu$}, otherwise $\fa$ would belong to $\ClosedEd\nu$ because of its completeness. Therefore, modulo replacing it by one of these $\fa_n$, we can always assume that \mbox{$\fa\in\Ed\mu\setminus\ClosedEd\nu$}. If it is so, then by \refdef{EdNet} there exists $\fx=\Arch{\Pillar{s}{i}}{\Pillar{t}{j}}\in\AllObsPaths{\mu}$ such that $\fa\in\EdPath\fx\subseteq\Ed\mu$, which means that $\fa=\Arch{\Pillar{sw_0}{i}}{\Pillar{tw_0}{j}}$ and, for every $w\in\Cc\times\Cc$, $\Arch{\Pillar{sw}{i}}{\Pillar{tw}{j}}\in\Ed\mu$. Now let $s'_1,\ldots,s'_n,\ldots$ be a sequence of prefixes of increasing length of $w_0$, and set, for all $n$, $s_n=ss'_n$ and $t_n=ts_n'$. Suppose that, for all $n$, there exist two pairs of infinite words $u_n,v_n$ such that $\fa_n=\Arch{\Pillar{s_nu_n}{i}}{\Pillar{t_nv_n}{j}}\in\ClosedEd\nu$; it is not hard to verify that the arches $\fa_n$ would form a Cauchy sequence of limit $\fa$, and thus, by the completeness of $\ClosedEd\nu$, we would obtain $\fa\in\ClosedEd\nu$, a contradiction. Therefore, there must exist an integer $n$ such that, for all $w$, $\Arch{\Pillar{s_nw}{i}}{\Pillar{t_nw}{j}}\in\Ed\mu\setminus\ClosedEd\nu$.\qed

To prove full abstraction for $\AxEq$, we first need the following separation result:
\begin{lem}
	\label{lemma:WireSep}
	Let $\W$ be a quasi-wire (\reffig{QuasiWire}), and let $\mu$ be a net with two free ports, such that $\Arch{\Pillar{s}{i}}{\Pillar{t}{j}}\in\AllObsPaths{\mu}$ implies $i=j$. Then, there exists a test $\theta$ such that $\Obs{\Ctxt\theta\W}$ and $\Blind{\Ctxt\theta\mu}$.
\end{lem}
\proof If $\Blind\mu$, the identity test suffices, so suppose $\Obs\mu$. By hypothesis, all observable paths appearing in the reducts of $\mu$ connect one of the free ports to itself. Therefore, there exists $\mu'$ such that $\mu\red\mu'$, and
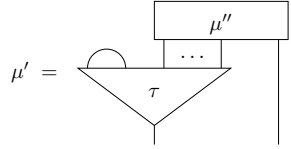
\begin{center}\scalebox{0.8}{\input{MuPrime.pstex_t}}\end{center}
In the above picture, we have supposed that the observable path connects the free port $1$ to itself, and that the leaves connected by the observable axiom are the two ``leftmost'' leaves of $\tau$. These are just graphically convenient assumptions, causing no loss of generality: the observable path may as well connect port $2$ to itself, and the leaves connected may be any two leaves of $\tau$. Now, if we define
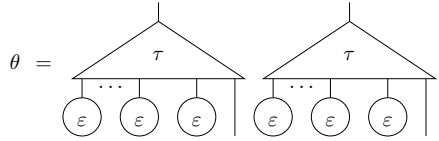
\begin{center}\scalebox{0.8}{\input{Theta.pstex_t}}\end{center}
we have that, thanks to \reflemma{TreeAnnihilation}, $\Ctxt{\theta}{\W}\red\W$, while $\Ctxt\theta\mu$ reduces to a net whose free port $1$ is connected to an $\E$ cell. If this net is blind, we are done; otherwise, there is a reduct of $\Ctxt\theta\mu$ containing an observable path between the free port $2$ and itself. This observable path can be ``eliminated'' with the same technique, while the $\E$ cell on port $1$ will ``eat'' any tree fed to it, so in the end we obtain a test $\theta'$ such that $\Ctxt{\theta'}{\W}\red\ImmObs\W$, while $\Blind{\Ctxt{\theta'}{\mu}}$, as desired.\qed

We are now ready to prove our second full abstraction theorem:
\begin{thm}[Full abstraction for $\AxEq$]
	\label{th:ObsEqFullAbs}
	For all nets $\mu,\nu$, $\mu\AxEq\nu$ iff $\ClosedEd\mu=\ClosedEd\nu$.
\end{thm}
\proof Once again, the adequacy property, \ie, the backward implication, is a consequence of \refcor{ClosedEdCong} and \reflemma{ObsSep}, so let us turn to the actual full abstraction property. For this, we consider the contrapositive statement, and assume $\ClosedEd\mu\neq\ClosedEd\nu$. Let $I$ be the interface of $\mu$ and $\nu$. By \reflemma{ComplDiff}, we know that there exist $i,j\in I$ and $\fx=\Arch{\Pillar{s}{i}}{\Pillar{t}{j}}\in\AllObsPaths{\mu}$ such that, for all $w$, $\Arch{\Pillar{sw}{i}}{\Pillar{tw}{j}}\in\EdPath{\fx}\setminus\ClosedEd\nu$ (it could actually be that these arches belong to $\EdPath\fy\setminus\ClosedEd\mu$, where $\fy\in\AllObsPaths\nu$, but obviously our assumption causes no loss of generality). We shall suppose $i\neq j$; the reader is invited to check that the argument can be adapted to the case $i=j$. Since $\fx\in\AllObsPaths{\mu}$, we have
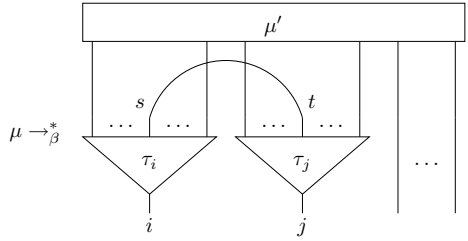
\begin{center}\scalebox{0.8}{\input{MuApxFullAbs.pstex_t}}\end{center}
where we have explicitly drawn the observable axiom of address $\fx$. On the other hand, by \reflemma{Equiv}, we have
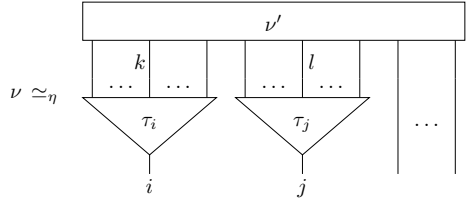
\begin{center}\scalebox{0.8}{\input{NuEtaFullAbs.pstex_t}}\end{center}
where we have called $k$ and $l$ the two free ports of $\nu'$ corresponding resp.\ to the addresses $t$ and $s$ in $\tau_i$ and $\tau_j$. Observe that, by the fact that closed edifices model $\BetaEpsilonEtaEq$, the edifice of the net on the right is still $\ClosedEd\nu$. Now if, in any reduct of $\nu'$, there appeared an observable path between $k$ and $l$, then we would contradict the fact that, for all $w$, $\Arch{\Pillar{sw}{i}}{\Pillar{tw}{j}}\not\in\ClosedEd\nu$. Therefore, no observable path ever appears between $k$ and $l$ in any reduct of $\nu'$.

Consider then the test
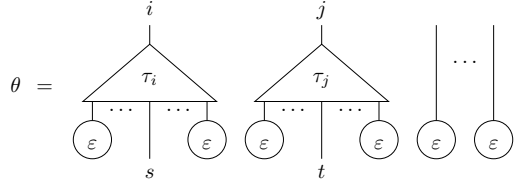
\begin{center}\scalebox{0.8}{\input{CtxtFullAbs.pstex_t}}\end{center}
where we have left free only the leaves corresponding to the addresses $s$ and $t$ of $\tau_i$ and $\tau_j$. Now, by \reflemma{TreeAnnihilation}, $\Ctxt\theta\mu$ $\beta$-reduces to a quasi-wire; on the other hand, we have
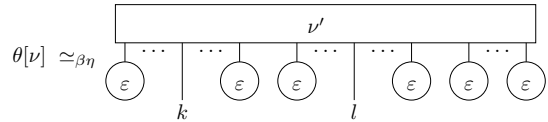
\begin{center}\scalebox{0.8}{\input{CtxtNuRedFullAbs.pstex_t}}\end{center}
But $\nu'$ never develops observable paths between $k$ and $l$, so \reflemma{WireSep} applies, and we obtain $\mu\not\AxEq\nu$.\qed

By inspecting the proofs of Theorems \ref{th:TotEqFullAbs} and \ref{th:ObsEqFullAbs}, we see that only tests are used to discriminate nets. Since those two results say precisely that equality of edifices and closed edifices coincides with finitary axiom-equivalence and axiom-equivalence, respectively, we get the following Context Lemma for free:
\begin{lem}[Context]
	\label{lemma:Context}
	$\mu\FinAxEq\nu$ (resp.\ $\mu\AxEq\nu$) iff, for every test $\theta$, $\StrObs{\Ctxt{\theta}{\mu}}$ iff $\StrObs{\Ctxt{\theta}{\nu}}$ (resp.\ $\Obs{\Ctxt{\theta}{\mu}}$ iff $\Obs{\Ctxt{\theta}{\nu}}$).\qed
\end{lem}

\sloppy{Furthermore, combined with \refprop{BetaEtaEpsModel}, \refth{TotEqFullAbs} gives us that $\beta\eta\E$-equivalence is included in finitary axiom-equivalence (and hence in axiom-equivalence, by \refprop{TotEqInObsEq});} in \refsect{Theories} we shall see that this inclusion is strict (\cf\ \reffig{PingPong}).
\begin{cor}
	\label{cor:BetaEtaEpsInFinAxEq}
	For every nets $\mu,\nu$, $\mu\BetaEpsilonEtaEq\nu$ implies $\mu\FinAxEq\nu$.\qed
\end{cor}

\begin{figure}[t]
	\begin{center}\scalebox{0.8}{\input{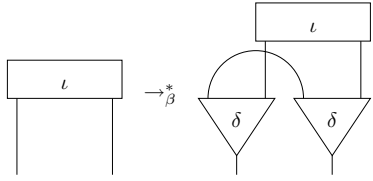}}\end{center}
	\caption{A non-$\beta\E$-normalizable net observationally equivalent to a wire.}
	\label{fig:Iota}
\end{figure}
On the other hand, as an application of \refth{ObsEqFullAbs} we give an example showing that the inclusion of \refprop{TotEqInObsEq} is strict, \ie, that there exist axiom-equivalent nets which are not finitarily axiom-equivalent. Such example is based on a net which is not $\beta\E$-normalizable, and yet is observationally equivalent to a wire. This is analogous to Wadsworth's ``infinitely $\eta$-expanding'' term $J=RR$, where \mbox{$R=\lambda xzy.z(xxy)$}, which is well known to be hnf-equivalent to $\lambda z.z$.

Consider a net $\iota$ reducing as in \reffig{Iota}. Such a net exists by what we have shown in \refsect{Expressiveness}; furthermore, after constructing it, one can see that $\iota$ is not immediately observable, and that $\fx\in\AllObsPaths{\iota}$ iff $\fx=\Arch{\Pillar{\BiWord{\q^n\p}{\Id}}{1}}{\Pillar{\BiWord{\q^n\p}{\Id}}{2}}$ for some non-negative integer $n$. On the other hand, if $\omega$ denotes a wire, we have
$$\ClosedEd\omega=\Ed\omega=\{\Arch{\Pillar{u}{1}}{\Pillar{u}{2}}~;~\forall u\in\Cc\times\Cc\}.$$
Now, if $\q^\infty$ denotes an infinite sequence of $\q$'s, all arches of the form
$$\fa_y=\Arch{\Pillar{\BiWord{\q^\infty}{y}}{1}}{\Pillar{\BiWord{\q^\infty}{y}}{2}}$$
are missing from $\Ed\iota$, hence \mbox{$\Ed\iota\varsubsetneq\Ed\omega$}. But these arches are all adherent to $\Ed\iota$: in fact, it is very easy to construct a Cauchy sequence in $\Ed\iota$ of limit $\fa_y$, for any $y$. Therefore, $\ClosedEd\iota=\ClosedEd\omega$, and $\iota\AxEq\W$. On the other hand, $\iota\not\FinAxEq\omega$, and we do not need \refth{TotEqFullAbs} to prove that: in fact, the identity is a context discriminating between the two nets.

Note that the reducts of $\iota$ are ``almost'' $\eta$-equivalent to a wire: there is just one missing connection. We can say that this connection forms ``in the limit'', when the reduction is carried on forever. When one interprets nets as edifices, this informal remark becomes a precise topological fact, \ie, we have a true limit.

\section{Concluding Remarks}
\label{sect:Conc}

\subsection{Comparison with previous work}
\label{sect:Comparison}
The first notion of observational equivalence for interaction nets introduced in the literature is due to Bechet \cite{Bechet:INPartialEval}. In his work, the author mentions a notion of behavioral equivalence based on Girard's coherence spaces \cite{Girard:LL}. However, we have not been able to reformulate this equivalence so as to compare it to the ones studied in the present paper.

The situation is different with Fern\'andez and Mackie's work \cite{FernandezMackie:OpEquiv}, the only other existing work on observational equivalence for interaction nets, for which we have precise results. First of all, we may note that Fern\'andez and Mackie's approach is more general, \ie, it applies to all systems of interaction nets, not just to the symmetric interaction combinators. However, we have already mentioned that our notion of observable and finitarily observable net can also be generalized to arbitrary systems of interaction nets, as shown in the author's Ph.D.~thesis \cite{Mazza:PhDThesis}; the details of this generalization are out of the scope of this paper though.

What is more interesting is to compare our notions of observational equivalence with the specialization of Fern\'andez and Mackie's observational equivalence to the symmetric interaction combinators, which we shall call here \emph{visible equivalence}. It can be formulated as follows:
\begin{figure}[t]
	\begin{center}\scalebox{0.8}{\input{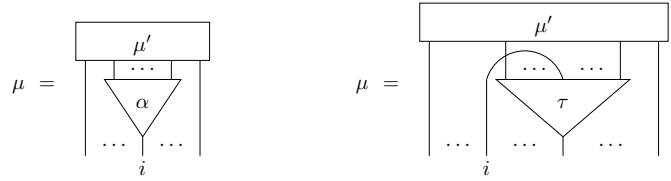}}\end{center}
	\caption{Visible ports; $\alpha$ is any symbol, and $\tau$ any tree.}
	\label{fig:VisiblePort}
\end{figure}
\begin{defi}[Visible port, visible equivalence \cite{FernandezMackie:OpEquiv}]
	Let $\mu$ be a net with $n$ free ports, and let $1\leq i\leq n$. We say that $i$ is \emph{immediately visible} in $\mu$ iff $\mu$ has one of the shapes given in \reffig{VisiblePort}. We say that $i$ is \emph{visible} in $\mu$, and we write $\FMObs{i}{\mu}$, iff $\mu\red\mu'$ such that $i$ is immediately visible in $\mu'$. We write $\FMBlind{i}{\mu}$ for the negation of $\FMObs{i}{\mu}$.

	Given two nets $\mu,\nu$ with $n$ free ports, we say that $\mu$ and $\nu$ are \emph{visibly equivalent}, and we write $\mu\FMEq\nu$, iff, for every $1\leq i\leq n$ and for every context $C$, $\FMObs{i}{\ctxt\mu}$ iff $\FMObs{i}{\ctxt\nu}$.
\end{defi}
Fern\'andez and Mackie~\cite{FernandezMackie:OpEquiv} also give, in case a port $i$ is visible, a notion of \emph{visible agent} at $i$, and require furthermore that either the visible agents at $i$ of $\ctxt\mu$ and $\ctxt\nu$ are the same, or that one of such visible agents is not a constructor. In interaction net systems, a constructor is simply a symbol declared to be such, \ie, it is not an intrinsic notion. Declaring symbols to be constructors may be useful from an ``intentional'' point view, when one has in mind a particular semantics for the given interaction net system. In the symmetric combinators, because cells with the same symbol may interact, there are arguably no constructors, hence the simplified definition we give here.

Note that, as defined above, visible equivalence does not quite fit in the general pattern of Morris-like observational equivalences discussed in \refsect{ObsAx}, because it is defined ``port-wise'', \ie, it takes free ports into account. In other words, visibility is a property of free ports, not of nets. However, Fern\'andez and Mackie's definition can easily be adjusted so as to conform to Morris' pattern.
\begin{defi}[Visible net]
	Let $\mu$ be a net with $n$ free ports. We say that $\mu$ is \emph{visible}, and we write $\FMObs{}{\mu}$, iff $\FMObs{i}{\mu}$ for some $1\leq i\leq n$.
\end{defi}
\begin{lem}
	\label{lemma:Morris}
	Let $\mu$ be a net with $n$ free ports, let $1\leq i\leq n$, and let
	\begin{center}\scalebox{0.8}{\input{MuWithEps.pstex_t}}\end{center}
	Then, $\FMObs{}{\nu}$ iff $\FMObs{i}{\mu}$.
\end{lem}
\proof For the backward implication, by definition $\mu$ reduces to a net of one of the shapes given in \reffig{VisiblePort}. In the case on the left, the appearance of the $\alpha$ cell is not modified by the $\E$ cells in $\nu$; in the case on the right, simply observe that the $\E$ cell plugged at the root of $\tau$ will ``eat'' the tree until arriving at free port $i$. For the forward implication, we have that only the left case of \reffig{VisiblePort} is possible, \ie, there is a reduct of $\nu$ in which the principal port of a cell $c$ appears at its only free port. Note that $\E$ cells only produce $\E$ cells through interaction. Hence, if $c$ is not an $\E$ cell, it already appears in $\mu$; if $c$ is an $\E$ cell, either it already appears in $\mu$, or $c$ ``comes from'' one of the $\E$ cells plugged to $\mu$ in $\nu$. Then, it is not hard to see that $\mu$ must reduce to a net of the shape at the right of \reffig{VisiblePort}, with the $\E$ cell ``producing'' $c$ being the one plugged to the root of $\tau$.\qed
\begin{prop}
	For all nets $\mu,\nu$, $\mu\FMEq\nu$ iff, for every context $C$, $\FMObs{}{\ctxt\mu}$ iff $\FMObs{}{\ctxt\nu}$.
\end{prop}
\proof The forward implication is trivial; for what concerns the converse, consider the contrapositive statement: there exist $C$ and $1\leq i\leq n$ such that, for instance, $\FMObs{i}{\ctxt\mu}$ and $\FMBlind{i}{\ctxt\nu}$. Then, let $E$ be the context plugging $\E$ cells to all free ports of $\ctxt\mu$ and $\ctxt\nu$ except~$i$; by \reflemma{Morris}, we have $\FMObs{}{\Ctxt{E}{\ctxt\mu}}$ and $\FMBlind{}{\Ctxt{E}{\ctxt\nu}}$, as desired.\qed

In their paper \cite{FernandezMackie:OpEquiv}, the authors prove that $\BetaEtaEq\,\subseteq\,\FMEq$, so, by \refprop{Collapse}, we have that $\FMEq$ coincides with $\BetaEtaEq$ on total nets, just like all the other equivalences introduced in this paper (except of course $\BetaEq$). However, the situation is quite different if we consider non-total nets; indeed, we can show that visible equivalence is strictly stronger than finitary axiom-equivalence (and, by \refprop{TotEqInObsEq}, than axiom-equivalence).
\begin{lem}
	\label{lemma:QWandPrinc}
	Let $\mu,\nu$ be two nets with the same interface, and let $C$ be a context such that $\ctxt\mu$ reduces to a quasi-wire, while $\ctxt\nu$ reduces to a net whose one of the two free ports is connected to a principal port. Then, $\mu\not\FMEq\nu$.
\end{lem}
\proof Simply consider the context
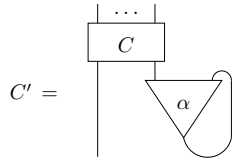
\begin{center}\scalebox{0.8}{\input{CtxtFM.pstex_t}}\end{center}
where $\alpha$ is any binary symbol, and we have supposed, without loss of generality, that the free port of the reduct of $\ctxt\nu$ which is connected to the principal port is the one on the left in the above picture (while the free ports drawn at the top of the picture are those that are connected to $\mu$ and $\nu$ in $\ctxt\mu$ and $\ctxt\nu$, respectively). Then, we have, for some net without interface $o$, some symbol $\beta$, and some net $\nu_0$,
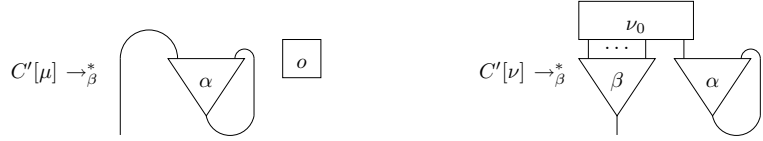
\begin{center}\scalebox{0.8}{\input{CtxtFMRed.pstex_t}}\end{center}
so that $\FMBlind{}{\Ctxt{C'}{\mu}}$, whereas $\FMObs{}{\Ctxt{C'}{\nu}}$.\qed
\begin{lem}
	\label{lemma:FMIsSemiSens}
	Let $\Obs\mu$ and $\Blind\nu$. Then, $\mu\not\FMEq\nu$.
\end{lem}
\proof By \refprop{Solvable}, there exists a test $\theta$ such that $\Ctxt{\theta}{\mu}$ reduced to a quasi-wire. On the other hand, by \reflemma{Blindness} we still have $\Blind{\Ctxt{\theta}{\nu}}$. Observe that, if $i$ is any of the two free ports of $\Ctxt{\theta}{\mu}$, we have $\FMObs{i}{\Ctxt{\theta}{\mu}}$. Now, if one of the two free ports of $\Ctxt{\theta}{\nu}$ is not visible, we are done. We may then assume $\Ctxt{\theta}{\nu}\red\nu'$, where both free ports of $\nu'$ are immediately visible; then, they must both be connected to a principal port, otherwise $\nu'$ would be immediately observable (\cf\ \reffig{VisiblePort}, right), whereas we know it to be blind. We may therefore conclude by applying \reflemma{QWandPrinc}.\qed

The fact that visible equivalence is stronger than finitary axiom-equivalence is a trivial corollary of the following:
\begin{prop}
	\label{prop:FinObsAndFM}
	Let $\StrObs\mu$ and $\WkBlind\nu$. Then, $\mu\not\FMEq\nu$.
\end{prop}
\proof Let $\mu$ and $\nu$ have $n$ free ports. We have two possibilities: either $\Blind\nu$, or $\AllObsPaths{\nu}$ is infinite. In the first case, observe that $\StrObs\mu$ implies $\Obs\mu$, so we conclude by \reflemma{FMIsSemiSens}. In the second case, since $\AllObsPaths{\mu}$ is finite, there must exist $1\leq i,j\leq n$ and pairs of words $s,t$ such that $\Arch{\Pillar{s}{i}}{\Pillar{t}{j}}\in\AllObsPaths{\nu}\setminus\AllObsPaths{\mu}$. Now, $s$ and $t$ describe two trees $\tau_s,\tau_t$ and a leaf in each of them, such that
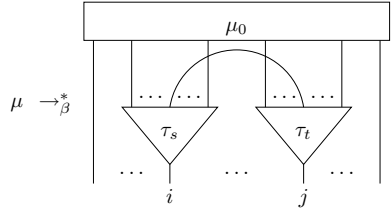
\begin{center}\scalebox{0.8}{\input{MuRedObs.pstex_t}}\end{center}
where the wire shown connects the two leaves of $\tau_s,\tau_t$ described by $s,t$, respectively. Consider then the test
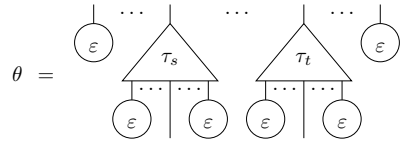
\begin{center}\scalebox{0.8}{\input{TestObsAxST.pstex_t}}\end{center}
where the only leaves of $\tau_s,\tau_t$ left free are again those described by $s,t$, respectively. We obviously have that $\Ctxt{\theta}{\nu}$ $\beta$-reduces to a quasi-wire; on the contrary, because of the way we have chosen $s$ and $t$, $\Ctxt{\theta}{\mu}$ is not $\beta\eta$-equivalent to a wire. Observe however that $\AllObsPaths{\Ctxt{\theta}{\mu}}$ is still finite (although it may now be empty). Then, we may consider the $\beta\E$-normal forms of $\Ctxt{\theta}{\nu}$ and $\Ctxt{\theta}{\mu}$, which exist by \refcor{BENormAndObsPaths}, and apply the Separation \refth{Separation} to them. We thus obtain a further test $\theta'$ such that $\Ctxt{\theta'}{\Ctxt{\theta}{\mu}}$ $\beta$-reduces to a quasi-wire and $\Blind{\Ctxt{\theta'}{\Ctxt{\theta}{\nu}}}$, or vice versa. In any case, we reason as in the proof of \reflemma{FMIsSemiSens}: if one of the ports of the blind net is not visible, we conclude; otherwise, we apply \reflemma{QWandPrinc}.\qed
\begin{cor}
	\label{cor:FMInFinAxEq}
	For all nets $\mu,\nu$, $\mu\FMEq\nu$ implies $\mu\FinAxEq\nu$.
\end{cor}
\proof Consider the contrapositive statement: $\mu\not\FinAxEq\nu$ implies that there exists $C$ such that, for example, $\StrObs{\ctxt\mu}$ and $\WkBlind{\ctxt\nu}$; by \refprop{FinObsAndFM}, we have $\ctxt{\mu}\not\FMEq\ctxt{\nu}$, so we conclude $\mu\not\FMEq\nu$ by using the fact that $\FMEq$ is a congruence.\qed

\begin{figure}[t]
	\begin{center}\scalebox{0.8}{\input{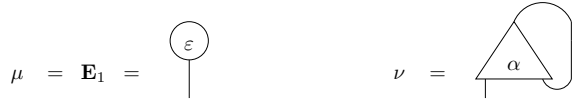}}\end{center}
	\caption{Two nets showing that visible equivalence is more discriminative than finitary axiom-equivalence; $\alpha$ is any binary symbol.}
	\label{fig:NotFMEq}
\end{figure}

To see that visible equivalence is \emph{strictly} stronger than finitary equivalence, consider the nets of \reffig{NotFMEq}: the only free port of $\mu=\Epsilon{1}$ is visible, while the only free port of $\nu$ is not visible, so $\mu\not\FMEq\nu$; on the contrary, the edifice of both nets is empty, so by \refth{TotEqFullAbs} we have $\mu\FinAxEq\nu$.

An intuitive justification to \refcor{FMInFinAxEq} and to the example of \reffig{NotFMEq} is that the difference between a visible and an observable net is seemingly akin to the difference between a head-normalizable and a \emph{weak}-head-normalizable \lat. In fact, the two cases of \reffig{VisiblePort} are strikingly similar to the cases $\lambda x.M$ (left) and $xM_1\ldots M_n$ (right) defining weak head normal forms: the latter case is a special case of observable net, just like $xM_1\ldots M_n$ is a special case of head-normal-form; the former case guarantees that a net visible on port $i$ is ``reactive'' when we plug the principal port of a cell to $i$ itself, \ie, an active pair is created, just like $\lambda x.M$ is ``reactive'' to application (a redex is created). In the \lac, whnf-equivalence is strictly stronger than nf- and hnf-equivalence~\cite{DezaniGiovannetti}; this is in accord with our intuition about visible equivalence and (finitary) axiom-equivalence.

\subsection{Approximations and the Genericity Lemma}
\label{sect:Meaningless}
In the \lac, unsolvable terms are important because they represent \emph{meaningless data}. One of the main formal arguments in favor of this intuition is the so-called Genericity Lemma~\cite{Barendregt}: let $M$ be an unsolvable \lat, and let $C$ be such that $C[M]$ is normalizable; then, $C[X]\BetaEq C[M]$ for every \lat\ $X$. In other words, if we see $C[\cdot]$ as a function, the only functions which are able to produce something meaningful (a normal form) out of unsolvable terms are the constant functions, confirming the fact that unsolvable terms are meaningless.

\begin{figure}[t]
	\begin{center}\scalebox{0.8}{\input{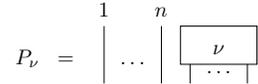}}\end{center}
	\caption{The parallelizing context for nets with $n$ free ports; $\nu$ is an arbitrary net.}
	\label{fig:Parallelizer}
\end{figure}
In the symmetric interaction combinators, a word-by-word rephrasing of the Genericity Lemma fails; this is because of two interesting differences with respect to the \lac:
\begin{enumerate}[$\bullet$]
	\item the intrinsic parallelism of interaction nets, which has no equivalent in the \lac;
	\item the fact that, for every $n\in\Nat$, there is an unsolvable net with $n$ free ports which is cut-free, hence normal (namely, the net we called $\Epsilon{n}$); by contrast, no normal \lat\ can be unsolvable.
\end{enumerate}
Indeed, consider the context $P_\nu$ given in \reffig{Parallelizer}. We may refer to this context as the ``parallelizing'' context: in fact, for every net $\mu$ with $n$ free ports, $\Ctxt{P_\nu}{\mu}=\mu\bullet\nu$, \ie, the juxtaposition of $\mu$ and $\nu$. Now, let $\nu$ be a net which is normalizable, or total, or finitarily observable, and take the unsolvable net $\Epsilon{n}$. Clearly $\Ctxt{P_\nu}{\Epsilon{n}}$ is normalizable, or total, or finitarily observable, but there always exists a net $\xi$ such that $\Ctxt{P_\nu}{\xi}$ need not be normalizable, or total, or finitarily observable, let alone $\beta$-equivalent to $\Ctxt{P_\nu}{\Epsilon{n}}$. This is because $\xi$ and $\nu$ do not interact, so the properties of $\Ctxt{P_\nu}{\xi}$ basically depend solely on $\xi$.

However, there is a reformulation of the Genericity Lemma which holds for the symmetric interaction combinators, and which supports the fact that our notion of unsolvable net coincides indeed with that of meaningless data. Take an unsolvable net $\mu$, and take a context $C$. Then, we can prove that, whenever
\begin{center}\scalebox{0.8}{\input{CtxtMuRed.pstex_t}}\end{center}
such that $C_0$ is cut-free and none of the wires connecting $C_0$ to $\mu_0$ is a cut, we have that, for every net $\xi$ with the same number of free ports as $\mu$, there exists $\xi_0$ such that
\begin{center}\scalebox{0.8}{\input{CtxtXiRed.pstex_t}}\end{center}
In other words, every bit of information in the result of the computation represented by $\ctxt\mu$ is also present in $\ctxt\xi$, for all $\xi$, which means that $\mu$ actually does not produce any information, and is thus meaningless.

The above concept of ``bit of information'' may be formalized by the notion of \emph{approximation}:
\begin{defi}[Approximation]
	Let $\mu$ be a net. An \emph{approximation} of $\mu$ is a cut-free net $\nu$ such that:
	\begin{enumerate}[$\bullet$]
		\item $\nu=\ctxt{\Epsilon n}$ for some context $C$ and $n\in\Nat$;
		\item $\mu\red\ctxt{\mu_0}$ for some $\mu_0$ with $n$ free ports.
	\end{enumerate}
	If $\nu$ is an approximation of $\mu$, we write $\nu\apx\mu$.
\end{defi}
Intuitively, an approximation of $\mu$ is a ``piece'' of the hypothetical cut-free form of $\mu$, \ie, it gives a partial information on the result of the computation represented by $\mu$. The least information, or the lack thereof, is $\Epsilon{n}$, which is an approximation of every net with $n$ free ports. If a net is total, then its cut-free form is also an approximation of it, the most complete one indeed.

Approximations and edifices are related by the following, whose proof is left to the reader:
\begin{prop}
	\label{prop:ApxAndEd}
	Let $\mu$ be a net, and let $\nu$ be a cut-free net with the same interface as $\mu$. Then, $\nu\apx\mu$ iff $\Ed\nu\subseteq\Ed\mu$.\qed
\end{prop}
The above result confirms in particular the idea that the edifice of a net may be seen as its ``infinite cut-free form''. Indeed, the relation $\apx$ can be made a partial order, and the set of approximations of a net can be shown to be a directed set. However, the order given by $\apx$ is not complete, so this set has no least upper bound in general; to make it complete, one should introduce infinite cut-free nets, which is more or less what edifices are.

We may then state the Genericity Lemma as follows:
\begin{lem}[Genericity]
	Let $\mu$ be an unsolvable net with $n$ free ports. Then, for every context $C$ and for every net $\xi$ with $n$ free ports, $\nu\apx\ctxt\mu$ implies $\nu\apx\ctxt\xi$.
\end{lem}
\proof By \refprop{Solvable}, we have $\Blind\mu$, so $\Ed\mu=\emptyset$. Hence, we can write, using \refcor{Trace}, \reflemma{TraceMono}, and the decomposition $\ctxt\mu=\Ctxt{\sigma}{C\bullet\mu}$ for a suitable feedback~$\sigma$,
$$\Ed{\ctxt\mu}=\Trace{\sigma}{\Ed C\cup\Ed\mu}=\Trace{\sigma}{\Ed C}\subseteq\Trace{\sigma}{\Ed C\cup\Ed\xi}=\Ed{\ctxt\xi}.$$
Now, by \refprop{ApxAndEd}, $\nu\apx\ctxt\mu$ implies $\Ed\nu\subseteq\Ed{\ctxt\mu}$, so $\Ed\nu\subseteq\Ed{\ctxt\xi}$, and we conclude $\nu\apx\ctxt\xi$ again by \refprop{ApxAndEd}.\qed

\subsection{Theories for the symmetric interaction combinators}
\label{sect:Theories}
In the foundational studies concerning the \lac, an important role is played by \emph{$\lambda$-theories}~\cite{Barendregt,LusinSalibra}. These can be straight-forwardly be reformulated in the context of the symmetric interaction combinators:
\begin{defi}[Theory]
	A \emph{theory} is a binary relation $\GenEq$ on nets such that:
	\begin{enumerate}
		\item $\GenEq$ relates nets with the same interface;
		\item $\GenEq$ is a congruence;
		\item $\BetaEq\,\subseteq\,\GenEq$.
	\end{enumerate}
\end{defi}

The set of theories $\fT$ is a complete bounded lattice with respect to inclusion: given any family of theories $(\GenEq_i)_{i\in I}$, the least upper bound (lub) is defined by $(\bigcup_{i\in I}\GenEq_i)^+$, and the greatest lower bound (glb) by $\bigcap_{i\in I}\GenEq_i$; the least element is $\BetaEq$, and the greatest element is the inconsistent theory $\top$, which equates all nets with the same interface.

Much effort has been put forth in order to understand the structure of the lattice of $\lambda$-theories; quite a few things are known about it~\cite{Visser,Barendregt,LusinSalibra}, and many more are the subject of ongoing research~\cite{BerlineManzonettoSalibra:1,BucciarelliSalibra,BerlineManzonettoSalibra:2,CarraroSalibra}. In the case of the symmetric combinators, we suspect the structure of $\fT$ to be at least as intricate as in the case of the \lac. In this section, we gather everything we presently know about it (which is arguably not much!), leaving several questions open for further work.

As in the case of the \lac, we may define \emph{sensible} and \emph{semi-sensible} theories, based on the fact that unsolvable nets are meaningless, and it is therefore sensible to identify all of them:
\begin{defi}[Sensible and semi-sensible theory]
	A theory $\GenEq$ is \emph{sensible} iff, for all $\mu,\nu$ unsolvable, $\mu\GenEq\nu$. A theory $\GenEq$ is \emph{semi-sensible} iff $\mu\GenEq\nu$ implies $\mu$ solvable iff $\nu$ solvable.
\end{defi}
Note that any theory containing a sensible theory is sensible, while any theory contained in a semi-sensible theory is semi-sensible (these are both immediate consequences of the definition). The two notions are related as follows.
\begin{lem}
	\label{lemma:Sensible}
	Let $\GenEq$ be a sensible theory. Then:
	\begin{enumerate}
		\item for every blind net $\mu$ with $n$ free ports, $\mu\GenEq\Epsilon n$;
		\item for every quasi-wire $W$, $W\GenEq\omega$, where $\omega$ is a wire.
	\end{enumerate}
\end{lem}
\proof Point (1) is obvious (modulo \refprop{Solvable}). For point (2), observe that $W=\Ctxt{\omega}{o}$, where $o$ is some net without interface, necessarily blind; moreover, note that $\omega=\Ctxt{\omega}{\Epsilon 0}$, where $\Epsilon{0}$ is the empty net. Now, by point (1), $o\GenEq\Epsilon{0}$; but then we can conclude, because $\GenEq$ is a congruence.\qed
\begin{prop}
	\label{prop:SensSemiSens}
	A consistent sensible theory is semi-sensible.
\end{prop}
\proof Let $\GenEq$ be a sensible theory, and let $\Obs\mu$ and $\Blind\nu$. We shall prove that $\mu\GenEq\nu$ implies $\GenEq\,=\top$. First of all, by \refprop{Solvable}, by the fact that $\GenEq$ includes $\beta$-equivalence, and by point (2) of \reflemma{Sensible}, there exists a test $\theta$ such that $\Ctxt{\theta}{\mu}\GenEq\omega$. On the other hand, by \reflemma{Blindness}, and by point (1) of \reflemma{Sensible}, we have $\Ctxt{\theta}{\nu}\GenEq\Epsilon{2}$. But $\GenEq$ is a congruence, so $\omega\GenEq\Epsilon{2}$, and we may conclude by \refprop{Collapse}.\qed

Apart from $\BetaEq$, in the course of this paper we introduced several theories: $\BetaEtaEq$, $\BetaEpsilonEq$, $\BetaEpsilonEtaEq$, $\FinAxEq$, $\AxEq$, and $\FMEq$ (the first and the last were actually introduced by Fern\'andez and Mackie, \cf\ \refsect{Comparison}). All of them are semi-sensible, because they are all included in $\AxEq$, which is semi-sensible by definition. Furthermore, since $\BetaEpsilonEq$ is sensible by definition ($\E$-reduction equates precisely all unsolvable nets), all the theories including it are also sensible, namely $\BetaEpsilonEtaEq$, $\FinAxEq$, and $\AxEq$. On the contrary, $\BetaEq$, $\BetaEtaEq$, and $\FMEq$ are not sensible: \reffig{NotFMEq} gives an example of two unsolvable nets which are distinguished by all of these theories.

Indeed, Fern\'andez and Mackie's equivalence is an example of non-sensible theory which strictly extends $\beta\eta$-equivalence, and stands quite on its own with respect to the other theories discussed in this paper. For instance, it is completely orthogonal to $\beta\E$-equivalence: this latter is not included in $\FMEq$, as shown again by the example of \reffig{NotFMEq}; and $\FMEq$ is not included in $\BetaEpsilonEq$, because the former includes $\eta$-equivalence. On the other hand, although the example of \reffig{NotFMEq} tells us that $\BetaEpsilonEtaEq$ is not included in $\FMEq$, we know nothing about the converse. All we know is that $\FMEq$ is strictly contained in $\FinAxEq$ (\refcor{FMInFinAxEq} and \reffig{NotFMEq}).

What about consistent sensible theories in general? First of all, observe that the lub and glb of a family of sensible theories is sensible, so the set of sensible theories is a complete sub-lattice of $\fT$, which is actually bounded. The least element is obviously $\BetaEpsilonEq$, because it is defined so as to validate exactly $\beta$-equivalence plus equality of every unsolvable net. The greatest element turns out to be $\AxEq$; in fact, this can be shown to be a coatom of $\fT$, \ie, a maximal consistent theory (so $\AxEq$ is also the greatest semi-sensible theory).
\begin{prop}
	Let $\GenEq$ be a theory such that $\AxEq\,\varsubsetneq\,\GenEq$. Then, $\GenEq\,=\top$.
\end{prop}
\proof We start by observing that $\GenEq$ is sensible, because it includes $\AxEq$. Now, let $\mu\GenEq\nu$, with $\mu\not\AxEq\nu$. We then have a context $C$ such that, for example, $\Obs{\ctxt\mu}$ and $\Blind{\ctxt\nu}$. But $\GenEq$ is a congruence, so $\ctxt\mu\GenEq\ctxt\nu$, which proves that $\GenEq$ is not semi-sensible. By \refprop{SensSemiSens}, the only sensible theory which is not semi-sensible is $\top$.\qed

So far, the situation is identical to the case of the \lac, in which the lattice of consistent sensible $\lambda$-theories has least element $\simeq_{\beta\Omega}$ (also known as $\cH$) and greatest element hnf-equivalence (also known as $\cH^\ast$).

\begin{figure}[t]
	\begin{center}\scalebox{0.8}{\input{PingPong.pstex_t}}\end{center}
	\caption{Nets showing that $\mu\FinAxEq\nu$ does not imply $\mu\BetaEpsilonEtaEq\nu$.}
	\label{fig:PingPong}
\end{figure}
In between the two, there is $\simeq_{\beta\eta\Omega}$, which coincides with nf-equivalence (also denoted by $\cH\eta$). Here we find the first sharp difference with respect to the \lac: $\BetaEpsilonEtaEq$, which is analogous to $\simeq_{\beta\eta\Omega}$, does not coincide with $\FinAxEq$, which, morally, is the counterpart of nf-equivalence. In fact, the converse of \refcor{BetaEtaEpsInFinAxEq} fails: the two nets of \reffig{PingPong} give an interesting example of this. They can be built by slightly twisting the constructions given in \refsect{Expressiveness}. It is not hard to show that $\mu_1\not\BetaEpsilonEtaEq\mu_2$; in some sense, the two nets endlessly ``chase'' each other in their reduction, never managing to meet. And yet, it is evident that they generate exactly the same observable axioms, \ie, $\AllObsPaths{\mu_1}=\AllObsPaths{\mu_2}$. Therefore, $\Ed{\mu_1}=\Ed{\mu_2}$, and $\mu_1\FinAxEq\mu_2$ by \refth{TotEqFullAbs}. Note how the parallelism of interaction nets, absent in the \lac, plays once again a crucial role in this example. If we use the analogy that observable axioms are head variables, here we are clearly exploiting the fact that interaction nets allow several head variables in parallel: although $\mu_1$ and $\mu_2$ have the same ``head variables'', they ``produce'' them in a different order.

In the \lac, it is possible to show that between $\cH\eta$ and $\cH^\ast$ there is a continuum of sensible theories~\cite{Barendregt}; we ignore whether this is the case for the symmetric interaction combinators. Indeed, a related open question is the existence of \emph{easy} nets, \ie, nets which can be consistently equated with any other net with the same interface. A first difference with the \lac\ is that $\Epsilon 2$, which is the prototypical unsolvable net with $2$ free ports, is not easy (\refprop{Collapse} shows that there is no consistent theory equating it with a wire); on the contrary, the \lat\ $\Omega$, which is the prototypical unsolvable term, can be shown to be easy.

Finally, we give an example of a consistent non-semi-sensible theory. Define \emph{total equivalence} as $\mu\TotEq\nu$ iff, for every context $C$, $\ctxt\mu$ is total iff $\ctxt\nu$ is total. It can be shown that $\eta$-equivalence does not alter totality, so $\BetaEtaEq\,\subseteq\,\TotEq$. Moreover, observe that total equivalence is the only theory we introduced which distinguishes between the empty net $\Epsilon{0}$ (which is total) from all other nets with no interface not reducing to $\Epsilon{0}$ (which are not total). This proves in particular that the theory is consistent.

However, consider the net $\xi$ obtained by juxtaposing two copies of the net $\nu$ given in \reffig{NotFMEq}, and let $\iota$ be the net of \reffig{Iota}. Both nets have $2$ free ports, and are thus comparable; furthermore, none of the two nets is total ($\iota$ is not normalizable, $\xi$ contains vicious circles), and we clearly have $\Obs\iota$ and $\Blind\xi$. Now, non-totality has a quite singular behavior if compared to non-normalizability in the \lac, in that it is ``resistant'' to contexts: in fact, by the locality of interaction rules, neither active pairs nor vicious circles can be eliminated through interaction; thus, if $\mu$ is not total, so is $\ctxt\mu$, for any $C$. Therefore, we have $\iota\TotEq\xi$, which proves that total equivalence is not semi-sensible. So the only possible relationship with the other known theories is $\FMEq\,\varsubsetneq\,\TotEq$; we ignore whether this is the case.

\subsection{More open questions and further work}
\label{sect:FurtherWork}
\begin{figure}[t]
	\begin{center}\scalebox{1}{\input{SalibrasPotato.pstex_t}}\end{center}
	\caption{The lattice of theories for the symmetric interaction combinators.}
	\label{fig:Theories}
\end{figure}
In \reffig{Theories} we graphically resume what we know about theories in the symmetric interaction combinators. A solid line represents inclusion, from bottom to top; a thick solid line represents atomic inclusion, \ie, there is no theory in between. The main open questions concerning \reffig{Theories} discussed up to now may be resumed as follows:
\begin{enumerate}[$\bullet$]
	\item Is $\FMEq$ included in $\BetaEpsilonEtaEq$? (We know that the converse does not hold).
	\item Is $\FMEq$ included in $\TotEq$? (We know that the converse does not hold).
	\item Does the lattice of consistent sensible theories have the cardinality of the continuum?
	\item Related to the above question: do easy nets exist?
\end{enumerate}
The list does not stop here, though: there are a few more open questions about theories for the symmetric interaction combinators, and, more in general, about the mathematical objects presented in this work.

In the light of the author's previous work on denotational semantics for the symmetric combinators~\cite{Mazza:CombSem}, and still drawing inspiration from the \lac, a first question we ask is: what about the theories generated by the models of the symmetric combinators? In particular, is any of the theories of \reffig{Theories} the theory of a model? Indeed, when we have a denotational semantics of the symmetric combinators (in the sense of \refsect{DenSem}), we automatically have a theory, given by denotational equality. Our full abstraction Theorems~\ref{th:TotEqFullAbs} and~\ref{th:ObsEqFullAbs} tell us for example that the theory of edifices and closed edifices is exactly $\FinAxEq$ and $\AxEq$, respectively. What about the denotational semantics based on interaction sets~\cite{Mazza:CombSem}? The examples we gave in that work can actually be shown to induce theories which are in between $\FinAxEq$ and $\AxEq$. However, we know that there exist interaction sets yielding fully abstract models for both of these theories; we prefer to keep this for further publication though.

What about the other theories of \reffig{Theories}? In the \lac, no non-syntactic model whose theory is $\beta$- or $\beta\eta$-equivalence is known. Ongoing work by Berline, Manzonetto and Salibra~\cite{BerlineManzonettoSalibra:1,BerlineManzonettoSalibra:2} suggests that there is a good reason for this: there is a sort of duality between the complexity of certain classes of non-syntactic models of the \lac\ and the complexity of the $\lambda$-theory that they generate: in particular, for such classes of models, recursively enumerable theories like $\beta$- or $\beta\eta$-equivalence may be obtained only by non-recursively-enumerable models. Of course we have no formal reason to believe that a similar phenomenon takes place in the case of the symmetric interaction combinators, but we suspect that finding non-syntactic fully abstract models of $\BetaEq$ and $\BetaEtaEq$ is not an easy task.

The situation may be different for $\BetaEpsilonEq$ (or $\BetaEpsilonEtaEq$). \reffig{PingPong} shows that there exists nets such that $\mu_1\not\BetaEpsilonEq\mu_2$ and yet $\AllObsPaths{\mu_1}=\AllObsPaths{\mu_2}$, so any semantics based on simply collecting observable axioms will not work. What would be needed is an additional structure to $\AllObsPaths{\mu}$, which takes into account the causal relationship between observable axioms. For instance, we may think of endowing $\AllObsPaths{\mu}$ with a poset structure, in the style of Winskel's event structures~\cite{Winskel}: given $\fx,\fy\in\AllObsPaths{\mu}$, $\fx\leq\fy$ iff $\mu\red\mu'$ and $\fy\in\ObsPaths{\mu'}$ imply $\fx\in\ObsPaths{\mu'}$. For example, consider the nets $\mu_1,\mu_2$ of \reffig{PingPong}. The observable axioms generated by these two nets fall within one of two categories: those based at free port $1$, whose addresses we denote by $\fx_1^1,\fx_2^1,\ldots$, and those based at free port $2$, whose addresses we denote by $\fx_1^2,\fx_2^2,\ldots$. Then, the structure of $\AllObsPaths{\mu_1}$ as a poset would be $\fx_1^1<\fx_1^2<\fx_2^1<\fx_2^2<\cdots$, while the structure of $\AllObsPaths{\mu_2}$ would be $\fx_1^2<\fx_1^1<\fx_2^2<\fx_2^1<\cdots$, which is enough to tell the two nets apart. We have not yet attempted to formalize these ideas, but we believe them to be a promising direction of research to obtain a full abstraction result for $\BetaEpsilonEq$ (or $\BetaEpsilonEtaEq$, which is perhaps more feasible---again, in the \lac, no non-syntactic model is known whose theory is $\cH$, which corresponds to our $\BetaEpsilonEq$).

Of course, there is also the question of semantically characterizing Fern\'andez and Mackie's $\FMEq$. We currently have no clue about this question, but if, as we discussed in the end of \refsect{Comparison}, this equivalence is akin to whnf-equivalence in the \lac, then we may be facing a difficult problem: in the \lac, no full abstraction result exists at present for such equivalence.

Concerning edifices, an aspect which should further be explored is their connection with games semantics and traced monoidal categories. Indeed, the trace operation on edifices is strikingly reminiscent of the notion of ``composition and hiding'' for composing strategies in games semantics, trace sequences representing plays. One may wonder whether the set of edifices presented in this paper can be seen as some sort of ``reflexive object'' in a traced monoidal category of edifices. This would be quite interesting, because it would open the way for introducing a typed version of the symmetric interaction combinators: types would be modeled by the objects of this category, and the set of edifices introduced here would appear as a special type capable of modeling untyped nets (in the context of the \lac, this would be like an object $D$ in a Cartesian closed category such that $D\Rightarrow D$ is a retract of $D$). Such considerations also bring forth the question of what is a categorical model of the symmetric interaction combinators, a question for which we have some clues, but which is still unsettled.

Finally, there is the intriguing possibility of using edifices as the basis for defining new non-deterministic extentions of the symmetric interaction combinators, or modeling existing non-deterministic systems. What we have in mind is something in the vein of Ehrhard and Regnier's differential interaction nets~\cite{EhrhardRegnier:DiffNets}. In fact, as soon as one considers non-simple edifices (\cf\ \refdef{SimpleEd}), several non-deterministic phenomena emerge: arches may superpose, \ie, they may share a pillar, which is reminiscent of additive slices in proof nets, or in differential interaction nets; and trace sequences, which represent computational paths in nets, are no longer uniquely determined by the arch they generate (\reflemma{UniqueSeq} fails).

\bibliographystyle{alpha}
\bibliography{Biblio}

\appendix
\section{Invariance of the Trace}
\label{app:TraceInv}
This appendix is devoted to the proof of \refprop{StabTrace}, which we recall below:

\smallskip
\noindent\textbf{\refprop{StabTrace}.} \emph{Let $\mu\onered\mu'$, and let $\mu=\Ctxt{\sigma}{\nu}$ and $\mu'=\Ctxt{\sigma'}{\nu'}$ according to the Decomposition \reflemma{Decomposition}. Then, $\Trace{\sigma}{\Ed\nu}=\Trace{\sigma'}{\Ed{\nu'}}$.}

\smallskip
The result is basically a corollary of the following:
\begin{lem}
	\label{lemma:StabTrace}
	Let $\nu$ and $\sigma$ be resp.\ the cut-free net and feedback drawn below:
	\begin{center}\scalebox{0.8}{\input{SimpleFeedback.pstex_t}}\end{center}
	Then:
	\begin{enumerate}
		\item if $\alpha=\beta$, and if $\nu'$ and $\sigma'$ are the following cut-free net and feedback
		\begin{center}\scalebox{0.8}{\input{SimpleFeedbackRed1.pstex_t}}\end{center}
		then $\Trace{\sigma'}{\nu'}=\Trace{\sigma}{\nu}$;
		\item if $\alpha\neq\beta$, and if $\nu'$ and $\sigma'$ are the following cut-free net and feedback
		\begin{center}\scalebox{0.8}{\input{SimpleFeedbackRed2.pstex_t}}\end{center}
		then $\Trace{\sigma'}{\nu'}=\Trace{\sigma}{\nu}$;
	\end{enumerate}
\end{lem}
\proof The proof is a bit technical, and not very interesting. We shall only address point (2), point (1) being analogous and technically simpler. We assume $\alpha=\D$ and $\beta=\Z$, the other configuration being perfectly symmetric, and start by showing the inclusion $\Trace{\sigma}{\nu}\subseteq\Trace{\sigma'}{\nu'}$. Let $\fs=(\Arch{\Pillar{u_k}{i_k}}{\Pillar{v_k}{j_k}})_{1\leq k\leq n}$ be a trace sequence of $\Ed\nu$ along $\sigma$, such that $i_1\not\in\dom\sigma$. We shall build by induction on $n$ a  trace sequence $\fs'=(\Arch{\Pillar{u_k'}{i_k'}}{\Pillar{v_k'}{j_k'}})_{1\leq k\leq n'}$ of $\Ed{\nu'}$ along $\sigma'$ such that:
\begin{enumerate}[$\bullet$]
	\item $\Pillar{u'_1}{i_1'}=\Pillar{u_1}{i_1}$;
	\item $\Pillar{v'_{n'}}{j'_{n'}}=\Pillar{v_n}{j_n}$ in case $j_n\not\in\dom\sigma$, \ie, in case $\fs$ is visible;
	\item otherwise, $j_n$ is one of the free ports denoted by $p_1,p_2$ in the above picture of $\Ctxt{\sigma}{\nu}$, and $v_n$ is of the form $t_hw$, where $1\leq h\leq 4$ and $t_1=\BiWord{\p}{\Id}$, $t_2=\BiWord{\q}{\Id}$, $t_3=\BiWord{\Id}{\p}$, $t_4=\BiWord{\Id}{\q}$, the value of $h$ depending on the cell and auxiliary port ``used'' by the observable axiom/path of $\nu$ inducing $\fs_n$. In that case, $\fs'$ will be such that $\Pillar{v'_{n'}}{j'_{n'}}=\Pillar{w}{q_h}$, where $q_h$ is one of the free ports of $\nu'$ as shown in the picture of $\Ctxt{\sigma'}{\nu'}$, point (2).
\end{enumerate}
It is obvious that the above is sufficient to prove the inclusion, because a visible trace sequences $\fs$ yield a visible trace sequence $\fs'$ such that $\fa(\fs')=\fa(\fs)$.

The base case is $n=1$, in which $\fs$ consists of a single arch $\Arch{\Pillar{u}{i}}{\Pillar{v}{j}}$. If $j\not\in\dom\sigma$, then $j$ is a free port of $\Ctxt{\sigma}{\nu}$, and $\fs$ is also a visible trace sequence of $\Ed{\nu'}$ along $\sigma'$, so we take $\fs'=\fs$. Otherwise, $j\in\{p_1,p_2\}$, and $v=t_hv'$, with $1\leq h\leq 4$ and $t_h$ as described above. In this case, the sequence $\fs'$ is defined to consist of the sole arch $\Arch{\Pillar{u}{i}}{\Pillar{\BiWord{x}{y}}{q_h}}$; this is clearly in $\Ed{\nu'}$, and $\fs'$ satisfies the desired requirements.

Let now $n>1$. We write $\fs_{n-1}=\Arch{\xi}{\Pillar{v_{n-1}}{j_{n-1}}}$, and observe that $j_{n-1}\in\dom\sigma$, because of the chain condition. Then, we have $j_{n-1}\in\{p_1,p_2\}$ and $v_n=t_hw$ for some $w\in\Cc\times\Cc$ and $1\leq h\leq 4$, with $t_h$ is as described above. We shall assume $h=1$; the other three cases are perfectly similar. So we have $j_{n-1}=p_1$, and the chain condition forces $i_n=\sigma(j_{n-1})=p_2$, so $\fs_n$ is of the form $\Arch{\Pillar{su}{p_2}}{\Pillar{v_n}{j_n}}$ for some $u\in\Cc\times\Cc$ and $s\in\{\BiWord{\Id}{\p},\BiWord{\Id}{\q}\}$. We make the choice $s=\BiWord{\Id}{\p}$, the other cases being again analogous. So, to resume, we know that the last two arches of $\fs$ are of the form
{\setlength\arraycolsep{2pt}\begin{eqnarray*}
	\fs_{n-1}&=&\Arch{\xi}{\Pillar{\BiWord{\p x}{y}}{p_1}},\\
	\fs_n&=&\Arch{\Pillar{\BiWord{x'}{\p y'}}{p_2}}{\Pillar{v_n}{j_n}},
\end{eqnarray*}}
for some $x,y,x',y'\in\Cc$ and $\xi\in\cP$; observe that the match condition implies $x'=\p x$ and $y=\p y'$.

Now, by the induction hypothesis applied to $\fs_1,\ldots,\fs_{n-1}$, we know how to build a sequence $\fs'=(\Arch{\Pillar{u_k'}{i_k'}}{\Pillar{v_k'}{j_k'}})_{1\leq k\leq n'}$ such that $\Pillar{u_1'}{i_1'}=\Pillar{u_1}{i_1}$ and such that the last arch is of the form
$$\fs'_{n'}=\Arch{\xi'}{\Pillar{\BiWord{x}{y}}{q_1}}$$
Remark that $\sigma'(q_1)=q_8$, $\sigma'(q_6)=q_3$, and that
$$\Arch{\Pillar{\BiWord{x''}{\p y''}}{q_8}}{\Pillar{\BiWord{\p x''}{y''}}{q_6}}\in\Ed{\nu'}$$
for all $x'',y''\in\Cc$.
We now have two possibilities:
\begin{enumerate}[$\bullet$]
	\item $j_n\not\in\dom\sigma$, \ie, $j_n$ is a free port of $\Ctxt{\sigma}{\nu}$ and $\Ctxt{\sigma'}{\nu'}$. In this case, $\fs_n\in\Ed\nu$ implies $\Arch{\Pillar{\BiWord{x'}{y'}}{q_3}}{\Pillar{v_n}{j_n}}\in\Ed{\nu'}$;
	\item $j_n\in\dom\sigma$, which implies $j_n\in\{p_1,p_2\}$. We assume $j_n=p_1$, again the case $j_n=p_2$ being perfectly similar. Then, we have $v_n=\BiWord{\p z}{z'}$ for some $z,z',\in\Cc$, which implies $\Arch{\Pillar{\BiWord{x'}{y'}}{q_3}}{\Pillar{\BiWord{z}{z'}}{p_1}}\in\Ed{\nu'}$.
\end{enumerate}
Then, define
$$\fs'_{n'+1}=\Arch{\Pillar{\BiWord{x}{\p y'}}{q_8}}{\Pillar{\BiWord{\p x}{y'}}{q_6}},$$
and
$$\fs'_{n'+2}=\Arch{\Pillar{\BiWord{x'}{y'}}{q_3}}{\Pillar{v_n}{j_n}}$$
in case $j_n\not\in\dom\sigma$, and
$$\fs'_{n'+2}=\Arch{\Pillar{\BiWord{x'}{y'}}{q_3}}{\Pillar{\BiWord{z}{z'}}{p_1}}$$
in case $j_n\in\dom\sigma$. In both cases, by the arguments given above we have $\fs'_{n'+1},\fs'_{n'+2}\in\Ed{\nu'}$, and $(\fs_k)_{1\leq k\leq n''+2}$ is a trace sequence of $\Ed{\nu'}$ along $\sigma'$ satisfying the desired requirements.

We are left with proving that $\Trace{\sigma'}{\Ed{\nu'}}\subseteq\Trace{\sigma}{\Ed\nu}$. We use a similar argument, but this time we build a trace sequence $\fs$ of $\Ed\nu$ along $\sigma$ only starting from a trace sequence $\fs'=(\Arch{\Pillar{u'_k}{i'_k}}{\Pillar{v'_k}{j'_k}})_{1\leq k\leq n'}$ of $\Ed{\nu'}$ along $\sigma'$ such that $j'_{n'}\not\in\{q_5,q_6,q_7,q_8\}$ (the induction is on $n'$). This will be enough for the inclusion to be proved, because visible trace sequences of $\Ed{\nu'}$ along $\sigma'$ do not end with any of those free ports of $\nu'$, as they are not free in $\Ctxt{\sigma'}{\nu'}$. The sequence $\fs=(\Arch{\Pillar{u_k}{i_k}}{\Pillar{v_k}{j_k}})_{1\leq k\leq n}$ will have to satisfy the following:
\begin{enumerate}[$\bullet$]
	\item $\Pillar{u_1}{i_1}=\Pillar{u'_1}{i'_1}$;
	\item $\Pillar{v_n}{j_n}=\Pillar{v'_{n'}}{j'_{n'}}$ if $j'_{n'}\not\in\dom\sigma'$, \ie, in case $\fs'$ is visible;
	\item otherwise, we must have $j'_{n'}\in\{q_1,q_2,q_3,q_4\}$; then, $\fs$ will satisfy $\Pillar{v_n}{j_n}=\Pillar{sv'_{n'}}{p_1}$ with $s=\BiWord{\p}{\Id}$ (resp.\ $s=\BiWord{\q}{\Id}$) if $j'_{n'}=q_1$ (resp.\ $j'_{n'}=q_2$), or $\Pillar{v_n}{j_n}=\Pillar{sv'_{n'}}{p_2}$ with $s=\BiWord{\Id}{\p}$ (resp.\ $s=\BiWord{\Id}{\q}$) if $j'_{n'}=q_3$ (resp.\ $j'_{n'}=q_4$).
\end{enumerate}

The base case is $n'=1$, in which $\fs'$ consists of exactly one arch $\Arch{\Pillar{u'}{i'}}{\Pillar{v'}{j'}}$, and obviously $j'\not\in\{q_5,q_6,q_7,q_8\}$, because $i'\not\in\dom\sigma'$. If $j'\not\in\dom\sigma'$, we take $\fs=\fs'$. Otherwise, supposing $j'=q_1$, we take $\fs$ to be made of the sole arch $\Arch{\Pillar{u'}{i'}}{\Pillar{(\BiWord{\p}{\Id})v'}{p_1}}$; the other three possible values of $j'$ are handled similarly, prefixing $v'$ with $\BiWord{\q}{\Id},\BiWord{\Id}{\p},\BiWord{\Id}{\q}$ as appropriate.

Let now $n'>1$. We put $j=j'_{n'}$, and observe that, by $j\not\in\{q_5,q_6,q_7,q_8\}$ and by the chain condition, we have $i'_{n'}\in\{q_1,q_2,q_3,q_4\}$. We have again four cases; as above, each time we shall have a choice in the sequel, we shall analyze only one arbitrary case, all cases being easily recoverable every time from each other. So we assume, for instance, $i'_{n'}=q_1$. We have $j'_{n'-1}=\sigma'(q_1)=q_8$, so $\fs'_{n'-1}$ is based at $q,q_8$, where $q\in\{q_5,q_6\}$; in both cases, $q$ is not a free port of $\Ctxt{\sigma'}{\nu'}$, so the sequence must contain a previous arch of the form $\fs'_{n'-2}=\Arch{\xi'}{\Pillar{v'}{q'}}$, with $q'\in\{q_1,q_2,q_3,q_4\}$. We choose $q=q_6$ and $q'=q_3$, so we can write, by using the match condition, that the last three arches of $\fs'$ are of the form
{\setlength\arraycolsep{2pt}\begin{eqnarray*}
	\fs'_{n'-2} &=& \Arch{\xi}{\Pillar{\BiWord{x}{y}}{q_3}}, \\
	\fs'_{n'-1} &=& \Arch{\Pillar{\BiWord{\p x'}{y'}}{q_6}}{\Pillar{\BiWord{x'}{\p y'}}{q_8}}, \\
	\fs'_{n'} &=& \Arch{\Pillar{\BiWord{x''}{y''}}{q_1}}{\Pillar{v}{j}},
\end{eqnarray*}}
for some $x,y,x',y',x'',y''\in\Cc$, $v\in\Cc\times\Cc$, and $\xi\in\cP$. Furthermore, by the match condition, we know that $x=\p x'$, $y=y'$, $x''=x'$, and $y''=\p y'$. If we apply the induction hypothesis to the sequence $\fs'_1,\ldots,\fs'_{n'-2}$, we obtain a sequence $\fs=(\Arch{\Pillar{u_1}{i_1}}{\Pillar{v_n}{j_n}})_{1\leq k\leq n}$ satisfying all the requirements mentioned above; in particular, we have
$$\fs_n=\Arch{\xi'}{\Pillar{\BiWord{x}{\p y}}{p_2}}.$$
Now, suppose $j\in\dom\sigma'$; we choose for example $j=q_1$. In that case, define
$$\fs_{n+1}=\Arch{\Pillar{\BiWord{\p x''}{y''}}{p_1}}{\Pillar{(\BiWord{\q}{\Id})v}{p_1}}.$$
Otherwise, $j$ is a free port of $\Ctxt{\sigma'}{\nu'}$; then, we set
$$\fs_{n+1}=\Arch{\Pillar{\BiWord{\p x''}{y''}}{p_1}}{\Pillar{v}{j}}.$$
In both cases, it is easy to see that $\fs_{n+1}\in\Ed\nu$, and that $\fs_1,\ldots,\fs_n,\fs_{n+1}$ is a trace sequence of $\Ed\nu$ along $\sigma$, which is visible iff $\fs'$ is.\qed

To prove \refprop{StabTrace}, observe first of all that trace sequences never use $\E$ cells, so the only interesting interaction rules are those addressed by \reflemma{StabTrace}. Then, we can always write $\Ctxt{\sigma}{\nu}=\Ctxt{\sigma_1}{\Ctxt{\sigma_0}{\nu}}$, where $\Ctxt{\sigma_0}{\nu}$ is of the form given in \reflemma{StabTrace}; similarly, we can write $\Ctxt{\sigma'}{\nu'}=\Ctxt{\sigma_1}{\Ctxt{\sigma'_0}{\nu'}}$, where $\Ctxt{\sigma_0'}{\nu'}$ is of one of the forms given in point (1) or (2) of \reflemma{StabTrace}, depending on whether the interaction rule is an annihilation or commutation. Note that $\sigma_1,\sigma_0$ and $\sigma_1,\sigma_0'$ are disjoint feedbacks. Then, by \reflemma{StabTrace} and the associativity of the trace (\reflemma{TraceAssoc}), we have
$$\Trace{\sigma}{\Ed\nu}=\Trace{\sigma_1}{\Trace{\sigma_0}{\Ed\nu}}=\Trace{\sigma_1}{\Trace{\sigma_0'}{\Ed{\nu'}}}=\Trace{\sigma'}{\Ed{\nu'}}.$$

\end{document}